\RequirePackage{ifpdf}

\documentclass[a4paper,11pt]{article}
\pdfoutput=1 % if your are submitting a pdflatex (i.e. if you have
\usepackage{jheppub} % for details on the use of the package, please
\usepackage[T1]{fontenc} % if needed

\usepackage{ifpdf} % to get it to work with pdflatex and not break for everything else

\title{Two-Loop Master Integrals for $q \bar{q} \to V V$: the Planar Topologies}

\author{Thomas Gehrmann,}
\author{Lorenzo Tancredi}
\author{and Erich Weihs}

\affiliation{Institut f\"ur Theoretische Physik, Universit\"at Z\"urich, Wintherturerstrasse 190,\\CH-8057 Z\"urich, Switzerland}

\emailAdd{thomas.gehrmann@uzh.ch}
\emailAdd{tancredi@physik.uzh.ch}
\emailAdd{erich.weihs@physik.uzh.ch}


\title{Two-Loop Master Integrals for $q \bar{q} \to V V$: the Planar Topologies}
\abstract{The two-loop QCD corrections to vector boson pair production at hadron colliders 
involve a new class of Feynman integrals: two-loop four-point functions with two off-shell 
external legs. We describe their reduction to a small set of master integrals
by solving linear relations among them. We then use differential equations 
in the external invariants to compute all master integrals that are relevant to planar Feynman 
amplitudes. Our results are expressed analytically in terms of generalized harmonic polylogarithms.
The calculation %of the master integrals 
relies heavily on techniques that 
exploit the algebraic structure of these functions, which we describe in detail.}

\ifpdf
\fi

\def\Q2{\left(Q^{2}\right)}

\def\d{{\rm d}}

\def\l({\left(} 
\def\r){\right)}

\def\d{\hbox{d}}

\newcommand{\be}{\begin{equation}}
\newcommand{\ee}{\end{equation}}
\newcommand{\bea}{\begin{eqnarray}}
\newcommand{\eea}{\end{eqnarray}}
\newcommand{\T}{\mathcal{T}}
\newcommand{\B}{\mathcal{B}}
\newcommand{\I}{\mathcal{I}}

\def\bsp#1\esp{\begin{split}#1\end{split}}
\newcommand{\li}{\textrm{Li}}
\newcommand{\id}{\textrm{id}}
\def\dt{dt}
\def\C{\mathbb{C}}

\def\S{\mathcal{S}}
\def\N{\mathbb{N}}

\newcommand{\cH}{{\cal H}}

\newcommand{\triangleoneloop}[3]{
\mbox{\parbox{3cm}{\hspace{0.25cm}
\begin{picture}(2.5,1.4)
\put(0.3,0.7){\vector(1,0){0.1}}
\put(1.9,0.2){\vector(1,0){0.1}}
\put(1.9,1.2){\vector(1,0){0.1}}
\put(0.5,0.7){\line(2,1){1}}
\put(0.5,0.7){\line(2,-1){1}}
\put(1.5,1.2){\line(0,-1){1}}
\linethickness{0.35mm}
\put(1.5,1.2){\line(1,0){0.5}}
\put(0,0.7){\line(1,0){0.5}}
\put(1.5,0.2){\line(1,0){0.5}}
\thinlines
\put(0.25,0.9){\makebox(0,0)[b]{$#1$}}
\put(2.05,1.2){\makebox(0,0)[l]{$#2$}}
\put(2.05,0.2){\makebox(0,0)[l]{$#3$}}
\end{picture}
}}
\hfill}

\newcommand{\bubbleoneloop}[1]{
\mbox{\parbox{2.5cm}{\hspace{0.25cm}
\begin{picture}(2,1)
\put(0.3,0.5){\vector(1,0){0.1}}
\linethickness{0.35mm}
\put(0,0.5){\line(1,0){0.5}}
\put(1.5,0.5){\line(1,0){0.5}}
\thinlines
\put(1,0.5){\circle{1}}
\put(0.25,0.7){\makebox(0,0)[b]{$#1$}}
\end{picture}
}}
\hfill}

\newcommand{\bubbleNLO}[1]{
\mbox{\parbox{2.5cm}{\hspace{0.25cm}
\begin{picture}(2,1)
\put(0.3,0.5){\vector(1,0){0.1}}
\put(0.5,0.5){\line(1,0){1}}
\linethickness{0.35mm}
\put(0,0.5){\line(1,0){0.5}}
\put(1.5,0.5){\line(1,0){0.5}}
\thinlines
\put(1,0.5){\circle{1}}
\put(0.25,0.7){\makebox(0,0)[b]{$#1$}}
\end{picture}
}}
\hfill}

\newcommand{\triangleOne}[3]{
\mbox{\parbox{3cm}{\hspace{0.25cm}
\begin{picture}(2.5,1.4)
\put(0.3,0.7){\vector(1,0){0.1}}
\put(1.7,0.2){\vector(1,0){0.1}}
\put(1.7,1.2){\vector(1,0){0.1}}
\linethickness{0.35mm}
\put(0,0.7){\line(1,0){0.5}}
\thinlines
\put(1,1.2){\line(0,-1){1}}
\put(1,1.2){\line(1,0){1}}
\put(1,0.2){\line(1,0){1}}
\put(1,0.7){\circle{1}}
\put(0.25,0.9){\makebox(0,0)[b]{$#1$}}
\put(2.05,1.2){\makebox(0,0)[l]{$#2$}}
\put(2.05,0.2){\makebox(0,0)[l]{$#3$}}
\end{picture}
}}
\hfill}

\newcommand{\triangleTwo}[3]{
\mbox{\parbox{3cm}{\hspace{0.25cm}
\begin{picture}(2.5,1.4)
\put(0.3,0.7){\vector(1,0){0.1}}
\put(1.7,0.2){\vector(1,0){0.1}}
\put(1.7,1.2){\vector(1,0){0.1}}
\put(1,1.2){\line(0,-1){1}}
\put(1,1.2){\line(1,0){1}}
\linethickness{0.35mm}
\put(0,0.7){\line(1,0){0.5}}
\put(1,0.2){\line(1,0){1}}
\thinlines
\put(1,0.7){\circle{1}}
\put(0.25,0.9){\makebox(0,0)[b]{$#1$}}
\put(2.05,1.2){\makebox(0,0)[l]{$#2$}}
\put(2.05,0.2){\makebox(0,0)[l]{$#3$}}
\end{picture}
}}
\hfill}

\newcommand{\triangleTwox}[3]{
\mbox{\parbox{3cm}{\hspace{0.25cm}
\begin{picture}(2.5,1.4)
\put(0.3,0.7){\vector(1,0){0.1}}
\put(1.7,0.2){\vector(1,0){0.1}}
\put(1.7,1.2){\vector(1,0){0.1}}
\put(0.5,1.2){\oval(1,1)[br]}
\put(1,1.2){\line(1,0){1}}
\linethickness{0.35mm}
\put(1,0.2){\line(1,0){1}}
\put(0,0.7){\line(1,0){0.5}}
\thinlines
\put(1,0.7){\circle{1}}
\put(0.25,0.9){\makebox(0,0)[b]{$#1$}}
\put(2.05,1.2){\makebox(0,0)[l]{$#2$}}
\put(2.05,0.2){\makebox(0,0)[l]{$#3$}}
\end{picture}
}}
\hfill}

\newcommand{\triangleThree}[3]{
\mbox{\parbox{3cm}{\hspace{0.25cm}
\begin{picture}(2.5,1.4)
\put(0.3,0.7){\vector(1,0){0.1}}
\put(1.7,0.2){\vector(1,0){0.1}}
\put(1.7,1.2){\vector(1,0){0.1}}
\linethickness{0.35mm}
\put(0,0.7){\line(1,0){0.5}}
\put(1,1.2){\line(1,0){1}}
\put(1,0.2){\line(1,0){1}}
\thinlines
\put(1,1.2){\line(0,-1){1}}
\put(1,0.7){\circle{1}}
\put(0.25,0.9){\makebox(0,0)[b]{$#1$}}
\put(2.05,1.2){\makebox(0,0)[l]{$#2$}}
\put(2.05,0.2){\makebox(0,0)[l]{$#3$}}
\end{picture}
}}
\hfill}

\newcommand{\triangleThreedot}[3]{
\mbox{\parbox{3cm}{\hspace{0.25cm}
\begin{picture}(2.5,1.4)
\put(0.3,0.7){\vector(1,0){0.1}}
\put(1.7,0.2){\vector(1,0){0.1}}
\put(1.7,1.2){\vector(1,0){0.1}}
\linethickness{0.35mm}
\put(0,0.7){\line(1,0){0.5}}
\put(1,1.2){\line(1,0){1}}
\put(1,0.2){\line(1,0){1}}
\thinlines
\put(1,1.2){\line(0,-1){1}}
\put(1.0,0.7){\circle*{0.2}}
\put(1,0.7){\circle{1}}
\put(0.25,0.9){\makebox(0,0)[b]{$#1$}}
\put(2.05,1.2){\makebox(0,0)[l]{$#2$}}
\put(2.05,0.2){\makebox(0,0)[l]{$#3$}}
\end{picture}
}}
\hfill}

\newcommand{\triangleThreex}[3]{
\mbox{\parbox{3cm}{\hspace{0.25cm}
\begin{picture}(2.5,1.4)
\put(0.3,0.7){\vector(1,0){0.1}}
\put(1.7,0.2){\vector(1,0){0.1}}
\put(1.7,1.2){\vector(1,0){0.1}}
\linethickness{0.35mm}
\put(0,0.7){\line(1,0){0.5}}
\put(1,1.2){\line(1,0){1}}
\put(1,0.2){\line(1,0){1}}
\thinlines
\put(0.5,1.2){\oval(1,1)[br]}
\put(1,0.7){\circle{1}}
\put(0.25,0.9){\makebox(0,0)[b]{$#1$}}
\put(2.05,1.2){\makebox(0,0)[l]{$#2$}}
\put(2.05,0.2){\makebox(0,0)[l]{$#3$}}
\end{picture}
}}
\hfill}

\newcommand{\triangleThreexdot}[3]{
\mbox{\parbox{3cm}{\hspace{0.25cm}
\begin{picture}(2.5,1.4)
\put(0.3,0.7){\vector(1,0){0.1}}
\put(1.7,0.2){\vector(1,0){0.1}}
\put(1.7,1.2){\vector(1,0){0.1}}
\linethickness{0.35mm}
\put(0,0.7){\line(1,0){0.5}}
\put(1,1.2){\line(1,0){1}}
\put(1,0.2){\line(1,0){1}}
\thinlines
\put(0.5,1.2){\oval(1,1)[br]}
\put(0.85,0.85){\circle*{0.2}}
\put(1,0.7){\circle{1}}
\put(0.25,0.9){\makebox(0,0)[b]{$#1$}}
\put(2.05,1.2){\makebox(0,0)[l]{$#2$}}
\put(2.05,0.2){\makebox(0,0)[l]{$#3$}}
\end{picture}
}}
\hfill}

\newcommand{\trianglebTwo}[3]{
\mbox{\parbox{3cm}{\hspace{0.25cm}
\begin{picture}(2.5,1.4)
\put(0.3,0.7){\vector(1,0){0.1}}
\put(1.9,0.2){\vector(1,0){0.1}}
\put(1.9,1.2){\vector(1,0){0.1}}
\put(0.5,0.7){\line(2,1){1}}
\put(0.5,0.7){\line(2,-1){1}}
\put(1.5,1.2){\line(0,-1){1}}
\put(1.5,1.2){\line(1,0){0.5}}
\linethickness{0.35mm}
\put(0,0.7){\line(1,0){0.5}}
\put(1.5,0.2){\line(1,0){0.5}}
\thinlines
\put(1.1,0.4){\line(1,2){0.4}}
\put(0.25,0.9){\makebox(0,0)[b]{$#1$}}
\put(2.05,1.2){\makebox(0,0)[l]{$#2$}}
\put(2.05,0.2){\makebox(0,0)[l]{$#3$}}
\end{picture}
}}
\hfill}

\newcommand{\trianglebThree}[3]{
\mbox{\parbox{3cm}{\hspace{0.25cm}
\begin{picture}(2.5,1.4)
\put(0.3,0.7){\vector(1,0){0.1}}
\put(1.9,0.2){\vector(1,0){0.1}}
\put(1.9,1.2){\vector(1,0){0.1}}
\put(0.5,0.7){\line(2,1){1}}
\put(0.5,0.7){\line(2,-1){1}}
\put(1.5,1.2){\line(0,-1){1}}
\linethickness{0.35mm}
\put(0,0.7){\line(1,0){0.5}}
\put(1.5,1.2){\line(1,0){0.5}}
\put(1.5,0.2){\line(1,0){0.5}}
\thinlines
\put(1.5,0.2){\line(-1,2){0.4}}
\put(0.25,0.9){\makebox(0,0)[b]{$#1$}}
\put(2.05,1.2){\makebox(0,0)[l]{$#2$}}
\put(2.05,0.2){\makebox(0,0)[l]{$#3$}}
\end{picture}
}}
\hfill}

\newcommand{\trianglebThreex}[3]{
\mbox{\parbox{3cm}{\hspace{0.25cm}
\begin{picture}(2.5,1.4)
\put(0.3,0.7){\vector(1,0){0.1}}
\put(1.9,0.2){\vector(1,0){0.1}}
\put(1.9,1.2){\vector(1,0){0.1}}
\put(0.5,0.7){\line(2,1){1}}
\put(0.5,0.7){\line(2,-1){1}}
\put(1.5,1.2){\line(0,-1){1}}
\linethickness{0.35mm}
\put(0,0.7){\line(1,0){0.5}}
\put(1.5,1.2){\line(1,0){0.5}}
\put(1.5,0.2){\line(1,0){0.5}}
\thinlines
\put(1.1,0.4){\line(1,2){0.4}}
\put(0.25,0.9){\makebox(0,0)[b]{$#1$}}
\put(2.05,1.2){\makebox(0,0)[l]{$#2$}}
\put(2.05,0.2){\makebox(0,0)[l]{$#3$}}
\end{picture}
}}
\hfill}

\newcommand{\boxbubblea}[4]{
\mbox{\parbox{3.5cm}{\hspace{0.25cm}
\begin{picture}(3,1.4)
\put(0.8,0.2){\vector(1,0){0.1}}
\put(2.3,0.2){\vector(1,0){0.1}}
\put(2.3,1.2){\vector(1,0){0.1}}
\put(0.8,1.2){\vector(1,0){0.1}}
\put(1.5,1.2){\oval(1,1)[b]}
\put(0.5,0.2){\line(1,0){1.5}}
\put(0.5,1.2){\line(1,0){1.5}}
\linethickness{0.35mm}
\put(2.0,0.2){\line(1,0){0.5}}
\put(2.0,1.2){\line(1,0){0.5}}
\thinlines
\put(1,0.2){\line(0,1){1}}
\put(2,0.2){\line(0,1){1}}
\put(0.45,1.2){\makebox(0,0)[r]{$#1$}}
\put(0.45,0.2){\makebox(0,0)[r]{$#2$}}
\put(2.55,1.2){\makebox(0,0)[l]{$#3$}}
\put(2.55,0.2){\makebox(0,0)[l]{$#4$}}
\end{picture}
}} 
\hfill}

\newcommand{\boxbubbleb}[4]{
\mbox{\parbox{3.5cm}{\hspace{0.25cm}
\begin{picture}(3,1.4)
\put(0.8,0.2){\vector(1,0){0.1}}
\put(2.3,0.2){\vector(1,0){0.1}}
\put(2.3,1.2){\vector(1,0){0.1}}
\put(0.8,1.2){\vector(1,0){0.1}}
\put(2.0,0.7){\oval(1,1)[l]}
\put(0.5,0.2){\line(1,0){1.5}}
\put(0.5,1.2){\line(1,0){1.5}}
\linethickness{0.35mm}
\put(2.0,0.2){\line(1,0){0.5}}
\put(2.0,1.2){\line(1,0){0.5}}
\thinlines
\put(1,0.2){\line(0,1){1}}
\put(2,0.2){\line(0,1){1}}
\put(0.45,1.2){\makebox(0,0)[r]{$#1$}}
\put(0.45,0.2){\makebox(0,0)[r]{$#2$}}
\put(2.55,1.2){\makebox(0,0)[l]{$#3$}}
\put(2.55,0.2){\makebox(0,0)[l]{$#4$}}
\end{picture}
}} 
\hfill}

\newcommand{\boxbubblec}[4]{
\mbox{\parbox{3.5cm}{\hspace{0.25cm}
\begin{picture}(3,1.4)
\put(0.8,0.2){\vector(1,0){0.1}}
\put(2.3,0.2){\vector(1,0){0.1}}
\put(2.3,1.2){\vector(1,0){0.1}}
\put(0.8,1.2){\vector(1,0){0.1}}
\put(1.0,0.7){\oval(1,1)[r]}
\put(0.5,0.2){\line(1,0){1.5}}
\put(0.5,1.2){\line(1,0){1.5}}
\linethickness{0.35mm}
\put(2.0,0.2){\line(1,0){0.5}}
\put(2.0,1.2){\line(1,0){0.5}}
\thinlines
\put(1,0.2){\line(0,1){1}}
\put(2,0.2){\line(0,1){1}}
\put(0.45,1.2){\makebox(0,0)[r]{$#1$}}
\put(0.45,0.2){\makebox(0,0)[r]{$#2$}}
\put(2.55,1.2){\makebox(0,0)[l]{$#3$}}
\put(2.55,0.2){\makebox(0,0)[l]{$#4$}}
\end{picture}
}} 
\hfill}

\newcommand{\boxbubblecdot}[4]{
\mbox{\parbox{3.5cm}{\hspace{0.25cm}
\begin{picture}(3,1.4)
\put(0.8,0.2){\vector(1,0){0.1}}
\put(2.3,0.2){\vector(1,0){0.1}}
\put(2.3,1.2){\vector(1,0){0.1}}
\put(0.8,1.2){\vector(1,0){0.1}}
\put(1.0,0.7){\oval(1,1)[r]}
\put(0.5,0.2){\line(1,0){1.5}}
\put(0.5,1.2){\line(1,0){1.5}}
\linethickness{0.35mm}
\put(2.0,0.2){\line(1,0){0.5}}
\put(2.0,1.2){\line(1,0){0.5}}
\thinlines
\put(1,0.2){\line(0,1){1}}
\put(2,0.2){\line(0,1){1}}
\put(1.5,0.7){\circle*{0.2}}
\put(0.45,1.2){\makebox(0,0)[r]{$#1$}}
\put(0.45,0.2){\makebox(0,0)[r]{$#2$}}
\put(2.55,1.2){\makebox(0,0)[l]{$#3$}}
\put(2.55,0.2){\makebox(0,0)[l]{$#4$}}
\end{picture}
}} 
\hfill}

\newcommand{\boxxbNLO}[4]{
\mbox{\parbox{3.5cm}{\hspace{0.25cm}
\begin{picture}(3,1.4)
\put(0.8,0.2){\vector(1,0){0.1}}
\put(2.3,0.2){\vector(1,0){0.1}}
\put(2.3,1.2){\vector(1,0){0.1}}
\put(0.8,1.2){\vector(1,0){0.1}}
\put(1,0.2){\line(1,1){1}}
\put(0.5,0.2){\line(1,0){1.5}}
\put(0.5,1.2){\line(1,0){1.5}}
\linethickness{0.35mm}
\put(2.0,0.2){\line(1,0){0.5}}
\put(2.0,1.2){\line(1,0){0.5}}
\thinlines
\put(1,0.2){\line(0,1){1}}
\put(2,0.2){\line(0,1){1}}
\put(0.45,1.2){\makebox(0,0)[r]{$#1$}}
\put(0.45,0.2){\makebox(0,0)[r]{$#2$}}
\put(2.55,1.2){\makebox(0,0)[l]{$#3$}}
\put(2.55,0.2){\makebox(0,0)[l]{$#4$}}
\end{picture}
}} 
\hfill}

\newcommand{\boxxbdotNLO}[4]{
\mbox{\parbox{3.5cm}{\hspace{0.25cm}
\begin{picture}(3,1.4)
\put(0.8,0.2){\vector(1,0){0.1}}
\put(2.3,0.2){\vector(1,0){0.1}}
\put(2.3,1.2){\vector(1,0){0.1}}
\put(0.8,1.2){\vector(1,0){0.1}}
\put(1,0.2){\line(1,1){1}}
\put(0.5,0.2){\line(1,0){1.5}}
\put(0.5,1.2){\line(1,0){1.5}}
\linethickness{0.35mm}
\put(2.0,0.2){\line(1,0){0.5}}
\put(2.0,1.2){\line(1,0){0.5}}
\thinlines
\put(2.0,0.7){\circle*{0.2}}
\put(1,0.2){\line(0,1){1}}
\put(2,0.2){\line(0,1){1}}
\put(0.45,1.2){\makebox(0,0)[r]{$#1$}}
\put(0.45,0.2){\makebox(0,0)[r]{$#2$}}
\put(2.55,1.2){\makebox(0,0)[l]{$#3$}}
\put(2.55,0.2){\makebox(0,0)[l]{$#4$}}
\end{picture}
}} 
\hfill}

\newcommand{\boxxbpaNLO}[4]{
\mbox{\parbox{3.5cm}{\hspace{0.25cm}
\begin{picture}(3,1.4)
\put(0.8,0.2){\vector(1,0){0.1}}
\put(2.3,0.2){\vector(1,0){0.1}}
\put(2.3,1.2){\vector(1,0){0.1}}
\put(0.8,1.2){\vector(1,0){0.1}}
\put(0.5,0.2){\line(1,0){1.5}}
\put(0.5,1.2){\line(1,0){1.5}}
\linethickness{0.35mm}
\put(2.0,0.2){\line(1,0){0.5}}
\put(2.0,1.2){\line(1,0){0.5}}
\thinlines
\put(1.5,1.2){\line(1,-2){0.5}}
\put(1,0.2){\line(0,1){1}}
\put(2,0.2){\line(0,1){1}}
\put(0.45,1.2){\makebox(0,0)[r]{$#1$}}
\put(0.45,0.2){\makebox(0,0)[r]{$#2$}}
\put(2.55,1.2){\makebox(0,0)[l]{$#3$}}
\put(2.55,0.2){\makebox(0,0)[l]{$#4$}}
\end{picture}
}} 
\hfill}

\newcommand{\boxxbpbNLO}[4]{
\mbox{\parbox{3.5cm}{\hspace{0.25cm}
\begin{picture}(3,1.4)
\put(0.8,0.2){\vector(1,0){0.1}}
\put(2.3,0.2){\vector(1,0){0.1}}
\put(2.3,1.2){\vector(1,0){0.1}}
\put(0.8,1.2){\vector(1,0){0.1}}
\put(0.5,0.2){\line(1,0){1.5}}
\put(0.5,1.2){\line(1,0){1.5}}
\linethickness{0.35mm}
\put(2.0,0.2){\line(1,0){0.5}}
\put(2.0,1.2){\line(1,0){0.5}}
\thinlines
\put(1.0,1.2){\line(2,-1){1}}
\put(1,0.2){\line(0,1){1}}
\put(2,0.2){\line(0,1){1}}
\put(0.45,1.2){\makebox(0,0)[r]{$#1$}}
\put(0.45,0.2){\makebox(0,0)[r]{$#2$}}
\put(2.55,1.2){\makebox(0,0)[l]{$#3$}}
\put(2.55,0.2){\makebox(0,0)[l]{$#4$}}
\end{picture}
}} 
\hfill}

\newcommand{\doubleboxa}[4]{
\mbox{\parbox{3.5cm}{\hspace{0.25cm}
\begin{picture}(3,1.4)
\put(0.8,0.2){\vector(1,0){0.1}}
\put(2.3,0.2){\vector(1,0){0.1}}
\put(2.3,1.2){\vector(1,0){0.1}}
\put(0.8,1.2){\vector(1,0){0.1}}
\put(0.5,0.2){\line(1,0){1.5}}
\put(0.5,1.2){\line(1,0){1.5}}
\linethickness{0.35mm}
\put(2.0,0.2){\line(1,0){0.5}}
\put(2.0,1.2){\line(1,0){0.5}}
\thinlines
\put(1,0.2){\line(0,1){1}}
\put(1.5,0.2){\line(0,1){1}}
\put(2,0.2){\line(0,1){1}}
\put(0.45,1.2){\makebox(0,0)[r]{$#1$}}
\put(0.45,0.2){\makebox(0,0)[r]{$#2$}}
\put(2.55,1.2){\makebox(0,0)[l]{$#3$}}
\put(2.55,0.2){\makebox(0,0)[l]{$#4$}}
\end{picture}
}}
\hfill}

\newcommand{\doubleboxatwo}[4]{
\mbox{\parbox{3.5cm}{\hspace{0.25cm}
\begin{picture}(3,1.4)
\put(0.8,0.2){\vector(1,0){0.1}}
\put(2.3,0.2){\vector(1,0){0.1}}
\put(2.3,1.2){\vector(1,0){0.1}}
\put(0.8,1.2){\vector(1,0){0.1}}
\put(0.5,0.2){\line(1,0){1.5}}
\put(0.5,1.2){\line(1,0){1.5}}
\linethickness{0.35mm}
\put(2.0,0.2){\line(1,0){0.5}}
\put(2.0,1.2){\line(1,0){0.5}}
\thinlines
\put(1,0.2){\line(0,1){1}}
\put(1.5,0.2){\line(0,1){1}}
\put(2,0.2){\line(0,1){1}}
\put(0.45,1.2){\makebox(0,0)[r]{$#1$}}
\put(0.45,0.2){\makebox(0,0)[r]{$#2$}}
\put(2.55,1.2){\makebox(0,0)[l]{$#3$}}
\put(2.55,0.2){\makebox(0,0)[l]{$#4$}}
\put(1.075,1.0){\makebox(0,0)[l]{$_{(2)}$}}
\end{picture}
}}
\hfill}

\newcommand{\doubleboxathree}[4]{
\mbox{\parbox{3.5cm}{\hspace{0.25cm}
\begin{picture}(3,1.4)
\put(0.8,0.2){\vector(1,0){0.1}}
\put(2.3,0.2){\vector(1,0){0.1}}
\put(2.3,1.2){\vector(1,0){0.1}}
\put(0.8,1.2){\vector(1,0){0.1}}
\put(0.5,0.2){\line(1,0){1.5}}
\put(0.5,1.2){\line(1,0){1.5}}
\linethickness{0.35mm}
\put(2.0,0.2){\line(1,0){0.5}}
\put(2.0,1.2){\line(1,0){0.5}}
\thinlines
\put(1,0.2){\line(0,1){1}}
\put(1.5,0.2){\line(0,1){1}}
\put(2,0.2){\line(0,1){1}}
\put(0.45,1.2){\makebox(0,0)[r]{$#1$}}
\put(0.45,0.2){\makebox(0,0)[r]{$#2$}}
\put(2.55,1.2){\makebox(0,0)[l]{$#3$}}
\put(2.55,0.2){\makebox(0,0)[l]{$#4$}}
\put(1.075,1.0){\makebox(0,0)[l]{$_{(3)}$}}
\end{picture}
}}
\hfill}

\newcommand{\doubleboxb}[4]{
\mbox{\parbox{3.5cm}{\hspace{0.25cm}
\begin{picture}(3,1.4)
\put(0.8,0.2){\vector(1,0){0.1}}
\put(2.3,0.2){\vector(1,0){0.1}}
\put(2.3,1.2){\vector(1,0){0.1}}
\put(0.8,1.2){\vector(1,0){0.1}}
\put(0.5,0.2){\line(1,0){1.5}}
\put(0.5,1.2){\line(1,0){1.5}}
\linethickness{0.35mm}
\put(2.0,0.2){\line(1,0){0.5}}
\put(2.0,1.2){\line(1,0){0.5}}
\thinlines
\put(1,0.2){\line(0,1){1}}
\put(1.0,0.7){\line(1,0){1}}
\put(2,0.2){\line(0,1){1}}
\put(0.45,1.2){\makebox(0,0)[r]{$#1$}}
\put(0.45,0.2){\makebox(0,0)[r]{$#2$}}
\put(2.55,1.2){\makebox(0,0)[l]{$#3$}}
\put(2.55,0.2){\makebox(0,0)[l]{$#4$}}
\end{picture}
}}
\hfill}

\newcommand{\doubleboxbtwo}[4]{
\mbox{\parbox{3.5cm}{\hspace{0.25cm}
\begin{picture}(3,1.4)
\put(0.8,0.2){\vector(1,0){0.1}}
\put(2.3,0.2){\vector(1,0){0.1}}
\put(2.3,1.2){\vector(1,0){0.1}}
\put(0.8,1.2){\vector(1,0){0.1}}
\put(0.5,0.2){\line(1,0){1.5}}
\put(0.5,1.2){\line(1,0){1.5}}
\linethickness{0.35mm}
\put(2.0,0.2){\line(1,0){0.5}}
\put(2.0,1.2){\line(1,0){0.5}}
\thinlines
\put(1,0.2){\line(0,1){1}}
\put(1.0,0.7){\line(1,0){1}}
\put(2,0.2){\line(0,1){1}}
\put(0.45,1.2){\makebox(0,0)[r]{$#1$}}
\put(0.45,0.2){\makebox(0,0)[r]{$#2$}}
\put(2.55,1.2){\makebox(0,0)[l]{$#3$}}
\put(2.55,0.2){\makebox(0,0)[l]{$#4$}}
\put(1.075,1.0){\makebox(0,0)[l]{$_{(2)}$}}
\end{picture}
}}
\hfill}

\newcommand{\boxbubbleaB}[4]{
\mbox{\parbox{3.5cm}{\hspace{0.25cm}
\begin{picture}(3,1.4)
\put(0.8,0.2){\vector(-1,0){0.1}}
\put(2.3,0.2){\vector(-1,0){0.1}}
\put(2.3,1.2){\vector(1,0){0.1}}
\put(0.8,1.2){\vector(1,0){0.1}}
\put(1.5,1.2){\oval(1,1)[b]}
\put(1.0,0.2){\line(1,0){1.5}}
\put(0.5,1.2){\line(1,0){1.5}}
\linethickness{0.35mm}
\put(0.5,0.2){\line(1,0){0.5}}
\put(2.0,1.2){\line(1,0){0.5}}
\thinlines
\put(1,0.2){\line(0,1){1}}
\put(2,0.2){\line(0,1){1}}
\put(0.45,1.2){\makebox(0,0)[r]{$#1$}}
\put(0.45,0.2){\makebox(0,0)[r]{$#2$}}
\put(2.55,1.2){\makebox(0,0)[l]{$#3$}}
\put(2.55,0.2){\makebox(0,0)[l]{$#4$}}
\end{picture}
}} 
\hfill}

\newcommand{\boxxaB}[4]{
\mbox{\parbox{3.5cm}{\hspace{0.25cm}
\begin{picture}(3,1.4)
\put(0.8,0.2){\vector(-1,0){0.1}}
\put(2.3,0.2){\vector(-1,0){0.1}}
\put(2.3,1.2){\vector(1,0){0.1}}
\put(0.8,1.2){\vector(1,0){0.1}}
\put(1,0.2){\line(1,1){1}}
\put(1.0,0.2){\line(1,0){1.5}}
\put(0.5,1.2){\line(1,0){1.5}}
\linethickness{0.35mm}
\put(0.5,0.2){\line(1,0){0.5}}
\put(2.0,1.2){\line(1,0){0.5}}
\thinlines
\put(1,0.2){\line(0,1){1}}
\put(2,0.2){\line(0,1){1}}
\put(0.45,1.2){\makebox(0,0)[r]{$#1$}}
\put(0.45,0.2){\makebox(0,0)[r]{$#2$}}
\put(2.55,1.2){\makebox(0,0)[l]{$#3$}}
\put(2.55,0.2){\makebox(0,0)[l]{$#4$}}
\end{picture}
}} 
\hfill}

\newcommand{\boxxbB}[4]{
\mbox{\parbox{3.5cm}{\hspace{0.25cm}
\begin{picture}(3,1.4)
\put(0.8,0.2){\vector(-1,0){0.1}}
\put(2.3,0.2){\vector(-1,0){0.1}}
\put(2.3,1.2){\vector(1,0){0.1}}
\put(0.8,1.2){\vector(1,0){0.1}}
\put(2,0.2){\line(-1,1){1}}
\put(1.0,0.2){\line(1,0){1.5}}
\put(0.5,1.2){\line(1,0){1.5}}
\linethickness{0.35mm}
\put(0.5,0.2){\line(1,0){0.5}}
\put(2.0,1.2){\line(1,0){0.5}}
\thinlines
\put(1,0.2){\line(0,1){1}}
\put(2,0.2){\line(0,1){1}}
\put(0.45,1.2){\makebox(0,0)[r]{$#1$}}
\put(0.45,0.2){\makebox(0,0)[r]{$#2$}}
\put(2.55,1.2){\makebox(0,0)[l]{$#3$}}
\put(2.55,0.2){\makebox(0,0)[l]{$#4$}}
\end{picture}
}} 
\hfill}

\newcommand{\boxxbbB}[4]{
\mbox{\parbox{3.5cm}{\hspace{0.25cm}
\begin{picture}(3,1.4)
\put(0.8,0.2){\vector(-1,0){0.1}}
\put(2.3,0.2){\vector(-1,0){0.1}}
\put(2.3,1.2){\vector(1,0){0.1}}
\put(0.8,1.2){\vector(1,0){0.1}}
\put(1.5,0.7){\circle*{0.2}}
\put(2,0.2){\line(-1,1){1}}
\put(1.0,0.2){\line(1,0){1.5}}
\put(0.5,1.2){\line(1,0){1.5}}
\linethickness{0.35mm}
\put(0.5,0.2){\line(1,0){0.5}}
\put(2.0,1.2){\line(1,0){0.5}}
\thinlines
\put(1,0.2){\line(0,1){1}}
\put(2,0.2){\line(0,1){1}}
\put(0.45,1.2){\makebox(0,0)[r]{$#1$}}
\put(0.45,0.2){\makebox(0,0)[r]{$#2$}}
\put(2.55,1.2){\makebox(0,0)[l]{$#3$}}
\put(2.55,0.2){\makebox(0,0)[l]{$#4$}}
\end{picture}
}} 
\hfill}

\newcommand{\boxxbbbB}[4]{
\mbox{\parbox{3.5cm}{\hspace{0.25cm}
\begin{picture}(3,1.4)
\put(0.8,0.2){\vector(-1,0){0.1}}
\put(2.3,0.2){\vector(-1,0){0.1}}
\put(2.3,1.2){\vector(1,0){0.1}}
\put(0.8,1.2){\vector(1,0){0.1}}
\put(1.5,0.2){\circle*{0.2}}
\put(2,0.2){\line(-1,1){1}}
\put(1.0,0.2){\line(1,0){1.5}}
\put(0.5,1.2){\line(1,0){1.5}}
\linethickness{0.35mm}
\put(0.5,0.2){\line(1,0){0.5}}
\put(2.0,1.2){\line(1,0){0.5}}
\thinlines
\put(1,0.2){\line(0,1){1}}
\put(2,0.2){\line(0,1){1}}
\put(0.45,1.2){\makebox(0,0)[r]{$#1$}}
\put(0.45,0.2){\makebox(0,0)[r]{$#2$}}
\put(2.55,1.2){\makebox(0,0)[l]{$#3$}}
\put(2.55,0.2){\makebox(0,0)[l]{$#4$}}
\end{picture}
}} 
\hfill}

\newcommand{\boxxbbbbbB}[4]{
\mbox{\parbox{3.5cm}{\hspace{0.25cm}
\begin{picture}(3,1.4)
\put(0.8,0.2){\vector(-1,0){0.1}}
\put(2.3,0.2){\vector(-1,0){0.1}}
\put(2.3,1.2){\vector(1,0){0.1}}
\put(0.8,1.2){\vector(1,0){0.1}}
\put(1.5,1.2){\circle*{0.2}}
\put(2,0.2){\line(-1,1){1}}
\put(1.0,0.2){\line(1,0){1.5}}
\put(0.5,1.2){\line(1,0){1.5}}
\linethickness{0.35mm}
\put(0.5,0.2){\line(1,0){0.5}}
\put(2.0,1.2){\line(1,0){0.5}}
\thinlines
\put(1,0.2){\line(0,1){1}}
\put(2,0.2){\line(0,1){1}}
\put(0.45,1.2){\makebox(0,0)[r]{$#1$}}
\put(0.45,0.2){\makebox(0,0)[r]{$#2$}}
\put(2.55,1.2){\makebox(0,0)[l]{$#3$}}
\put(2.55,0.2){\makebox(0,0)[l]{$#4$}}
\end{picture}
}} 
\hfill}

\newcommand{\boxxbbbbB}[4]{
\mbox{\parbox{3.5cm}{\hspace{0.25cm}
\begin{picture}(3,1.4)
\put(0.8,0.2){\vector(-1,0){0.1}}
\put(2.3,0.2){\vector(-1,0){0.1}}
\put(2.3,1.2){\vector(1,0){0.1}}
\put(0.8,1.2){\vector(1,0){0.1}}
\put(2.0,0.7){\circle*{0.2}}
\put(2,0.2){\line(-1,1){1}}
\put(1.0,0.2){\line(1,0){1.5}}
\put(0.5,1.2){\line(1,0){1.5}}
\linethickness{0.35mm}
\put(0.5,0.2){\line(1,0){0.5}}
\put(2.0,1.2){\line(1,0){0.5}}
\thinlines
\put(1,0.2){\line(0,1){1}}
\put(2,0.2){\line(0,1){1}}
\put(0.45,1.2){\makebox(0,0)[r]{$#1$}}
\put(0.45,0.2){\makebox(0,0)[r]{$#2$}}
\put(2.55,1.2){\makebox(0,0)[l]{$#3$}}
\put(2.55,0.2){\makebox(0,0)[l]{$#4$}}
\end{picture}
}} 
\hfill}

\newcommand{\boxxbbbbbbB}[4]{
\mbox{\parbox{3.5cm}{\hspace{0.25cm}
\begin{picture}(3,1.4)
\put(0.8,0.2){\vector(-1,0){0.1}}
\put(2.3,0.2){\vector(-1,0){0.1}}
\put(2.3,1.2){\vector(1,0){0.1}}
\put(0.8,1.2){\vector(1,0){0.1}}
\put(1.0,0.7){\circle*{0.2}}
\put(2,0.2){\line(-1,1){1}}
\put(1.0,0.2){\line(1,0){1.5}}
\put(0.5,1.2){\line(1,0){1.5}}
\linethickness{0.35mm}
\put(0.5,0.2){\line(1,0){0.5}}
\put(2.0,1.2){\line(1,0){0.5}}
\thinlines
\put(1,0.2){\line(0,1){1}}
\put(2,0.2){\line(0,1){1}}
\put(0.45,1.2){\makebox(0,0)[r]{$#1$}}
\put(0.45,0.2){\makebox(0,0)[r]{$#2$}}
\put(2.55,1.2){\makebox(0,0)[l]{$#3$}}
\put(2.55,0.2){\makebox(0,0)[l]{$#4$}}
\end{picture}
}} 
\hfill}

\newcommand{\boxxbpaB}[4]{
\mbox{\parbox{3.5cm}{\hspace{0.25cm}
\begin{picture}(3,1.4)
\put(0.8,0.2){\vector(-1,0){0.1}}
\put(2.3,0.2){\vector(-1,0){0.1}}
\put(2.3,1.2){\vector(1,0){0.1}}
\put(0.8,1.2){\vector(1,0){0.1}}
\put(1.0,0.2){\line(1,0){1.5}}
\put(0.5,1.2){\line(1,0){1.5}}
\linethickness{0.35mm}
\put(0.5,0.2){\line(1,0){0.5}}
\put(2.0,1.2){\line(1,0){0.5}}
\thinlines
\put(1.5,1.2){\line(1,-2){0.5}}
\put(1,0.2){\line(0,1){1}}
\put(2,0.2){\line(0,1){1}}
\put(0.45,1.2){\makebox(0,0)[r]{$#1$}}
\put(0.45,0.2){\makebox(0,0)[r]{$#2$}}
\put(2.55,1.2){\makebox(0,0)[l]{$#3$}}
\put(2.55,0.2){\makebox(0,0)[l]{$#4$}}
\end{picture}
}} 
\hfill}

\newcommand{\doubleboxB}[4]{
\mbox{\parbox{3.5cm}{\hspace{0.25cm}
\begin{picture}(3,1.4)
\put(0.8,0.2){\vector(-1,0){0.1}}
\put(2.3,0.2){\vector(-1,0){0.1}}
\put(2.3,1.2){\vector(1,0){0.1}}
\put(0.8,1.2){\vector(1,0){0.1}}
\put(1.0,0.2){\line(1,0){1.5}}
\put(0.5,1.2){\line(1,0){1.5}}
\linethickness{0.35mm}
\put(0.5,0.2){\line(1,0){0.5}}
\put(2.0,1.2){\line(1,0){0.5}}
\thinlines
\put(1,0.2){\line(0,1){1}}
\put(1.5,0.2){\line(0,1){1}}
\put(2,0.2){\line(0,1){1}}
\put(0.45,1.2){\makebox(0,0)[r]{$#1$}}
\put(0.45,0.2){\makebox(0,0)[r]{$#2$}}
\put(2.55,1.2){\makebox(0,0)[l]{$#3$}}
\put(2.55,0.2){\makebox(0,0)[l]{$#4$}}
\end{picture}
}}
\hfill}

\newcommand{\doubleboxBtwo}[4]{
\mbox{\parbox{3.5cm}{\hspace{0.25cm}
\begin{picture}(3,1.4)
\put(0.8,0.2){\vector(-1,0){0.1}}
\put(2.3,0.2){\vector(-1,0){0.1}}
\put(2.3,1.2){\vector(1,0){0.1}}
\put(0.8,1.2){\vector(1,0){0.1}}
\put(1.0,0.2){\line(1,0){1.5}}
\put(0.5,1.2){\line(1,0){1.5}}
\linethickness{0.35mm}
\put(0.5,0.2){\line(1,0){0.5}}
\put(2.0,1.2){\line(1,0){0.5}}
\thinlines
\put(1,0.2){\line(0,1){1}}
\put(1.5,0.2){\line(0,1){1}}
\put(2,0.2){\line(0,1){1}}
\put(0.45,1.2){\makebox(0,0)[r]{$#1$}}
\put(0.45,0.2){\makebox(0,0)[r]{$#2$}}
\put(2.55,1.2){\makebox(0,0)[l]{$#3$}}
\put(2.55,0.2){\makebox(0,0)[l]{$#4$}}
\put(1.075,1.0){\makebox(0,0)[l]{$_{(2)}$}}
\end{picture}
}}
\hfill}

\preprint{{ZU-TH 12/13, LPN13-039}}
\begin{document}
\unitlength1cm
\maketitle
\allowdisplaybreaks

\section{Introduction} 
Vector boson pair production ($\gamma\, \gamma$, $Z\, \gamma\,/\,W\, \gamma$,
 $Z\, Z$, $W\, W$, $W\, Z$) is a key process in studying the dynamics of the electroweak theory 
at the LHC. 
It enters as background not only for Higgs production, but also
for many other new-physics searches. 
It offers in fact a large number of observables which allow precise
tests of the electroweak symmetry-breaking and in general of the non-abelian gauge 
structure of the group $SU(2)\times U(1)$, for example of the triple gauge-boson couplings.
With large production rates to be expected from the future data taking at the LHC,
vector boson pair production processes will become electroweak precision observables. The high 
experimental precision must be matched by a comparably high accuracy of the theoretical predictions, 
typically requiring next-to-leading order (NLO) electroweak and next-to-next-to-leading order
(NNLO) QCD corrections. 

At present large parts of the NLO electroweak corrections~\cite{Accomando:2005xp,Accomando:2004de,Accomando:2005ra,Bierweiler:2013dja} 
and the NLO massless QCD corrections~\cite{Ohnemus:1992jn,Baur:1993ir,Baur:1997kz,Dixon:1998py,Dixon:1999di} 
are known for vector boson pair production, usually 
including the vector boson decays to leptons. The massless NNLO QCD corrections are known only for 
$\gamma\,\gamma$ production~\cite{Catani:2011qz}.

A full NNLO computation requires three different ingredients:
the two-loop double-virtual corrections to the partonic $2 \to 2$ process, the one-loop real-virtual
corrections to the $2 \to 3$ process for the production of the vector boson pair plus an additional 
parton, and the tree-level corrections to the $2 \to 4$ process involving two extra partons.

In the case of vector boson pair production, the $2 \to 3$ and $2 \to 4$ ingredients 
have already been known in the literature for some time in the context of NLO calculations
with higher final-state multiplicity~\cite{DelDuca:2003uz,Gehrmann:2013aga,Dittmaier:2007th,
Dittmaier:2009un,Binoth:2009wk,Campanario:2009um,Campanario:2010hv,
Campanario:2010hp}. 
On the other hand, 
the two-loop parton-level matrix elements
are known only for $\gamma\,\gamma$~\cite{Bern:2001df} 
and $V\,\gamma$ production~\cite{Gehrmann:2011ab,Gehrmann:2013vga}. 
Finally, in the case of $W\,W$ production the two-loop virtual 
corrections are known in the high-energy approximation~\cite{Chachamis:2008yb}.

The computation of the two-loop matrix elements for two massive vector boson production is still 
an outstanding task and it constitutes the bottleneck for having a complete NNLO 
description of the process.
When computing two-loop corrections to four-point functions in quantum 
field theory a large number of apparently different integrals appears. 
In particular, increasing the number of loops and/or the number of external legs, more and 
more different scales are added, increasing the complexity of analytically 
evaluating the integrals. 

In the framework of dimensional regularization,
many powerful techniques have been developed in order to make the computation 
of two-loop corrections to three- and four-point functions feasible.
Employing integration-by-parts (IBPs), Lorentz and 
symmetry identities~\cite{Tkachov:1981wb,Chetyrkin:1981qh}  a large number of relations among the integrals can be established.
The latter turn out to be simple linear equations which involve the integrals and only rational 
functions of the invariants and of the dimensional regularization parameter $d$.
Solving this system of equations allows one to express most of the integrals in terms of a 
relatively small subset of irreducible integrals, the so-called Master Integrals (MIs).

In non-trivial applications to two-loop corrections to four-point functions 
the system of equations can easily grow  to include tens or hundreds of thousands of equations, 
so that one must resort to the use of computer algebra.
In the last years many public and private implementations for the automatic reduction 
to master integrals using the Laporta algorithm~\cite{Laporta:2001dd} have become
available~\cite{Anastasiou:2004vj,Smirnov:2008iw,Studerus:2009ye,vonManteuffel:2012np}.
Symmetry relations can often give new equations that must be 
consistently included in the system in order to ensure a full reduction to a minimal set of MIs. 

Once the reduction is completed, we are left with the problem of computing the MIs.
While, quite in general, there exists no algorithm that allows to compute two-loop corrections 
to four-point functions with arbitrary configuration of external and internal masses, 
the differential equation method~\cite{Kotikov:1990kg,Remiddi:1997ny,Caffo:1998du,Gehrmann:1999as} has
proven to be very powerful in a large number of computations, 
including two-loop four-point 
functions with  massless and massive internal 
propagators~\cite{Gehrmann:2000zt,Gehrmann:2001ck,Bonciani:2008az,Bonciani:2009nb,Bonciani:2010mn,vonManteuffel:2013uoa}. 
In this method, differential equations for the integrals under consideration are derived at the 
basis of the integrands. The master integrals are then determined by solving 
these differential equations, matched to appropriate boundary conditions (that usually 
correspond to integrals in special points or with fewer scales). 
 In order to determine those boundary conditions and to perform necessary
transformations on the differential equations,
techniques that make use of the algebraic structure of the underlying
functions are used extensively
\cite{Brown:2008um,Duhr:2012fh,Anastasiou:2013srw}.
 
In this paper we make use of the differential equations method 
to compute all MIs appearing in the reduction of planar two-loop 
four-point functions with two legs off-shell with the same mass.
These masters constitute a fundamental ingredient towards the computation of the 
two-loop QCD corrections 
to $q\,\bar{q} \to Z\,Z\,/\,W\,W$ and $g\,g \to Z\,Z\,/\,W\,W$.
In all steps of the computation we made extensive use of FORM~\cite{Vermaseren:2000nd} and Mathematica~\cite{mathematica}.

The paper is organized as follows. In section~\ref{sec:def} we discuss the notation and the method employed.
Section~\ref{sec:algtools} shortly introduces the algebraic tools used
and describes the transformation algorithm.
The differential equations method is further explained in section~\ref{sec:TopoA}, where we describe the MIs 
that appear in the reduction of planar two-loop corrections to 
four-point functions with two adjacent massive legs. In section~\ref{sec:TopoB} we focus on the MIs 
needed for the two-loop corrections to four-point functions with two non-adjacent massive legs.
Extensive analytical and numerical checks to validate the results are documented in 
section~\ref{sec:checks}.
We conclude in section~\ref{sec:conc} and 
enclose two appendices with the analytic results for all planar masters.

\section{Definitions, Notation and Method}
\label{sec:def}
Four-point functions depend in general on three linearly independent momenta, 
which we will call  $p_1$, $p_2$ and $q_1$. 
In the scattering kinematics the fourth momentum is given by $q_2 = p_1+p_2-q_1$.
We take two of the momenta on-shell and the remaining two
momenta off-shell at the same invariant mass, such that 
\begin{equation}
 p_1^2 = p_2^2 = 0, \qquad q_1^2 = q_2^2 = Q^2, \label{eq:kinematics}
\end{equation}
where for physical applications $Q^2$ represents the mass of the vector boson. 

We define the usual Mandelstam variables as:
\begin{align}
 s = (p_1+p_2)^2 \,, \quad t = (p_1-q_1)^2\,, \quad u = (p_2-q_1)^2\, 
\quad \mbox{with} \quad s+t+u = 2\,Q^2. \label{eq:mandelstamP}
\end{align}
In the physical region relevant for vector boson pair production we have
$$Q^2 > 0\,, \quad s > 4\,Q^2 \,, \quad t < 0\,, \quad u < 0\,,$$
and the MIs are complex-valued functions.

In the general case of four-point functions there are up to six independent invariants 
which can be identified with the scalar products among the three external momenta: 
\begin{equation}s_1 = p_1^2\,,\quad s_2 = p_2^2\,,\quad s_3 = q_1^2\,,\quad 
  s_4 = p_1 \cdot p_2\,,\quad s_5 = p_1 \cdot q_1\,,\quad s_6 = p_2 \cdot q_1\,. \label{eq:invariants}
\end{equation}
Using the kinematical constraints in~\eqref{eq:kinematics} they reduce to the three independent invariants in~\eqref{eq:mandelstamP}. 
Due to Lorentz invariance all integrals can only depend on combinations of the latter. 

Feynman integrals can be classified in terms of their topology, i.e.\ in terms of the propagator
denominators appearing in them.
Any integral with the same set of denominators raised to any powers and involving any combination of
scalar products among external and internal momenta in the numerator belongs to the same topology.
Starting from those integrals with the largest number of denominators, we can define
their sub-topologies as all possible sets of denominators that can be built removing one or more
denominators.
For each topology one can derive a set of IBP identities, 
Lorentz invariance identities and (sometimes) symmetry relations 
which allow to express all integrals belonging to that topology in terms of a small number of MIs.

Once the MIs have been identified, 
we can derive differential equations for them with respect to the 
external invariants~\eqref{eq:invariants} as follows.
Starting from their very definition, one can easily see that the derivatives in the invariants 
can be expressed by suitable combinations of derivatives in the external momenta.
Let us consider for simplicity the case where all integrals in one given topology are reduced to one single MI.
Differentiating the MI with respect to any of the external invariants will generate
a linear combination of new integrals, all with the same subset of denominators as the starting integral,
but in general raised to different powers.
Using the IBP identities all these integrals can be reduced again to the MI itself plus 
integrals from its 
sub-topologies,  which in a bottom-up approach are considered as known.
We are thus left with a single first order non-homogeneous 
differential equation for the MI in each of the external invariants.

Equipped with a suitable boundary condition, the differential equation can be easily integrated 
by means of Euler's method of the variation of constants. 
This reduces the problem of performing one or more loop integrations
to a single one-dimensional integration.
In most cases a boundary condition can be found studying the behavior of the integrals 
in some well defined kinematical limit where the integral is known to be regular.

The method immediately generalizes to the case where $N$ master integrals are present. 
The single first order differential equation will be in general substituted by a system of $N$ 
coupled differential equations for the $N$ master integrals, which can in turn be rephrased 
as a $N$-th order differential equation for any of the masters.
Finding the solution to a $N$-th order differential equation requires fixing $N$ boundary conditions. 
Even though no general method is available for solving a system of $N$ coupled 
differential equations with non-constant coefficients, experience shows that 
by an appropriate choice of the basis of master integrals the system can usually  be 
diagonalized (or at least put in triangular form in the limit $d \to 4$), 
allowing a recursive solution by direct integration. A proposal towards  systematizing the computation 
of master integrals from differential equations has been put forward recently in~\cite{Henn:2013pwa}.
Especially for topologies with more than one master integral, finding boundary conditions is in general a non-trivial task 
and represents often the bottleneck of the whole procedure.

\subsection{Auxiliary topologies and reduction to Master Integrals}\label{sec:auxtopo}
Two-loop corrections to four-point functions involve integrals with up to seven different propagators and 
up to nine independent scalar products among the external and the loop momenta in the numerator.
This implies that two out of the nine scalar products are irreducible, 
namely they cannot be rewritten as linear combinations of the seven denominators.
One way to perform a complete reduction is then to define a larger set of nine denominators, 
that we will refer to as auxiliary topology. 
In the case of pair production of equal-mass vector bosons, three independent auxiliary topologies 
are needed in order to account for all the integrals appearing.

We define two planar topologies, named \textbf{Topo A} and \textbf{Topo B}, 
needed to represent respectively the double-boxes with adjacent and non-adjacent off-shell legs.
A third topology, named \textbf{Topo C}, is sufficient to represent all non-planar integrals.
We choose the propagators of the three topologies as listed in Table~\ref{tab:auxtopo}.
The reduction to MIs of the three auxiliary topologies above has been performed 
using Reduze1 and Reduze2~\cite{Studerus:2009ye,vonManteuffel:2012np}.

\begin{table}[h!]
\begin{center}
\begin{tabular}[H]{lll}
\textbf{Topo A}\hspace{5mm} & \textbf{Topo B}\hspace{5mm} & \textbf{Topo C}\\[2mm]
$k^2$               & $k^2$               & $k^2$               \\
$l^2$               & $l^2$               & $l^2$               \\
$(k-l)^2$           & $(k-l)^2$           & $(k-l)^2$           \\
$(k-p_1)^2$         & $(k-p_1)^2$         & $(k-p_1)^2$         \\
$(l-p_1)^2$         & $(l-p_1)^2$         & $(l-p_1)^2$         \\
$(k-p_1-p_2)^2$     & $(k-p_1+q_1)^2$     & $(k-p_1-p_2)^2$     \\
$(l-p_1-p_2)^2$     & $(l-p_1+q_1)^2$     & $(k-l  -q_1)^2$     \\
$(k-p_1-p_2+q_1)^2$ & $(k-p_1-p_2+q_1)^2$ & $(l-p_1-p_2+q_1)^2$ \\
$(l-p_1-p_2+q_1)^2$ & $(l-p_1-p_2+q_1)^2$ & $(k-l-p_1-p_2)^2$  
\end{tabular}
\end{center}
\caption{Propagators in the three different auxiliary topologies used to represent all two-loop 4-point integrals 
with two massless and two massive legs with the same mass.}
\label{tab:auxtopo}
\end{table}

We refer to the MIs in these topologies 
as $\mathcal{I}^{(T)}_n$, where $T$ is the auxiliary topology ($A$, $B$ or $C$), 
while $n$ is a decimal number which corresponds to the selection of propagator momenta that 
appear as denominators in the integral.
Any integral is first  mapped to one of the topologies above. Its set of denominators 
is then identified with a binary number, containing a 1 for each propagator momentum 
appearing in the integral, and a 0 for those momenta not appearing as propagators. 
The minimum binary number to which the integral can be mapped in the 
topology under consideration is then converted 
to a decimal number and used as label for the integral itself.
To give an explicit example consider the following integral belonging to \textbf{Topo A}:
\begin{align*}
\int \frac{d^d k}{(2 \pi)^d} \frac{d^d l}{(2 \pi)^d} \frac{1}{l^2(k-l)^2(k-p_{1})^2(k-p_1-p_2+q_1)^2}.
\end{align*}
The binary number associated to its denominators is $ 010001110 = 142 $, so that with the 
definitions above this integral will be labeled as $\mathcal{I}^{(A)}_{142}$.
In the following we will often use the terminology \textsl{sector} \textsl{`n'}
interchangeably with that of \textsl{topology} \textsl{`n'},
referring to the set of all those integrals whose label according to this procedure is \textsl{`n'}.

With the Mandelstam variables defined above, one can see
from the arrangement of the momenta in the three topologies that the planar
integrals belonging to \textbf{Topo A} have cuts only in $s$ and $u$ while those belonging 
to \textbf{Topo B} have cuts only in $t$ and $u$. 
On the other hand, as it is well known~\cite{Tausk:1999vh}, the non-planar integrals belonging to \textbf{Topo C}
are expected to have cuts in all three Mandelstam variables.
In what follows we will focus on the explicit evaluation of the Master Integrals 
that emerge from the reduction of the
two planar topologies, referring to a future work for the computation of the 
non-planar master integrals belonging to \textbf{Topo C}.

The arrangement of the cuts in the two planar topologies suggests that, in order to express 
the final results in compact form, 
the natural variables 
 should be 
($s$,$u$) for \textbf{Topo A} and ($t$,$u$) for \textbf{Topo B}. 
The kinematical constraints are those that determine the analyticity structure of the 
scattering amplitude and hence of the functions used to describe the result.
In the following two sub-sections we describe separately 
the kinematics for vector boson pair production in the two different sets of variables.

\subsection{Kinematics and analytic continuation in \texorpdfstring{$s,u$}{s,u}}
\label{sec:kinematicsA}
\begin{figure}[h]
\center
\includegraphics[width=11cm]{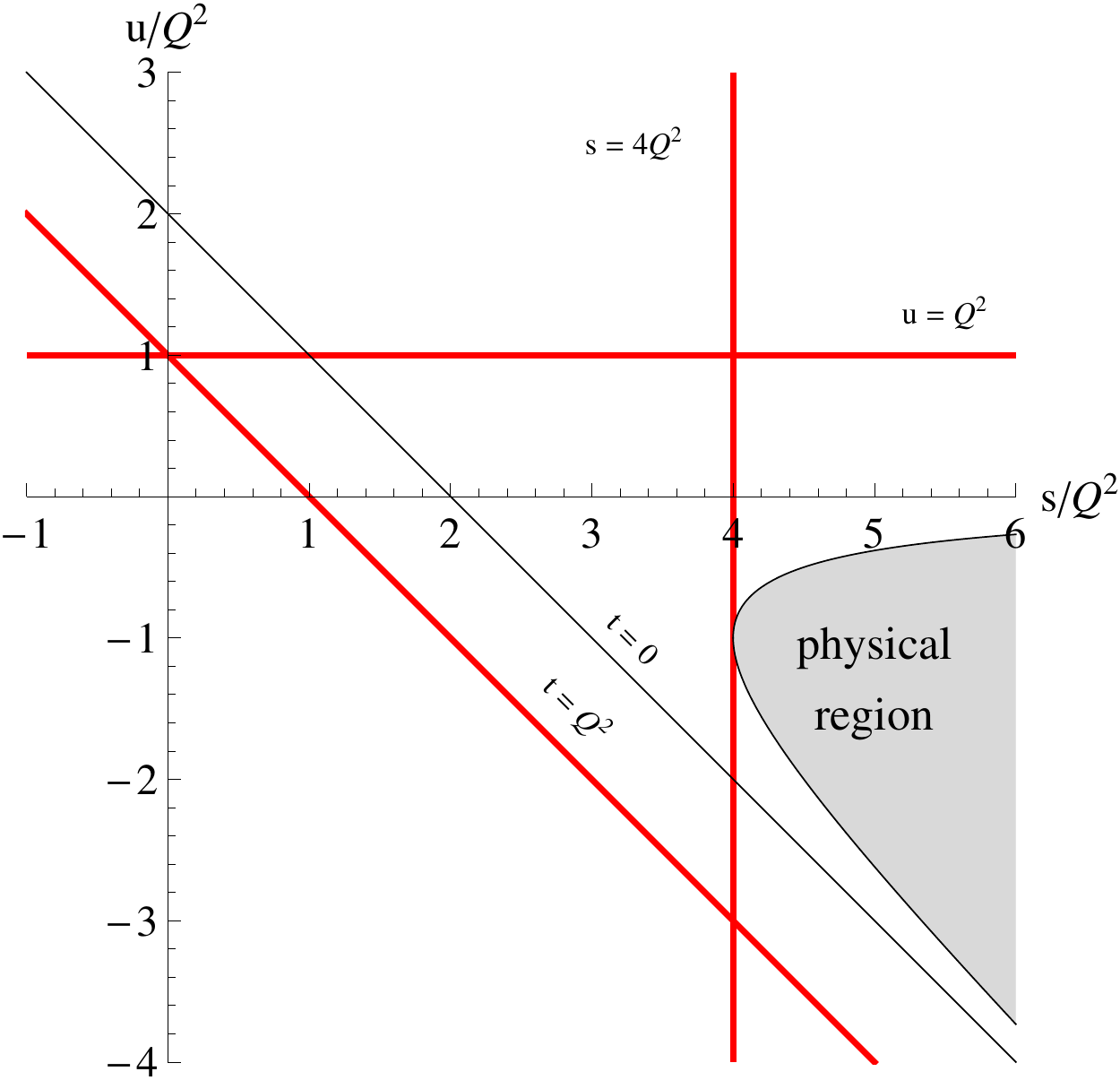}
\caption{Dalitz plot for the $2\to2$ scattering kinematics.}
\label{fig:Dalitz}
\end{figure}

In the physical region we have  $Q^2 > 0$ and $s > 4 \, Q^2$.
Using $s$ and $u$ as independent variables we can express the
kinematical constrains in $u$ in function of $s$ and $Q^2$. In the center-of-mass frame of 
$p_1,p_2$ we find
\begin{equation}
u = Q^2 - \frac{s}{2} \left[\,  1 - \sqrt{1 - \frac{4 Q^2}{s}}   \cos{\theta} \,\right]\,,
\end{equation}
where $\theta$ is the angle between $\vec{p}_2$ and $\vec{q}_1$. 
In Fig.~\ref{fig:Dalitz} we show the kinematical plane in $s,u$.

It is convenient to introduce the dimensionless variables
\begin{align}
s &= M^2{ (1+\xi)^2 \over \xi }\,,\quad
u = -M^2 \zeta\,, \quad
Q^2= M^2\,,
\end{align}
in  which the physical region is given by 
$$M^2 > 0\,, \qquad 0<\xi\leq1\,, \qquad \xi \leq \zeta \leq {1 \over \xi}\,.$$

With the parametrization above we obtain: 
\begin{equation}
 t = 2 Q^2 - s - u = - M^2 \left( {1 + \xi^2 \over \xi} - \zeta \right) < 0\,.
\end{equation}

In the physical region all integrals with a cut in $s$ and $Q^2$ will be complex-valued.
They can be expressed as analytical continuations of real-valued functions 
defined in a non-physical region where $s<0$, $u<0$ and $Q^2<0$, expressed as:
\begin{align}
 &s = - m^2 {(1+x)^2 \over x } < 0\,, \qquad
 u = - m^2 z < 0\,, \qquad
 Q^2 = -m^2 < 0 \,,
\end{align}
with 
\begin{align}
 x = {\sqrt{s} - \sqrt{s-4 Q^2} \over \sqrt{s} + \sqrt{s - 4 Q^2}}\,, \qquad  z =  {u\over Q^2}.
\end{align}
Now taking
\begin{align*}
 m^2 > 0\,, \quad 0<x<1\,, \quad z > 0\,,
\end{align*}
all MIs are real functions of the dimensionless variables $x$ and $z$.

The analytic continuation to the physical region can be obtained as:
\begin{align}\label{eq:TopoAAC}
 &m^2 \to -M^2 - i \eta\,, \quad x \to \xi \,,\quad z \to -\zeta + i \, \eta
\end{align}
such that the Mandelstam variables which become positive acquire the correct imaginary part
\begin{align}
 &s = M^2 {(1+\xi)^2 \over \xi } + i \eta \,,\qquad
 u = - M^2 \zeta\,, \qquad
 Q^2 = M^2 + i \eta \,.
\end{align}

In this context, a subtle issue about the analytical continuations should be 
recalled~\cite{Tausk:1999vh,Gehrmann:2002zr}.
If one tries to express all three Mandelstam variables $s,t,u$ in terms of $x$, $z$ and $m^2$ only,
one easily realizes that no
real value of $x$, $z$ and $m^2$ can make at the same time $s<0\,,\,t<0\,,\,u<0\,,\,Q^2<0$.
For example with our choice we find:
\begin{align}
 &s = - m^2 {(1+x)^2 \over x } < 0 \,,\qquad
 u = - m^2 z < 0 \,,\qquad
 Q^2 = -m^2 < 0 \,,\nonumber \\
 &t = + m^2\left( {1+x^2 \over x} + z \right) > 0\,, \qquad \mbox{with} \quad x,z \in \mathbb{R}^+
\end{align}
which means that if we tried to express in $x$ and $z$ MIs with a cut in $t$, 
they would have an imaginary part different from zero.
One could argue that this comes from having imposed the on-shell condition $s+t+u = 2Q^2$. 
This problem could be avoided by giving up the on-shell condition, 
i.e.\ computing the MIs \textsl{off-shell}, however at the 
expense of  introducing one more
independent scale in the computation.

This particular choice of variables was motivated by the cut structure of the first planar topology,
and will be used only for representing the integrals in this topology in a compact form.

\subsection{Kinematics and analytic continuation in \texorpdfstring{$t,u$}{t,u}}
\label{sec:kinematicsB}
Using $t$, $u$ and $Q^2$ as independent variables we can repeat the reasoning above.
Defining:
\begin{align}
t &= -{M^2\over w}\,,\quad
u = -M^2 v \,, \quad
Q^2= M^2\,,
\end{align}
the physical region is given by 
$$M^2 > 0\,, \qquad 0<v<\infty \,, \qquad 0 < w < v\,,$$
with 
\begin{equation}
 s = 2Q^2 -t-u = M^2 \left( 2 + v + {1 \over w} \right) > 0 \,.
\end{equation}

In order to make all integrals in \textbf{Topo B} real functions we define a non-physical
region where $Q^2<0$, $t<0$, $u<0$:
\begin{align}
  &t = - m^2 y < 0\,, \qquad
   u = - m^2 z < 0\,, \qquad
 Q^2 = - m^2 < 0 \,,
\end{align}
with
\begin{align}
y =  {t\over Q^2}\,, \qquad
z =  {u\over Q^2}\,.
\end{align}
Taking now $$m^2 > 0\,, \quad z >0\,, \quad y>0\,, \quad \mbox{with} \quad y+z < 1\,,$$
all MIs are real functions of $y$ and $z$.

In continuing to the physical region only $Q^2$ becomes positive,
while $t$ and $u$ remain negative.
The analytic continuation can be obtained through:
\begin{align}\label{eq:TopoBAC}
 &m^2 \to -M^2 - i \eta\,, \quad y \to - 1/w + i \eta  \,,\quad z \to -v+ i \, \eta
\end{align}
which gives 
 \begin{align}
 &t  = - {M^2 \over w}\,, \quad u = - {M^2 \, v}\,,  \quad Q^2 = M^2 + i \eta\,.
\end{align}

Note that even though $t$ and $u$ remain negative we still need to fix 
an imaginary part for $y$ and $z$, which is determined as:
\begin{align}
&y \to {t \over Q^2 + i \eta} = {t \over Q^2} + i \eta = - {1 \over w} + i \, \eta\,,\\
&z \to {u \over Q^2 + i \eta} = {u \over Q^2} + i \eta = - {v} + i \, \eta\,.
\end{align}

\section{Algebraic Tools for Feynman Integrals}\label{sec:algtools}
As outlined in the introduction, each extra scale in a multi-loop integral results in a more 
complicated structure of the functions that are required to express the analytical form of the integrals. 
For many multi-loop integrals, analytical expressions of their Laurent expansion in dimensional regularization 
can be expressed in terms of functions that are algebraic generalizations of polylogarithms.
These functions are called Generalized Harmonic PolyLogarithms (GHPLs) or Multiple PolyLogarithms (MPLs) in the literature. 

In the analytical integration of multi-loop Feynman integrals the following three  tasks have to be 
preformed repeatedly on these 
polylogarithmic functions: analytical continuation, determination of 
limiting values in special points and variable transformations. Finally, a simplification of the 
analytical results to a compact form is often desirable. 

Due to the higher number of different scales in the integrals these functions become increasingly 
difficult to describe and to handle. In the last years, improved understanding of their algebraic structure
has led to the development of the \emph{symbol formalism} and its generalization, 
the \emph{coproduct formalism} \cite{Zagier,Goncharov,Goncharov:2010jf,Goncharov:2005sla,Duhr:2012fh,Brown:2008um,Anastasiou:2013srw,Duhr:2011zq}.
The general idea of this formalism is to translate the expressions to a certain tensor space, 
where the complicated functional relations among GHPLs reduce to simple algebraic operations. 
These tensors can then be `integrated' back to the function space.

In the following, we will describe how we applied and implemented this algorithm, first described 
in~\cite{Anastasiou:2013srw}, to tackle the tasks above (analytical continuation, variable transformations and limiting behavior)
efficiently. In order to do this, the GHPLs are shortly introduced. 
Then, the most important aspects of the coproduct are discussed and the manipulation procedure described. 
An example application is discussed in section~\ref{sec:Int213}.

\subsection{Generalized Harmonic Polylogarithms (GHPLs)}

The generalized harmonic polylogarithms, or multiple polylogarithms, are defined as iterated integrals by
\be\bsp\label{MPLdef}
G(a_1,\ldots, a_n;x)=&\int_0^x \frac{\dt}{t-a_1} G(a_2,\ldots,a_n;t)\;,\\
&\text{with} \; G(x)=G(;x)=1;\quad G(0)=\int_0^0\dt =0 \,.
\esp\ee
The $a_i \in \mathbb{C} $ are constants and are called the \emph{index (vector)} $\vec{a}$ or 
\emph{vector of singularities} whereas $x \in \mathbb{C}$ will be called the \emph{argument} of the GHPL in the following. 
The length of the index vector $n$ is called the \emph{(transcendental) weight} of the function. 
In the special case that the index vector consists of all zeros $\vec{a}=\vec{0}_n=\{0,\ldots,0\}$, we define 
\be\label{MPLlogdef}
G(\vec{0}_n,x)= \frac{1}{n!} \log^n x \,.
\ee

In mathematics a \emph{transcendental} number is a number that is not the root of a polynomial equation 
with rational coefficients, or, the opposite to an algebraic number. 
In the context of repeated integrals it has been proven consistent and useful to define a 
\emph{(degree of) transcendentality} of a function by the number of iterated integrations it contains, 
which in the case of GHPLs corresponds to the length of the index vector. 
Therefore we will use the notion of weight and transcendentality interchangeably in the following.
Since 
\be
\pi = i \log(-1- i \epsilon ) = \int_1^{-1} \frac{1}{x- i \epsilon}\d x\;,
\ee
we have consistently that the transcendentality of $\pi$ is 1. 

In physics GHPLs usually show up with the entries of the index vector chosen from a limited set,
often called the \emph{alphabet} of the problem under consideration.
 In the special case that $a_i \in \{-1,0,1\}$ these functions are called 
\emph{Harmonic Polylogarithms (HPLs)} and were first studied in \cite{Remiddi:1999ew}. 
In multi-scale integrals the $a_i$ often depend on another variable, e.g. $a_i \in \{0,1,z,1-z\}$, 
in which case one speaks of  \emph{two-dimensional HPLs (2dHPLs)} which 
were studied first in \cite{Gehrmann:2000zt}. A  flexible numerical implementation of GHPLs 
exists~\cite{Vollinga:2004sn}. 
In the present context, it is useful to think of the entries of the index vector as functions dependent 
on the kinematic variables of the process themselves, but nevertheless belonging to a finite set.
See for example sections \ref{sec:TopoAGHPLs} and \ref{sec:TopoBGHPLs}.

Some closed expressions for special index choices are, besides eq.~(\ref{MPLlogdef}), for $a\neq0$,
\be \bsp\label{eq:MPLclosedform}
G(\vec{a}_n;x)  = & \frac{1}{n!} \log^n \left( 1-\frac{x}{a}\right)\quad\text{with}\;\vec{a}_n = \overbrace{(a,\ldots,a)}^{n\,\text{times}}\\
G(\vec{0}_{n-1},a;x)  = &-\li_n\left(\frac{x}{a}\right)\;,\\
G(\vec{0}_{n},\vec{a}_p;x)  = & (-1)^p S_{n,p} \left(\frac{x}{a}\right)\;,
\esp \ee
where $\li_n(x)$ is the ordinary and $S_{n,p}(x)$ is the Nielsen polylogarithm. 
From the series representation of the GHPLs it can be seen that 
\be\label{eq:zetavalues}
\li_n(1) = \zeta(n) \equiv \zeta_n\;,
\ee
where $\zeta(n)$ is the Riemann zeta function. Therefore, $\zeta(n)$ can be assigned 
the transcendentality $n$ for integer values $n\in \N$.
Up to weight three, GHPLs can be expressed in terms of ordinary logarithms and polylogarithms. 
For example at weight two we have, for $0\neq a \neq b$,
\be
G(a,b;x)=\li_2 \left(\frac{b-x}{b-a}\right) -\li_2 \left(\frac{b}{b-a}\right) 
+\log \left(1- \frac{x}{b}\right)\log \left(\frac{x-a}{b-a} \right)\;.
\ee

Repeated integrals form a shuffle algebra: the product of two GHPLs of weights $n_1$, $n_2$ can be expressed 
in terms of functions of weight $n_1+n_2$:
\be\label{eq:shufflerelation}
G(a_1,\ldots, a_{n_1};x)G(a_{n_1+1},\ldots,a_{n_1+n_2};x)= \sum_{\sigma \in \Sigma(n_1,n_2)} G(a_{\sigma{1}},\ldots,a_{\sigma{n_1+n_2}};x).
\ee
Formally, $\Sigma(n_1,n_2)$ is defined as the subset of the symmetric group $S_{n_1+n_2}$ defined by
\be\bsp
\Sigma(n_1,&n_2) = \\
&\big\{ \sigma \in S_{n_1+n_2} | \sigma^{-1}(1) < \ldots < \sigma^{-1}(n_1) \; \text{and} \; 
\sigma^{-1}(n_1+1) < \ldots < \sigma^{-1}(n_1+n_2) \big\} \,.
\esp\ee
This is the set of all possibilities of riffle shuffling two decks of cards with $n_1$ and $n_2$ cards, 
which means that the respective order of the cards belonging to one of the original decks remains intact.
A consequence of this property is that the GHPLs of a given weight with a certain set of indices are 
linearly dependent modulo products of lower weight functions.

\subsection{An intuitive introduction of the coproduct}
This section is a very short summary of the concepts presented in \cite{Duhr:2012fh}. 
Its aim is not to give a precise introduction and definition of the required mathematical concepts but 
to provide an intuitive understanding of the coproduct,
in view of its application in our calculation. 

The space of GHPLs $\cH$ is an algebra with the shuffle product as multiplication. 
This algebra is \emph{graded}, which means that can be written as the direct sum of subspaces 
which all contain the functions of a specific weight
\be
\cH = \bigoplus_{n=0}^\infty \cH_n
\ee
and the multiplication preserves the grading such that
\be
\cH_i \cdot \cH_j \subset \cH_{i+j=n}\;.
\ee
Note that the rational numbers are embedded in $\cH_0$.

Let us now introduce a bilinear tensor product $(\,.\otimes.\,)$ and a map $\mu$ which maps this tensor product 
onto the algebra multiplication
\be\bsp
\mu:\cH \otimes \cH & \rightarrow \cH \\
\mu(a \otimes b) & \mapsto a \cdot b\;.
\esp\ee
In words, $\mu$ `pulls two tensor product factors together'. This can be repeated and the associativity 
of the algebra multiplication assures that
\be
\mu(\id\otimes\mu)=\mu(\mu\otimes\id)\;,
\ee
i.e. that the order of pulling the factors together is irrelevant.

In \cite{Goncharov:2005sla} Goncharov defined a \emph{coproduct} $\Delta$ for $\cH$ and showed that 
it was compatible with its algebra structure, thereby promoting it to a \emph{bialgebra}. 
Since $\cH$ is connected because the rational numbers are included in it, the GHPLs form a \emph{Hopf algebra} $\cH$. 
The comultiplication $\Delta$ acts in the opposite direction: it is a way of `pulling apart' an element of the algebra, 
\be\bsp
\Delta:\cH & \rightarrow \cH \otimes \cH \\
a & \mapsto \Delta(a)\;,
\esp\ee
which is compatible with the algebra structure of $\cH$. This means that $\Delta$ is an algebra homomorphism:
\be
\Delta(a\cdot b)=\Delta(a)\cdot\Delta(b)\;,
\ee
where the multiplication on the right hand side is defined entry-by-entry as
\be
(a_1 \otimes a_2) \cdot (b_1 \otimes b_2) \equiv (a_1 \cdot b_1)\otimes (a_2 \cdot b_2)\;.
\ee
An important requirement on the coproduct is that it is \emph{coassiociative}:
\be
(\Delta\otimes\id)\Delta = (\id \otimes \Delta)\Delta\;.
\ee
This means that when applying the coproduct to one of the factors inside another coproduct, 
the order in which this is done is irrelevant.
Therefore, there is a unique way of iterating $\Delta$:
\be
\cH \stackrel{\Delta}{\longrightarrow} \cH \otimes \cH \stackrel{\id \otimes \Delta}{\longrightarrow} 
\cH \otimes \cH \otimes \cH \stackrel{ \id\otimes\id\otimes\Delta}{\longrightarrow} \ldots 
\ee

Goncharov has shown that $\cH$ can be equipped with a coproduct and therefore be turned into a Hopf algebra. 
The coproduct for mutually different indices $a_i$ is given by
\be\bsp\label{eq:coproduct_def}
\Delta(I(a_0;a_1, & \ldots,a_n;a_{n+1})) = \\
 \sum_{0=i_1 < i_2 <\ldots< i_k < i_{k+1} = n}  I&(a_0;a_{i_1},\ldots,a_{i_k};a_{n+1})  
\otimes \left[\prod_{p=0}^k I(a_{i_p};a_{i_{p+1}},\ldots a_{i_{p+1}-1};a_{i_{p+1}}) \right]\;,
\esp
\ee
where 
\be
I(a_0;a_1, \ldots,a_n;a_{n+1}) = \int_{a_0}^{a_{n+1}} \frac{\dt}{t-a_n} I(a_0;a_1,\ldots,a_{n-1};t) \;.
\ee
is the GHPL with $a_0$ as a starting point of each integration instead of zero. It can be converted to the $G$-notation 
easily (see, for example~\cite{Duhr:2012fh}).
When some indices are equal, regularization has to be performed, as has been discussed 
in~\cite{Goncharov:2005sla,Duhr:2012fh,2001math.3059G}.

As an example, let us just quote here the coproduct of the ordinary (poly-)logarithm:
\be\bsp\label{eq:coproduct_classical}
\Delta(\ln z) &\,= 1\otimes \ln z + \ln z\otimes 1\,,\\
\Delta(\li_n(z)) &\, = 1\otimes \li_n(z) + \li_n(z)\otimes 1 + \sum_{k=1}^{n-1}\li_{n-k}(z)\otimes {\ln^kz\over k!}\,.
\esp\ee
We observe that the coproduct $\Delta$ also preserves the weight in the sense below:
\be\label{eq:copr_decomp}
\cH \stackrel{\Delta}{\longrightarrow} \bigoplus_{p+q=n} \cH_p \otimes \cH_q\;,
\ee
where $\cH_n$ denotes the space of GHPLs of weight $n$. We can therefore write
\be
\Delta = \sum_{p+q=n} \Delta_{p,q}\;,
\ee
where $\Delta_{p,q}$ is the operator whose image are the tensor products with the weights $p$, $q$ in the respective factors.

To extend the coproduct to include constant transcendental factors, 
its action on $\zeta$ values needs to be defined as \cite{Brown:2011ik}
\be\bsp\label{eq:zetas}
\Delta(\zeta_2) & = \zeta_2 \otimes 1 ,\\
\Delta(\zeta_{2n} ) &  = \zeta_{2n} \otimes 1 \,,\quad n \in \N\;,\\
\Delta(\zeta_{2n+1}) & = \zeta_{2n+1} \otimes 1 + 1 \otimes \zeta_{2n+1} \,,\quad n \in \N\;.
\esp
\ee
Furthermore, it has been conjectured in \cite{Duhr:2012fh} that
\be\label{eq:coprPi}
\Delta(\pi) = \pi \otimes 1\;.
\ee
The two equations (\ref{eq:zetas}) and (\ref{eq:coprPi}) lead to the redefinition of the coproduct. It actually maps
\be\label{eq:coproductchainrefined}
\cH \stackrel{\Delta}{\longrightarrow} \cH \otimes \cH^\pi \stackrel{\Delta\otimes \id }{\longrightarrow} 
\cH \otimes \cH^\pi \otimes \cH^\pi\stackrel{\Delta\otimes \id\otimes \id }{\longrightarrow} \cH \otimes 
\cH^\pi \otimes \cH^\pi\otimes \cH^\pi \stackrel{\Delta\otimes \id\otimes \id \otimes \id}{\longrightarrow}\ldots
\ee
where we have defined $\cH^\pi$ as the quotient of $\cH$ by the (two-sided) ideal generated by $\pi$, 
or, practically speaking, that we drop all factors of $\pi$ in the tensor product `slots', except in the first one.

Let us also define the \emph{reduced coproduct}
\be
\Delta(a)=1 \otimes a + a \otimes 1 + \Delta'(a)\;.
\ee
Any element that vanishes under the action of $\Delta'$ is called a \emph{primitive element} of $\cH$.

In analogy to eq.~(\ref{eq:copr_decomp}), we introduce the notation
\be\label{eq:Deltaoperators}
\Delta_{i_1,\ldots,i_n}(a) \equiv \sum_{k_1,\ldots ,k_n} c_{k_1\ldots k_n} a_{k_1}^{(i_1)}
\otimes \ldots \otimes a_{k_n}^{(i_n)} \;,
\ee
which means that we apply the coproduct $n-1$ times and select only the tensor products 
whose factors have the weights $i_1,\ldots,i_n$, respectively. 

It has been argued in \cite{Duhr:2012fh} that $\Delta_{1,\ldots,1}$ is actually the symbol 
introduced in \cite{Zagier, Goncharov,Goncharov:2010jf}, 
modulo factors of $\pi$ in the first entry of the tensor products.
We will therefore refer to the \emph{symbol} of an expression $E$ when we mean $\Delta_{1,\ldots,1}(E)$, 
where the number of $1$s in the subscript denotes the expression's weight.

The point of departure of the rewriting procedure is that two equal expressions involving GHPLs 
also agree under the action of the reduced coproduct and of any 
of the operators defined by eq.~(\ref{eq:Deltaoperators}), even though they may look different.
However, these equalities only contain functions of lower weight, for which more functional 
equations might be already known, therefore making the agreement possibly explicit.
The general idea of~\cite{Duhr:2012fh} is now to proceed from the bottom up and compute 
the expression's symbol, which we then `integrate back' in the desired form using 
the algorithm described in the next subsection. 
Then we apply different operators $\Delta_{i_1,\ldots,i_k}$ to the difference between original 
and reconstructed expression to obtain the parts lost by the symbol map. 
At the last step the primitive elements of the expression have to be determined.

\subsection{Reconstructing a function from its symbol}\label{sec:symbolintegrationalternative}

In \cite{Duhr:2011zq} an algorithm was described to obtain an expression in terms 
of a pre-selected set of functions (the `basis') from a known symbol (its `integration'). 
It is very useful when dealing with a low number of basis functions and when one is interested 
in writing the result in a short form using functions with complicated arguments. 
However, often we would like to express it as combination of GHPLs with certain (simple) arguments. 
In this case an alternative way to integrate the symbol, first described in~\cite{Anastasiou:2013srw}, 
is advantageous and will be used in the following.

The general idea is to bring the symbol into a form where information about the GHPLs 
of the integrated version can be read off directly. 
As every GHPL of weight one can be written as a logarithm (see eq.~(\ref{eq:MPLclosedform})), 
every factor of the tensor products in $\Delta_{1,\ldots,1}(F)$ is a logarithm, 
which can be manipulated using the known rules. 
Factors of $\pi$ can be ignored everywhere, except in the first entry of the tensor products 
due to (\ref{eq:coproductchainrefined}), where they are obtained by proper analytical continuation 
of the occurring expressions using the $\epsilon$ prescriptions for the variables of the problem.
All the terms containing $\pi$ in the first entry can be interpreted separately as
\be
i\, c\, (\pi \otimes a_2 \otimes \ldots \otimes a_n) \rightarrow i \pi\, c\, (a_2 \otimes \ldots \otimes a_n)\;,
\ee
i.e. as terms $\pi$ times a lower weight symbol, which can be integrated separately. 
Therefore we will leave them out and assume that the symbol we want to integrate is free of such terms.

Let us suppose for the moment that the symbol of our expression $F$ with weight $n$ depends on one variable $x$. 
We would like to integrate it as GHPLs with $x$ in the argument and constants in the index. 
First, we need to bring the tensor product factors into a unique form, 
that is into a form where there are no relations between the factors like
\be
\sum_i c_i \log(f_i(x)) = 1\;,
\ee
which we choose as 
\be
\log\left(1-\frac{x}{c_i}\right) \quad \text{or} \quad \log(c_i) \quad  \text{with} \; c_{i} \in \C^{*}\;,
\ee

The algorithm is based on two observations. 
Firstly, it is a well known result that an expression in which all ``shuffles have been eliminated'' 
(i.e.\ products of GHPLs with the same argument have been replaced by the sum of their shuffled 
GHPLs using  (\ref{eq:shufflerelation})) is 
unique~\cite{Remiddi:1999ew} in the sense that there is no alternative way of writing it using GHPLs 
of the same index set without using products of GHPLs with the same argument.
Secondly, it can be seen that the symbol of a generic GHPL contains exactly one term of the form
\be\label{eq:gpldecompform}
\Delta_{1,\ldots,1} (G(a_1,\ldots,a_n;x)) = \ldots + \log\left(1-\frac{x}{a_n}\right)
\otimes \ldots \otimes \log\left(1-\frac{x}{a_1}\right) + \ldots \;,
\ee
whereas all the other terms contain at least one factor independent of $x$. 
Therefore, each term of that form present in the symbol comes from one GHPL with $x$ in the argument.

The idea is now to rewrite the symbol of the expression at hand in a way such that we can read off 
the terms of the form (\ref{eq:gpldecompform}) and use them to construct the ``shuffle free'' part of the integrated expression. 
Then, we proceed iteratively. 
We subtract from the symbol $\S(F_0)$, which we seek to integrate, the symbol of the newly found terms. 
The result should now contain no more terms of form (\ref{eq:gpldecompform}). 
In order to determine the parts proportional to $G(a_1,\ldots,a_{n-1},x)\,\log(a_n)$ we now focus on the terms
\be 
\Delta_{1,\ldots,1}( G(a_1,\ldots,a_{n-1};x)\log a_n )  = \ldots + \log (a_n) 
\otimes \log \left(1-\frac{x}{a_{n-1}}\right)\otimes \ldots \otimes \log\left(1-\frac{x}{a_1}\right) + \ldots \;,
\ee
and construct that part analogously. We then subtract that symbol from the ``left over'' symbol above.

Proceeding further we face a small subtlety: terms of the form 
$$G(a_1,\ldots,a_{n-2},x)\,\log(a_{n-1})\,\log(a_n)$$ 
produce two tensor products of the desired form due to the shuffle identities:
\be\bsp
\Delta_{1,\ldots,1}( G(a_1,\ldots,a_{n-2} &  ;x) \log a_{n-1} \log a_n ) =\\
= \ldots + & \log(a_n) \otimes \log (a_{n-1}) \otimes \log\left(1-\frac{x}{a_{n-1}}\right)\otimes \ldots \otimes\log \left(1-\frac{x}{a_1}\right) \\
+ & \log(a_{n-1}) \otimes \log(a_n) \otimes \log \left(1-\frac{x}{a_{n-1}}\right)\otimes \ldots \otimes \log\left(1-\frac{x}{a_1}\right) + \ldots \;.
\esp\ee
We should therefore only pick products where the constant factors $\log(a_i)$ in the beginning follow a certain order e.g.\
\be\label{eq:constantordering}
\log(a_i) \otimes\log( a_j) \otimes \ldots \qquad \text{with}\; a_i \leq a_j
\ee
to construct the next part of the result. 
In principle, also multiple polylogarithms with only constants in argument and index could produce such terms, for example
\be
\Delta_{1,\ldots,1}(G(-\frac{1}{2},-1;1))= \log(2) \otimes \log( 3) - \log(3)\otimes \log(2) \;,
\ee
but in the cases encountered in this work it was sufficient to only consider products of logarithms.

Repeating the above steps, all dependence of $x$ can be integrated and only tensor products consisting 
of constant terms should be left. In the cases met in this work these consisted always of powers of logarithms of constants.
These terms can again be integrated by using the ordered tensor products 
(in the sense of eq.\ (\ref{eq:constantordering}) above)  to construct the integrated expression.

In the case that the expression at hand depends on more than one variable, 
we can proceed in a completely analogous manner. For the case of two, $x_1$ and $x_2$, 
it is often desired to rewrite it in the form of GHPLs
\be
G(\{f_i(x_2)\};x_1) \quad \text{and} \quad G(\{c_i\};x_2)\quad c_i \in \C\;.
\ee
We cast the tensor product factors in $\Delta_{1,\ldots,1}( F(x_1,x_2))$ into the form
\be
\log\left(1-\frac{x_1}{f_i(x_2)}\right) \quad \text{or} \quad \log\left(1-\frac{x_2}{c_i}\right) 
\quad \text{or} \quad \log(c_i) \quad  \text{with} \; c_{i} \in \C^*\;,
\ee
and focus on the terms containing factors dependent on $x_1$ first. We proceed exactly as above, i.e. 
at step $i \leq n$ we then pick the terms
\be
\left( \bigotimes_{j=0}^{i-1} \log\left(1-\frac{x_2}{c_j}\right)\right) 
\otimes \left(\bigotimes_{k=i}^{n} \log \left(1-\frac{x_1}{f_k(x_2)}\right)\right)
\ee
which can be used to construct the parts proportional to
\be
G( c_{i-1} ,\ldots, c_1;x_2)\,G( f_n (x_2) ,\ldots, f_i (x_2) ;x_1)  \;
\ee
using the relation from equation (\ref{eq:gpldecompform}).
This way, we can integrate all tensor products containing $x_1$ first and are then left with a symbol 
depending on $x_2$ only, which can be treated as above.

\subsection{Coproduct simplification procedure}
We now have all the tools necessary to use the coproduct to perform transformations of GHPLs. 
Let us suppose that the expression under consideration is uniform in transcendentality. 
Parts with different transcendentalities can be treated separately.
For reasons that will become apparent shortly, we applied this algorithm to single GHPLs, with the entries of the index vector
 chosen from a specific set, as input, starting at weight one up to weight four.

Parting from an expression $F$, we first compute its symbol $\Delta_{1,\ldots,1}(F)$ and use 
the techniques described in the previous subsection to integrate it.
Then we reconstruct the parts of $F$ that lie in $\cH_{i_1 \ldots i_k}$ by applying 
the different operators $\Delta_{i_1\ldots i_k}$ to the difference of the original expression and 
the expression integrated from previous steps. 
Again, we work from bottom up, e.g. at weight 3 we have the an ordering of operators
\be
\Delta_{1,1,1} \succ \Delta_{2,1} , \Delta_{1,2}\;,
\ee
and at weight 4
\be
\Delta_{1,1,1,1} \succ \Delta_{2,1,1}, \Delta_{1,1,2}, \Delta_{1,2,1} \succ \Delta_{3,1}, \Delta_{2,2}, \Delta_{1,3}\;.
\ee
which we apply in consecutive order. 

To bring the expression to its unique form, i.e. everything is expressed in a unique set of GHPLs, we need additional input. 
For example, let us suppose that the expression we want to re-express contains $G(a,b,c;x)$. Computing
\be\bsp\label{eq:211example}
\Delta_{2,1,1}& G(a,b,c;x) = G(a, b; x)\otimes \log\left(1 - \frac{b}{c}\right) + G(a, c; x)
\otimes \log\left(1 - \frac{a}{b}\right) \\
& - G(a, c; x)\otimes \log\left(1 - \frac{c}{b}\right) - G(b, c; x)
\otimes \log\left(1 - \frac{b}{a}\right) + G(b, c; x)\otimes \log\left(1 - \frac{x}{a}\right)\;,
\esp\ee
we note that we now also need the identities re-expressing the lower weight functions 
$G(a,b;x)$, $G(a,c;x)$ and $G(b,c;x)$ as well as the limits  $\log(1 - b/c)$, $\log(1 - a/b)$, $\log(1 - c/b)$ and $\log(1- x/a)$. 
The lower weight identities have to be derived in a previous step whereas the limits have to be computed separately. 
In principle, these could be ill-defined since we do not have a cut prescription how these limits should be taken. 
However, this ambiguity is only up to the imaginary part, i.e. parts proportional to $\pi$. 
As we can see in the definition of the coproduct, eq. (\ref{eq:coproduct_def}), these limits only show up in the second factor, 
whereas the integration bounds of the first factor are the one of the original expression. 
Therefore we only need a cut prescription for the original expression.
On the other hand we know that tensor products containing $\pi$ anywhere but in the first factor vanish, 
which allows us to ignore the ambiguous terms from taking these limits.

Applying all these identities, after lots of cancellations, 
we obtain a part of the original expression that would have been lost in the symbol map. 
At weight three, $\Delta_{2,1}$ yields the parts proportional to $\pi^2$, and at weight four
\be\bsp
\Delta_{2,1,1} \quad & \text{yields parts } \propto \pi^2 \;,\\
\Delta_{3,1} \quad &  \text{yields parts } \propto \zeta_3,\,i \pi^3\;, \\
\Delta_{2,2} \quad &  \text{yields parts } \propto \pi^2\;, \\
\Delta_{1,3} \quad & \text{yields parts } \propto \zeta_3,\;.
\esp\ee
For example, let us suppose that we have an expression $F$ of weight four whose integrated symbol is also 
including the integrated symbol of the imaginary part, $F_{1,1,1,1}$. 
Applying $\Delta_{2,1,1}$ to the difference and using all identities of lower weight needed yields
\be
\Delta_{2,1,1}(F-F_{1,1,1,1})=\sum c_{i_1 i_2} \pi^2 \otimes a_{i_1} \otimes a_{i_2},
\ee
which can be integrated interpreting each summand as $\pi^2 (a_{i_1} \otimes a_{i_2})$, i.e. as $\pi^2$ times a symbol of weight 2. 
There cannot be anything other than $\pi^2$ in the first factor since any other such factor 
would also have produced a contribution to the symbol of the expression.
For the same reasons the operators $\Delta_{1,2,1}$ and $\Delta_{1,1,2}$ do not yield additional information.
It was found that in the cases treated in this work the operators
\be\bsp
\Delta_{1,2} & \quad \text{at weight 3 and }\\
\Delta_{1,2,1},\Delta_{1,1,2},\Delta_{2,2} &\quad\text{at weight 4}
\esp\ee
could be ignored for this reason \cite{Anastasiou:2013srw}.

Due to the symmetry of $\Delta(\zeta_3)=\zeta_3\otimes 1 + 1\otimes \zeta_3$ it was also not necessary 
to study the expression under the action of $\Delta_{1,3}$ either, once the result of the application 
of $\Delta_{3,1}$ was known. During the preparation of this work the only object encountered that 
did not obey this symmetry was one transcendental constant,
\be\label{eq:thetadef}
\theta_4 = \frac{\log^4 2}{24}+\li_4 \left(\frac{1}{2}\right)-\frac{1}{24} \pi^2 \log^2 (2) +\frac{7}{8} \log(2) \zeta_3
\ee
which vanishes under $\Delta_{3,1}$ but $\Delta_{1,3}(\theta_4) = \frac{7}{8}(\log( 2) \otimes \zeta_3)$. 
Curiously, one obtains $\theta_4$ when attempting to extend the familiar combination $\frac{\log^4 (2)}{24}+\li_4 \left(\frac{1}{2}\right)$, which has vanishing symbol, to a constant that also vanishes under the action of $\Delta_{2,1,1}$ and $\Delta_{3,1}$.

In summary, it was found to be sufficient to only use the operators $\Delta_{1,1,1}$, $\Delta_{2,1}$ 
at weight three and $\Delta_{1,1,1,1}$, $\Delta_{2,1,1}$ and $\Delta_{3,1}$ at weight four for 
the computation and to leave determination of the prefactor of the constant $\theta_4$ to the next step.
This has the advantage of speeding up the computations considerably, since the action of fewer operators has to be computed. 
Furthermore, it also lowers the number of required additional identities that have to be computed separately. 
As it was shown in eq.\ (\ref{eq:211example}), limit identities are needed in order to simplify all tensor product factors but the first ones. 
With the simplifications above these identities are only needed at weight one where they become simple and can be computed automatically in many cases.

\subsection{Reconstructing the primitive elements of the coproduct}
After the previous steps we are almost done transforming the expression in the desired way. 
The only parts that we are still missing are the \emph{primitive elements}, e.g. the elements $a$ 
such that $\Delta'(a)=0$. In the cases encountered during preparation of this work 
these were multiples of transcendental constants. 
They can again be derived evaluating the expression for a limit where the value of the original expression is known exactly. 
Alternatively, one can also evaluate both sides numerically to high precision and apply 
the PSLQ algorithm on the difference to find the constants. In the cases treated in this work, 
we found the following constants to be sufficient:
\be\bsp
\zeta_2 = \frac{\pi^2}{6} \quad & \text{at weight 2,}\\
\zeta_3 \quad & \text{at weight 3,}\\
\zeta_4 = \frac{\pi^4}{90},\;\theta_4 \quad &\text{at weight 4.}
\esp\ee
As mentioned above, $\theta_4$ (\ref{eq:thetadef}) does not vanish under the coproduct. 
Nevertheless it  was the only expression ever encountered that was asymmetric under $\Delta_{1,3}$ 
and $\Delta_{3,1}$ and therefore it is the most efficient way to compute the contribution from 
the application of $\Delta_{1,3}$ after the contribution from $\Delta_{3,1}$ is known.

After this step the transformed expression has been constructed in its entirety. 
As it can be seen, the coproduct provides an extension of the symbol formalism which allows to keep almost 
all the information contained in the original expression and to keep the `guessing' to a minimum. 
An example application of this algorithm can be found in section~\ref{sec:Int213}.

\section{Topo A}
\label{sec:TopoA}
In the following section we will focus on the first planar topology, \textbf{Topo A}.
We will use this first case also to describe all details of the differential equation method.
As discussed above, the first planar topology has cuts in $s$ and $u$, so that 
we expect the MIs to be naturally expressed in functions of these two Mandelstam variables.
We can then get rid of $t$ using $s+t+u = 2Q^2$ and exploit what we know \textsl{a priori} 
deriving directly differential equations for the MIs in $s$, $u$ and $Q^2$.

In the next sections we derive the differential operators and we work out explicitly a one-loop
example, which will also allow us to discuss the method in detail.
The common normalization factor of all master integrals is
\begin{equation}
 S_\epsilon = \left[ (4 \pi)^\epsilon \, {\Gamma(1+ \epsilon) \, \Gamma^2(1-\epsilon) \over \Gamma(1-2\epsilon)} \right]\,.
\end{equation}

\subsection{Differential equations}

In order to derive the differential equations in $s$ and $u$
we choose the following set of invariants as linear combinations of~\eqref{eq:invariants}:
\begin{equation*}s_1 = p_1^2\,,\quad s_2 = p_2^2\,,\quad s_3 = q_1^2 - q_2^2\,,\quad 
  s_4 = q_1^2\,,\quad s_5 = (p_1+p_2)^2\,,\quad s_6 = (p_2 - q_1)^2, \label{eq:invariantsA}
\end{equation*}
with these definitions, applying the on-shell conditions we find:
$$s_1 = 0 \,, \quad s_2 = 0 \,, \quad s_3 = 0\,, \quad s_4 = Q^2 \,, \quad s_5 = s\,, \quad s_6 = u.$$

Expressing the derivatives with respect to the three non-zero invariants  as
linear combinations of derivatives with respect to the external momenta we obtain:

\begin{align}
 s\,{\partial \over \partial s} &=  \frac{s}{Q^4 - 2 u Q^2 + u^2 + s u}  \, \Bigg\{
           \frac{(Q^4 - 2 u Q^2 + u^2 + 2 s u)}{2 s}\,  p_1^\mu {\partial \over \partial \, p_1^\mu} \nonumber \\
          &+ \frac{(u-Q^2)}{2}\, \Bigg[ \frac{(s+u-Q^2)}{s}  p_2^\mu {\partial \over \partial \, p_1^\mu}
          -   q_1^\mu {\partial \over \partial \, p_1^\mu} \, \Bigg]\, \Bigg\}\,, \nonumber \\ 
&\nonumber \nonumber\\
u\,{\partial \over \partial u} &=   \frac{u}{Q^4 - 2 u Q^2 + u^2 + s u} \, \Bigg\{
           \frac{(u-Q^2)}{2} \, \Bigg[\, p_1^\mu {\partial \over \partial \, p_1^\mu}
          -   p_1^\mu {\partial \over \partial \, p_2^\mu} \,\Bigg]\nonumber  \\
          &+ \frac{(s+u-Q^2)}{2}\, \Bigg[\, p_2^\mu {\partial \over \partial \, p_2^\mu}
          -  p_2^\mu {\partial \over \partial \, p_1^\mu}\, \Bigg]
          + \frac{s}{2}\, \Bigg[\,  q_1^\mu {\partial \over \partial \, p_1^\mu}
          -  q_1^\mu {\partial \over \partial \, p_2^\mu} \,\Bigg] \, \Bigg\}\,, \nonumber \\ 
  &\nonumber \nonumber\\
 Q^2\,{\partial \over \partial Q^2} &=  \frac{Q^2}{Q^4 - 2 u Q^2 + u^2 + s u}  \, \Bigg\{
          \frac{(Q^4 - 2 u Q^2 + u^2 + 2 s u)}{2 s}\,  p_1^\mu {\partial \over \partial \, q_1^\mu} \nonumber \\
         &+ \frac{(u-Q^2)}{2}\,\Bigg[ \frac{(s+u-Q^2)}{s} \, p_2^\mu {\partial \over \partial \, q_1^\mu}
          -   q_1^\mu {\partial \over \partial \, q_1^\mu}
          -   p_1^\mu {\partial \over \partial \, p_1^\mu}\,\Bigg]\nonumber \\ 
         &+ \frac{(s+u-Q^2)}{2} \, p_2^\mu {\partial \over \partial \, p_1^\mu}          
          - \frac{s}{2} \, q_1^\mu {\partial \over \partial \, p_1^\mu} \, \Bigg\}\,. \label{eq:diffop}
\end{align}

We recall here that, due to Lorentz invariance identities~\cite{Gehrmann:1999as},
 these relations are not unique. 
Scalar Feynman integrals, in fact, must be Lorentz invariant. 
This implies that given any integral $I(s_j)$ the following relation must be fulfilled:

\begin{equation} \label{eq:Lorentz}
  \left\{\,\left( p_1^\mu \, {\partial \over \partial p_1^\nu} - p_1^\nu \, {\partial \over \partial p_1^\mu} \right)
+ \left( p_2^\mu \, {\partial \over \partial p_2^\nu} - p_2^\nu \, {\partial \over \partial p_2^\mu} \right) 
+ \left( q_1^\mu \, {\partial \over \partial q_1^\nu} - q_1^\nu \, {\partial \over \partial q_1^\mu} \right)\,\right\}I(s_j)  = 0.
\end{equation}

Contracting~\eqref{eq:Lorentz} with any antisymmetric combination of the external momenta 
we find relations which connect the different derivative operators 
when applied on a Feynman integral.
Having three independent external momenta we can build up three antisymmetric combinations:
\begin{align*}
 p_1^\mu p_2^\nu - p_1^\nu p_2^\mu\,, \qquad p_1^\mu q_1^\nu - p_1^\nu q_1^\mu \,, \qquad p_2^\mu q_1^\nu - p_2^\nu q_1^\mu.
\end{align*}

Starting from the 6 scalar products in~\eqref{eq:invariants}, the three Lorentz invariance relations 
state what we already know, i.e. that only three of them are really independent.
As an explicit example, upon contracting~\eqref{eq:Lorentz} with $p_1^\mu p_2^\nu - p_1^\nu p_2^\mu$ we find
\begin{equation*} 
\left\{\,s\left[ p_1^\mu \, {\partial \over \partial p_1^\mu} - p_2^\mu \, {\partial \over \partial p_2^\mu} 
      - p_2^\mu \, {\partial \over \partial q_1^\mu} \right]
+ (u-Q^2)\left[ p_1^\mu \, {\partial \over \partial q_1^\mu} - p_2^\mu \, {\partial \over \partial q_1^\mu}  \right] \,\right\} I(s_j)  = 0.
\end{equation*}

Furthermore, the differential operators~\eqref{eq:diffop} when applied on a Feynman
integral are not independent due to the scaling properties of the integrals.
Assuming for the mass dimension of the integral
\begin{equation}
I(\lambda^2\,s,\lambda^2\,t,\lambda^2\,u) = \lambda^{\alpha} \, I( s, t, u) 
\end{equation}
one finds the Euler scaling relation

\begin{equation}
\left(\, s\,{\partial \over \partial s}  + u\,{\partial \over \partial u} + Q^2\,{\partial \over \partial Q^2} 
\,\right) \,  I( s, t, u)  = {\alpha \over 2}\, I( s, t, u)\,.
\end{equation}

\subsection{A one-loop example}

As illustration of the method let us consider the one-loop triangle with 
three legs off-shell. The example is interesting for at least two different reasons.
First, it appears naturally as subtopology of some of the two-loop integrals considered in the following.
Second, in spite of its simplicity, it will also allow us to discuss some of the features
of the computation of the more involved two-loop integrals.

We consider the one-loop triangle with three legs off-shell, two of which with the same mass.
With our notation it is a function of the ratio $\tilde{s} = s/Q^2$ and can be defined as:

\begin{equation}
\T(\tilde{s}) = \triangleoneloop{p_{12}}{q_1}{q_2} = \int \frac{d^d k}{(2 \pi)^d} \frac{1}{k^2(k-q_1)^2(k-p_1-p_2)^2} \label{eq:1ltri}
\end{equation}

We define also the one-loop bubble, which is its only subtopology, as:
\begin{equation}
\B(p^2) = \bubbleoneloop{p} = \int \frac{d^d k}{(2 \pi)^d} \frac{1}{k^2(k-p)^2}.
\end{equation}

Applying the differential operators~\eqref{eq:diffop} directly on the definition~\eqref{eq:1ltri},
and using the IBPs to re-express the resulting integrals in terms of the master $T(\tilde{s})$ itself and the bubbles, we find:

\begin{align}
  s\,{\partial \over \partial s} \T(\tilde{s}) &=  
-\left[ {d-2 \over 2} + {2 Q^2 (d-3) \over s-4 Q^2}\right]\,\T(\tilde{s})
+ \frac{2\,(d-3)}{s-4 Q^2} \left[ \,\B(Q^2) - \B(s)\, \right]\,, \\
 u\,{\partial \over \partial u} \T(\tilde{s}) &=  0\,, \\
 Q^2\,{\partial \over \partial Q^2} \T(\tilde{s})&=  \left[ {d-4} + {2 Q^2 (d-3) \over s-4 Q^2}\right]\,\T(\tilde{s})
+ {2 (d-3) \over s-4 Q^2}\,\left[ \B(s) - \B(Q^2) \, \right]\,.
\end{align}

Note that summing the three equation we find the expected Euler scaling-relation:
\begin{equation}
\left( s\,{\partial \over \partial s} + u\,{\partial \over \partial u} + Q^2\,{\partial \over \partial Q^2} \right) \,T(\tilde{s}) 
= {(d-6) \over 2} \,  T(\tilde{s}) \,.
\end{equation}

We focus now on the differential equation in $s$ and we attempt to solve it 
as a series expansion in $\epsilon = (4-d)/2$.
Since the one-loop bubble develops a $1/\epsilon$ pole,

\begin{align*}
 \B(p^2) &= i\,\left( S_\epsilon \over 16 \pi^2 \right)\, (-p^2)^{-\epsilon}\, {1 \over \epsilon (1-2\epsilon)}\\
 &= i\, \left( S_\epsilon \over 16 \pi^2 \right)\, (-p^2)^{-\epsilon}\,\left[{1 \over \epsilon} + 2 + 4\,\epsilon + \mathcal{O}(\epsilon^2) \right]\,,
\end{align*}
we should consistently expand also the function $\T(\tilde{s})$ starting from $1/\epsilon$,

\begin{align}\label{eq:1loopTriangle}
 \T(\tilde{s},\epsilon) = i\,\left( S_\epsilon \over 16 \pi^2 \right)\,(-Q^2)^{-\epsilon}
  \left[ \,{1 \over \epsilon}\, \T^{(-1)}(\tilde{s}) + \T^{(0)}(\tilde{s}) + \epsilon\,\T^{(1)}(\tilde{s}) + \mathcal{O}(\epsilon^2)\,\right] \,.
\end{align}

Inserting the expansions on both sides of the differential equation and keeping only 
the first two terms in the expansion we get the following differential equations for the Laurent 
coefficients of the first two orders:
\begin{align}
 &{\partial \over \partial s} \T^{(-1)}(\tilde{s}) = 
  - {1 \over 2} \,\left[ {1 \over s} + {1 \over s-4 Q^2}\right]\, \T^{(-1)}(\tilde{s})\,, \label{eq:ord-1}\\
 &{\partial \over \partial s} \T^{(0)}(\tilde{s})  = - {1 \over 2} \,\left[ {1 \over s} + {1 \over s-4 Q^2}\right]\, \T^{(0)}(\tilde{s}) \nonumber\\
 & \qquad +  {1 \over s-4 Q^2}\,\T^{(-1)}(\tilde{s}) 
   + {1 \over 2 Q^2}\, \left[ {1 \over s} - {1 \over s-4 Q^2}\right]\, \left( \, \ln{(-Q^2-i \, \eta)} - \ln{(-s-i \, \eta)} \,\right)\,, \label{eq:ord0}
\end{align}
where $\eta$ is a small positive real number which comes from the usual prescription 
$$s\to s+ i \eta\,,\quad Q^2 \to Q^2 + i\eta\,.$$
Note that the homogeneous part of the equation is the same at any order in $\epsilon$.

Since in the equation for order $\mathcal{T}^{(n)}(s)$ the previous orders appear as
inhomogeneous terms,
we must solve the equation bottom up in $\epsilon$ starting from the leading singularity term.
At order $1/\epsilon$ the equation is purely homogeneous and its solution is easily found as
\begin{equation}
 \T^{(-1)}(\tilde{s}) =  {C^{(-1)} \over \sqrt{s\,(s-4 Q^2)}} \,, \label{eq:squareroot}
\end{equation}
where $C^{(j)}$ is the integration constant at order $\epsilon^j$.

Upon matching this solution with an appropriate boundary condition one finds that 
in this particular case $C^{(-1)} = 0$, so that $$\T^{(-1)}(\tilde{s}) = 0\,.$$

Inserting this result into~\eqref{eq:ord0} we find a linear first-order 
non-homogeneous differential equation for $\T^{(0)}(\tilde{s})$, which can be solved 
using Euler's method of the variation of the constants:

\begin{align}
 \T^{(0)}(\tilde{s}) &=  {2 \over \sqrt{s\,(s-4 Q^2)}} \Bigg\{\int\, ds\, 
 {1 \over \sqrt{s\,(s-4 Q^2)}}\, \Big[ \, \ln{(-s-i \, \eta)}  - \ln{(-Q^2-i \, \eta)}\,\Big]
 + C^{(0)}
\Bigg\}
\end{align}

The square root $\sqrt{s\,(s-4 Q^2)}$ shows up in all computations of three- and four-point functions with respectively 3 and 2 legs 
off-shell (see for example~\cite{Aglietti:2004tq,Bonciani:2003te,Bonciani:2003cj})
and physically it represents the threshold for the production of the two massive particles of equal masses $Q^2$.

In general, integrating a square root with a combination of (poly-)logarithmic functions 
can be a quite non-trivial task and we expect the integrals to
become more and more involved as the order in $\epsilon$ increases.
Nevertheless, as it is well known, in this specific case one can get rid of the square-root 
through the usual Landau variable $\xi$ defined above.
In order to obtain a real result we use the parametrization defined in section~\ref{sec:kinematicsA}
as function of $x,z$ finding:

\begin{equation} \label{eq:Landau}
s \to -m^2{ (1+x)^2 \over x }\,, \qquad ds \to  m^2 \frac{(1-x^2)}{x^2}\,dx \,, \qquad \mbox{with} \quad Q^2 = -m^2\,, 
\end{equation}
so that
\begin{equation*}
 \int\, ds\, {1 \over \sqrt{s\,(s-4 Q^2)}} \to  \int {d x \over x }\,.
\end{equation*}

The arguments of the logarithms are now positive so that the explicit imaginary part can be dropped and we obtain:
\begin{align}
 \T^{(0)}(x) &=  - {2\,x \over m^2\,(1-x^2)} \left\{\int\, dx\, 
  \left[ \, {\ln{(x)} \over x }\, - 2\,{\ln{(1+x)} \over x }\, \right]
 + C^{(0)}
\right\}\,,
\end{align}
which can be now easily integrated in terms of HPLs~\cite{Remiddi:1999ew}. Fixing the boundary condition we find:
\begin{align}
 \T^{(0)}(x) &=  -{2\,x \over m^2\,(1-x^2)} \left\{ 
   \,  G(0,0,x) - 2\,G(0,-1,x)\,  
 + {\pi^2 \over 6} \,\right\}\,.
\end{align}
Iterating this procedure one can compute the function $\T(\tilde{s})$ up to any order in $\epsilon$.

Given the result in the non-physical region, the analytic continuation to the physical region can be
achieved as described above, i.e. with
$$m^2 \to - M^2 - i \eta\,, \quad x \to \xi\,\,$$
which has to be consistently performed also on the pre-factors in~\eqref{eq:1loopTriangle}.

\subsection{GHPLs as functions of \texorpdfstring{$(x,z)$}{(x,z)}}\label{sec:TopoAGHPLs}
Proceeding as outlined in the case of the one-loop triangle, we can derive the set of differential
equations in $(x,z)$ fulfilled by the MIs in \textbf{Topo A}.
By direct inspection of the denominators in the differential equations 
we can read off what the alphabet of all possible indices 
appearing in the GHPLs will be. 
We find that in the non-physical region only the following functions can appear:
\be\bsp
G(a_1,\ldots,a_n;x) \quad & \text{with} \;a_i 
\in \left\{ 0,\pm 1, -c, -\bar{c}\right\}\;,\\
G(a_1,\ldots,a_n;z) \quad & \text{with} \;a_i \in \{0,1, -x, -\frac{1}{x},-I_x,-\frac{1}{I_x} \}\;.
\esp\ee
where
\be\bsp
c &= \frac{1}{2}\left(1+i \sqrt{3}\right)\;,\\
I_\alpha &=\frac{\alpha}{1+\alpha+\alpha^2}\;.
\esp\ee
After analytical continuation to the physical region (\ref{eq:TopoAAC}) this becomes:
\be\bsp\label{eq:TopoAalphabet}
G(a_1,\ldots,a_n;\xi) \quad & \text{with} \;a_i 
\in \left\{ 0,\pm 1, -c, -\bar{c}\right\}\;,\\
G(a_1,\ldots,a_n;\zeta) \quad & \text{with} \;a_i \in \{0\,,\,-1\,,\, \xi\,, \frac{1}{\xi}\,,\,I_\xi\,,\,\frac{1}{I_\xi} \}\;.
\esp\ee
Note that this does not imply that \textsl{all} indices in \eqref{eq:TopoAalphabet}
will necessarily be needed in order to represent the physical result, it only provides us with the
largest set possible of allowed indices. Nevertheless, by direct computation we verified that indeed 
the full set of indices is needed in order to represent all MIs up to transcendentality 4.

\subsection{Master Integrals}
We list here all the genuine 2-loop master integrals that appear in the reduction, giving an explicit form in terms of propagators
for those topologies that have more than one master. 
In order to write down the integrals we introduce the following notation for the momenta:
$$p_{12} = p_1 + p_2\,, \qquad p_{13} = p_1 - q_1\,,\qquad p_{23} = p_2 - q_1\,, 
\qquad p_{123} = p_1 + p_2 - q_1 \,.$$

\subsubsection*{One-scale integrals}
There are three trivial one-scale topologies, whose values can be computed directly from their Feynman parameters representation
and have been known for a long time. 
%Following the notation in~\cite{Gehrmann:2000zt,Gehrmann:2001ck} we have
We have
\begin{eqnarray}
\mathcal{I}_{38}^{(A)}(s)  = \bubbleNLO{p_{12}} \qquad
\mathcal{I}_{53}^{(A)}(s)  = \triangleOne{p_{12}}{p_1}{p_2}
\end{eqnarray}
\begin{eqnarray}
\mathcal{I}_{134}^{(A)}(Q^2)=\mathcal{I}_{38}^{(A)}(Q^2)   \qquad &
\mathcal{I}_{148}^{(A)}(u)=\mathcal{I}_{38}^{(A)}(u) 
\end{eqnarray}

\subsubsection*{Two-Scale integrals}
All seven two-scale vertex functions have already been computed in~\cite{Gehrmann:2000zt,Gehrmann:2001ck} for two off-shell 
legs and in~\cite{Birthwright:2004kk,Chavez:2012kn} for three off-shell legs. 
In the latter papers
all three-point functions with three different external masses have been computed. 
In the context of this work we re-computed the planar ones in the case of two equal masses.
Note that they can all be conveniently expressed as functions of 
one of the two ratios $\tilde{s} = s/Q^2$ and $\tilde{u} = u/Q^2$.

Among the irreducible topologies with four denominators two have a single master integral:
\begin{eqnarray}
\mathcal{I}^{(A)}_{142}(\tilde{u}) = \triangleTwox{q_2}{p_1}{p_{23}} \qquad
\mathcal{I}^{(A)}_{149}(\tilde{u}) = \triangleTwo{q_2}{p_1}{p_{23}} 
\end{eqnarray}
while two have two master integrals each. We choose the following basis:
\begin{eqnarray}
\mathcal{I}^{(A)}_{166,1}(\tilde{s}) &=&\triangleThreex{p_{12}}{q_2}{q_1} =
\int \frac{d^d k}{(2 \pi)^d} \frac{d^d l}{(2 \pi)^d} \frac{1}{l^2(k-l)^2(k-p_{12})^2(k-p_{123})^2}\,, \\
\mathcal{I}^{(A)}_{166,2}(\tilde{s}) &=&\triangleThreexdot{p_{12}}{q_2}{q_1} =
\int \frac{d^d k}{(2 \pi)^d} \frac{d^d l}{(2 \pi)^d} \frac{1}{l^2(k-l)^4(k-p_{12})^2(k-p_{123})^2}\,
\end{eqnarray}
and 
\begin{eqnarray}
\mathcal{I}^{(A)}_{198,1}(\tilde{s}) &=& \triangleThree{p_{12}}{q_2}{q_1} =
\int \frac{d^d k}{(2 \pi)^d} \frac{d^d l}{(2 \pi)^d} \frac{1}{l^2(k-l)^2(l-p_{12})^2(k-p_{123})^2}\,, \\
\mathcal{I}^{(A)}_{198,2}(\tilde{s}) &=& \triangleThreedot {p_{12}}{q_2}{q_1} =
\int \frac{d^d k}{(2 \pi)^d} \frac{d^d l}{(2 \pi)^d} \frac{1}{l^2(k-l)^4(l-p_{12})^2(k-p_{123})^2}\,.
\end{eqnarray}

Finally, there are three different irreducible two-scale topologies with 5 denominators, 
all with one single master integral:
\begin{eqnarray}
\mathcal{I}^{(A)}_{398}(\tilde{u}) = \trianglebTwo{q_2}{p_1}{p_{23}}\,,
\end{eqnarray}
\begin{eqnarray}
\mathcal{I}^{(A)}_{199}(\tilde{s}) =\trianglebThree{q_2}{p_{12}}{q_1}\,, \qquad 
\mathcal{I}^{(A)}_{422}(\tilde{s}) = \trianglebThreex{q_2}{p_{12}}{q_1}\,. 
\end{eqnarray}

\subsubsection*{Three-Scale Integrals}
We found eight nontrivial three-scale topologies
which depend on both ratios $(\tilde{s},\tilde{u})$.

There are four irreducible topologies with 5 denominators, out of which two have one master integral:
\begin{eqnarray}
\mathcal{I}^{(A)}_{174} &=& \boxbubblea{p_1}{p_2}{q_2}{q_1}\,,  \qquad
\mathcal{I}^{(A)}_{214} = \boxbubbleb{p_1}{p_2}{q_2}{q_1} \,,
\end{eqnarray}
and two have two master integrals each, that we choose to be:
\begin{eqnarray}
\mathcal{I}^{(A)}_{181,1} &=& \boxbubblec{p_1}{p_2}{q_2}{q_1} \nonumber \\ &=& 
\int \frac{d^d k}{(2 \pi)^d} \frac{d^d l}{(2 \pi)^d} \frac{1}{k^2(k-l)^2(l-p_1)^2(k-p_{12})^2(k-p_{123})^2}\,, \\
\mathcal{I}^{(A)}_{181,2} &=& \boxbubblecdot{p_1}{p_2}{q_2}{q_1} \nonumber \\ &=& 
\int \frac{d^d k}{(2 \pi)^d} \frac{d^d l}{(2 \pi)^d} \frac{1}{k^2(k-l)^4(l-p_1)^2(k-p_{12})^2(k-p_{123})^2}\,,
\end{eqnarray}
\begin{eqnarray}
\mathcal{I}^{(A)}_{182,1} &=& \boxxbNLO{p_1}{p_2}{q_2}{q_1}  \nonumber \\ &=&
\int \frac{d^d k}{(2 \pi)^d} \frac{d^d l}{(2 \pi)^d} \frac{1}{l^2(k-l)^2(l-p_1)^2(k-p_{12})^2(k-p_{123})^2}\,, \\
\mathcal{I}^{(A)}_{182,2} &=& \boxxbdotNLO{p_1}{p_2}{q_2}{q_1} \nonumber \\ &=& 
\int \frac{d^d k}{(2 \pi)^d} \frac{d^d l}{(2 \pi)^d} \frac{1}{l^2(k-l)^2(l-p_1)^2(k-p_{12})^2(k-p_{123})^4}\,.
\end{eqnarray}

The two irreducible 6-denominator topologies have both one single master integral:
\begin{eqnarray}
\mathcal{I}^{(A)}_{215} &=& \boxxbpaNLO{p_1}{p_2}{q_2}{q_1}\,, \qquad 
\mathcal{I}^{(A)}_{430} = \boxxbpbNLO{p_1}{p_2}{q_2}{q_1}\,.
\end{eqnarray}

Finally there are two irreducible 7-denominator topologies,
one with three master integrals:
\begin{eqnarray}
\mathcal{I}^{(A)}_{247,1} &=& \doubleboxa{p_1}{p_2}{q_2}{q_1}  \nonumber \\ &=&
\int \frac{d^d k}{(2 \pi)^d} \frac{d^d l}{(2 \pi)^d} \frac{1}{k^2l^2 (k-l)^2(l-p_1)^2(k-p_{12})^2(l-p_{12})^2(k-p_{123})^2}\,, \\
\mathcal{I}^{(A)}_{247,2} &=& \doubleboxatwo{p_1}{p_2}{q_2}{q_1} \nonumber \\ &=&
\int \frac{d^d k}{(2 \pi)^d} \frac{d^d l}{(2 \pi)^d} \frac{(k-p_1)^2}{k^2l^2 (k-l)^2(l-p_1)^2(k-p_{12})^2(l-p_{12})^2(k-p_{123})^2}\,, \\
\mathcal{I}^{(A)}_{247,3} &=& \doubleboxathree{p_1}{p_2}{q_2}{q_1} \nonumber \\ &=&
\int \frac{d^d k}{(2 \pi)^d} \frac{d^d l}{(2 \pi)^d} \frac{(l-p_{123})^2}{k^2l^2 (k-l)^2(l-p_1)^2(k-p_{12})^2(l-p_{12})^2(k-p_{123})^2}\,
\end{eqnarray}
and one with two master integrals:
{\small
\begin{eqnarray}
\mathcal{I}^{(A)}_{446,1} &=& \doubleboxb{p_1}{p_2}{q_2}{q_1} \nonumber \\ &=&
\int \frac{d^d k}{(2 \pi)^d} 
\frac{d^d l}{(2 \pi)^d} \frac{1}{l^2 (k-l)^2(k-p_1)^2(l-p_1)^2(k-p_{12})^2(k-p_{123})^2(l-p_{123})^2}\,, \\
\mathcal{I}^{(A)}_{446,2} &=& \doubleboxbtwo{p_1}{p_2}{q_2}{q_1}\nonumber \\ &=&
\int \frac{d^d k}{(2 \pi)^d} \frac{d^d l}{(2 \pi)^d} \frac{k^2}{l^2 (k-l)^2(k-p_1)^2(l-p_1)^2(k-p_{12})^2(k-p_{123})^2(l-p_{123})^2}\,.
\end{eqnarray}
}

As it is well known the basis is not unique. 
The choice that we made was only motivated by the fact that with it we obtain systems of differential
equations that decouple in $d\to 4$ and allow a direct integration of the MIs.

The results for the masters computed in the non-physical region 
up to weight 3 can be found in Appendix~\ref{A:MIsTopoA}.
The full result up to weight 4 and
the analytical continuation to the physical region relevant for vector boson pair production
is attached to the arXiv submission of this paper.
In addition, we also provide the one-loop $\times$ one-loop integrals required for the full reduction in the same set of variables.
Their naming convention follows the same scheme outlined in section~\ref{sec:auxtopo}.

\section{Topo B}
\label{sec:TopoB}
We consider now the second topology which collects all planar masters with two 
non-adjacent massive legs. As already discussed, \textbf{Topo B} has cuts in $t$ and $u$. 
This suggests as first attempt to try to derive and solve
the differential equations in these variables. The kinematics and the analytical continuation
to the proper physical region has been described in section~\ref{sec:kinematicsB}.

Proceeding in the same way as outlined in section~\ref{sec:TopoA} we start from
the following set of invariants:
\begin{equation*}s_1 = p_1^2\,,\quad s_2 = p_2^2\,,\quad s_3 = q_1^2 - q_2^2\,,\quad 
  s_4 = q_1^2\,,\quad s_5 = (p_1-q_1)^2\,,\quad s_6 = (p_2 - q_1)^2, \label{eq:invariantsB}
\end{equation*}
where, applying the on-shell conditions we find:
$$s_1 = 0 \,, \quad s_2 = 0 \,, \quad s_3 = 0\,, \quad s_4 = Q^2 \,, \quad s_5 = t\,, \quad s_6 = u.$$
All the discussion done for \textbf{Topo A} applies also here, 
but for brevity we omit the explicit form of the differential operators in this case.

As discussed already in section~\ref{sec:TopoA}, the kinematics for
vector boson pair production generates naturally the square root
$\sqrt{s(s-4Q^2)}$.
This means that, if in order to highlight the symmetries of this topology
we choose not to use the Landau variable~\eqref{eq:Landau},
we can expect that in turn there will appear this same square root
in the homogeneous solutions of some of the differential equations.
On the other hand, introducing the Landau variable would oblige us to get rid of
either $t$ or $u$ in favor of $s$, and so to give up
the symmetry in $(t,u)$.

Nevertheless, limiting ourselves to considering 
only the MIs in \textbf{Topo B}, because
of their cut structure, 
this square root appears in only one single master integral. 
As it will be discussed in more detail below, 
this integral belongs to sector 213 which has five denominators and is reduced to four MIs.
This could easily generate a problem since, in the differential equation approach, 
the integral can be expected to affect the differential equations of the other three
MIs of its sector, plus all the topologies with a larger number of denominators 
that contain it as sub-topology.
In spite of that, by direct computation we found that with an appropriate choice
of the basis, the integral above 
``decouples''  from the rest of the integrals in its sector and also from all the integrals 
with a larger number of denominators, in a sense more clearly specified below.
This allows us to compute all the other integrals in \textbf{Topo B} 
without having to explicitly carry out the integration of this integral.
The computation of this last integral has been performed in the physical region reintroducing
the Landau variable $\xi$ as described in the following.

\subsection{GHPLs as functions of \texorpdfstring{$(y,z)$}{(y,z)}}\label{sec:TopoBGHPLs}
As for \textbf{Topo A}, the alphabet of indices for the GHPLs needed to describe all
MIs in \textbf{Topo B} can be read off directly from the differential equations.
By direct inspection
we find that all integrals except 
one can be written through the following set of GHPLs:
\be\bsp
G(a_1,\ldots,a_n;z) \quad & \text{with} \;a_i 
\in \{0,1,2\}\;,\\
G(a_1,\ldots,a_n;y) \quad & \text{with} \;a_i \in \{0,1,2-z,\frac{1}{z}\}\;,
\esp\ee
which becomes after analytical continuation (\ref{eq:TopoBAC})
\be\bsp
G(a_1,\ldots,a_n;v) \quad & \text{with} \;a_i 
\in \{0,-1,-2\}\;,\\
G(a_1,\ldots,a_n;w) \quad & \text{with} \;a_i \in \{0,-1,v,-\frac{1}{2+v}\}\;.
\esp\ee

One of the MIs, $\mathcal{I}^{(B)}_{213,1}$,
had to be integrated in $\xi$ and $\zeta$ (see section~\ref{sec:Int213}). 
Again, we can predict the full set of indices needed to describe the result deriving 
the differential equations fulfilled by the MI in these variables.
In order to gain a more thorough understanding, instead of doing it only for this MI,
we derived the differential equations in $(\xi,\zeta)$ for 
the full set of MIs in \textbf{Topo B}. We find that they can be expressed through
the following set of GHPLs:
\be\bsp
G(a_1,\ldots,a_n;\xi) \quad & \text{with} \;a_i 
\in \{0,-1,\pm i, -c, -\bar{c} \}\;,\\
G(a_1,\ldots,a_n;\zeta) \quad & \text{with} \;a_i \in 
\left\{ 0,-1,\xi,\frac{1}{\xi},\frac{1}{I_\xi},\frac{1}{J_\xi}\right\}\;.
\esp\ee
Note the appearance of the additional indices
\begin{align}
&\frac{1}{J_\xi} = \frac{1+\xi^2}{\xi}\,, \quad \mbox{in GHPLs with argument }\, \zeta\,, \nonumber \\
&\pm i \,, \quad \mbox{in GHPLs with argument }\, \xi\,,
\end{align}
compared to the alphabet of \textbf{Topo A}, eq.~(\ref{eq:TopoAalphabet}).
These indices are indeed needed in order to describe the cut in $t$ using instead $s$ and $u$
(expressed through $\xi$ and $\zeta$) as independent variables.

\subsection{Master Integrals}
As for the first topology, a catalog of all master integrals appearing in the 
reduction of this topology can be established. 
Since all one- and two-scale integrals are identical to those appearing in \textbf{Topo A} 
up to a permutation of the external legs,
we do not give their explicit expressions here for brevity. 
Nevertheless, depending on which set of variables we choose to express these MIs in, 
the identities to perform the transformation can be highly nontrivial.
In the differential equation method, in order to derive a sensible differential equation for a MI
we need to express all its sub-topologies in the same set of variables.
We introduce then a notation for these masters and give their explicit expression.
In the appendices and the attached Mathematica and FORM files the variables used are the 
ones that produce the most compact expressions; 
the results in all sets of variables can be obtained by the authors.

\subsubsection*{One-scale Integrals}
The one-scale integrals needed for the full reduction of \textbf{Topo B} are simply
\be
\mathcal{I}^{(A)}_{38}(t)\,,\quad\mathcal{I}^{(A)}_{38}(u)\,.
\ee
% $$A_3(t)\,,\quad A_3(u)\,,\quad A_{22}(t)\,,\quad A_{22}(u)\,.$$

\subsubsection*{Two-scale Integrals}
The two-scale integrals needed are
$$\mathcal{I}^{(A)}_{149}(u) \,,\quad \mathcal{I}^{(A)}_{142}(u)\,, $$
\begin{eqnarray}
\mathcal{I}^{(B)}_{46}(t) = \mathcal{I}^{(A)}_{149}(u \leftrightarrow t)\,, \qquad 
\mathcal{I}^{(B)}_{53}(t) = \mathcal{I}^{(A)}_{142}(u \leftrightarrow t)\,.
\end{eqnarray}
\begin{eqnarray}
\mathcal{I}^{(B)}_{110}(t) = \mathcal{I}^{(A)}_{398}(u \leftrightarrow t)\,.
\end{eqnarray}

\subsubsection*{Three-scale Integrals}
We find three irreducible 5-denominator topologies, two with one single master integral:
\begin{eqnarray}
\mathcal{I}^{(B)}_{174} &=& \boxbubbleaB{p_1}{q_1}{q_2}{p_2}\,,  \qquad
\mathcal{I}^{(B)}_{182} = \boxxaB{p_1}{q_1}{q_2}{p_2}\,,
\end{eqnarray}
and one with four master integrals. We choose them as:
\begin{eqnarray}
\mathcal{I}^{(B)}_{213,1} &=& \boxxbB{p_1}{q_1}{q_2}{p_2}  \nonumber \\ &=&
\int \frac{d^d k}{(2 \pi)^d} \frac{d^d l}{(2 \pi)^d} \frac{1}{k^2 (k-l)^2(l-p_1)^2(l-p_{13})^2(k-p_{123})^2}\,, \\ 
\mathcal{I}^{(B)}_{213,2} &=& \boxxbbB{p_1}{q_1}{q_2}{p_2} \nonumber \\ &=&
\int \frac{d^d k}{(2 \pi)^d} \frac{d^d l}{(2 \pi)^d} \frac{1}{k^2 (k-l)^4(l-p_1)^2(l-p_{13})^2(k-p_{123})^2}\,, \\ 
\mathcal{I}^{(B)}_{213,3} &=& \boxxbbbB{p_1}{q_1}{q_2}{p_2} = \boxxbbbbbB{p_1}{q_1}{q_2}{p_2}
\nonumber \\ &=&
\int \frac{d^d k}{(2 \pi)^d} \frac{d^d l}{(2 \pi)^d} \frac{1}{k^2 (k-l)^2(l-p_1)^2(l-p_{13})^4(k-p_{123})^2}\,, \\ 
\mathcal{I}^{(B)}_{213,4} &=& \boxxbbbbB{p_1}{q_1}{q_2}{p_2} = \boxxbbbbbbB{p_1}{q_1}{q_2}{p_2} \nonumber \\ &=&
\int \frac{d^d k}{(2 \pi)^d} \frac{d^d l}{(2 \pi)^d} \frac{1}{k^2 (k-l)^2(l-p_1)^2(l-p_{13})^2(k-p_{123})^4} \,,\label{topo213}
\end{eqnarray}

Note that, in the case of sector $213$, a naive reduction
using only IBP identities would indeed produce five independent MIs. 
Nevertheless, using the symmetry relations of the topology under shift of the integration
variables we realize that one of these masters can be re-expressed as 
linear combination of the others, leaving four independent MIs.
Our choice for the basis of MIs makes this symmetry manifest.
This is an explicit example of a case where symmetry relations do contribute to 
reduce the number of independent masters.
The consistent inclusion of symmetry relations in the reduction to MIs
has been recently automated inside Reduze2.

There is only one irreducible 6-denominator topology, with one single master integral
\begin{eqnarray}
\mathcal{I}^{(B)}_{215} = \boxxbpaB{p_1}{q_1}{q_2}{p_2}\,.
\end{eqnarray}

Finally, we find only one irreducible 7-denominator topology, with two master integrals,
\begin{eqnarray}
\mathcal{I}^{(B)}_{247,1} &=& \doubleboxB{p_1}{q_1}{q_2}{p_2}  \nonumber \\ &=&
\int \frac{d^d k}{(2 \pi)^d} \frac{d^d l}{(2 \pi)^d} \frac{1}{k^2 l^2 (k-l)^2(l-p_1)^2(k-p_{13})^2(l-p_{13})^2(k-p_{123})^2}\,, \\ 
\mathcal{I}^{(B)}_{247,2} &=& \doubleboxBtwo{p_1}{q_1}{q_2}{p_2} \nonumber \\ &=&
\int \frac{d^d k}{(2 \pi)^d} \frac{d^d l}{(2 \pi)^d} \frac{(k-p_1)^2}{k^2 l^2 (k-l)^2(l-p_1)^2(k-p_{13})^2(l-p_{13})^2(k-p_{123})^2}\,.
\end{eqnarray}

We want to stress here also another aspect. 
Very often in the case of a sector with a large number of masters, 
an appropriate choice of basis also helps simplify substantially 
the identities needed to express all integrals of that topology through the set of masters themselves. 
In treating four-point functions the reduction identities often attain an enormous size, 
and we found that using this basis helped speed up 
considerably the intermediate stages of the computations.

In Appendix~\ref{A:MIsTopoB} we report the results for all masters except $\I_{(213,1)}^B$ 
computed in the non-physical region up to weight 3. 
This last integral has been computed in the physical region, as discussed below. 
Since its expansion starts only at weight 4, 
we decided to include its explicit expression in the same Appendix.
The full result up to weight 4 and the analytic continuation to the physical region relevant 
for vector boson pair production for all MIs is attached to the arXiv submission of this paper.
In addition, we also provide the one-loop $\times$ one-loop integrals required for the full reduction in the same set of variables.
Their naming convention follows the same scheme outlined in section~\ref{sec:auxtopo}.

\subsection{The computation of the 4 MIs in sector 213.}\label{sec:Int213}
In the following section we describe in detail the computation of the four masters 
in sector $213$. We use for the 4 MIs the following notation:
$$M_j(y,z,\epsilon) = \mathcal{I}^{(B)}_{213,j}\quad \mbox{with} \quad j=1,2,3,4\,$$
where we made explicit the dependence on $y,z$ and on the dimensional regulator $\epsilon.$
Although the symmetries of \textbf{Topo B} would favor the choice of $(y,z)$ as variables,
one of the masters in this sector could not be expressed explicitly in terms of 
standard GHPLs only using these variables and required hence a special treatment.

Deriving the differential equations satisfied by the four MIs
in $(y,z)$ we find a system of four coupled differential equations. With the choice described
above the system assumes a triangular form in the limit $d\to4$, allowing at least in principle
its solution as series expansion in $\epsilon = (4-d)/2$.
Due to the symmetry of the sector under the exchange $y \leftrightarrow z$ we can limit
the discussion to the system of equations in $\partial/\partial y$.
Also, for the sake of argument, we look just at its homogeneous part. Highlighting the
dependence on $\epsilon$ the system takes the symbolic form
\begin{align}\label{eq:M213d}
% \left\{ \begin{array}{cccc}
 {\partial \over \partial y}\, M_1 \,&=\, a_{11}\, M_1 +a_{12}\, M_2 +a_{13}\, M_3 +a_{14}\, M_4 \nonumber \\
 {\partial \over \partial y}\, M_2 \,&=\,               a_{22}\, M_2 
                                    +\epsilon \,\left[a_{23}\, M_3 +a_{24}\, M_4 \right] \nonumber \\
 {\partial \over \partial y}\, M_3 \,&=\, \epsilon^2\,\left[ a_{31}\, M_1 \right] 
                                  +\epsilon \,\left[ a_{32}\, M_2 +a_{33}\, M_3 + a_{34}\, M_4\right] \nonumber \\
 {\partial \over \partial y}\, M_4 \,&=\, \epsilon^2\,\left[ a_{41}\, M_1 \right] 
                                         +\epsilon \,\left[ a_{42}\, M_2 +a_{43}\, M_3\right] + a_{44}\, M_4\,
% \end{array} \right.
\end{align}
where both the $M_j$ and the $a_{ij}$ are functions of $(y,z)$ and of the 
regularization parameter $\epsilon$, such that the $a_{ij}$ do not have any poles in $1/ \epsilon$.
The triangular form in $\epsilon \to 0$ is easy to see. 
In particular the equations for masters $M_2$, $M_3$, $M_4$ are 
decoupled from that for $M_1$  and from each other in this limit.
Our strategy will then be that of expanding both sides of the equations
in $\epsilon$ and attempt a solution in a bottom up approach.
Upon expanding in $\epsilon$ one obtains, order by order, a system of four
differential equations in triangular form. At order $\epsilon^n$ the system takes the form

\begin{align*}
   {\partial   \over \partial y} M^{(n)}_1(y,z)\,&=\, 
 {1 \over 2}\left[ {1 \over 2-y-z } - {1 \over 2+y+z} \right]\, M^{(n)}_1(y,z)\\
         &+ \left[ {m^2 ( 1-yz )\over (2+y+z)(2-y-z) }\right]\, M^{(n)}_2(y,z) \\
         &- \left[ {m^2 ( 1-y )( z^2 + 4yz - 5) \over (1-yz)(2+y+z)(2-y-z) }\right]\, M^{(n)}_3(y,z)\\
         &+ \left[ {m^2 ( 1-z )( z^2 - 2yz + 1) \over (1-yz)(2+y+z)(2-y-z) }\right]\, M^{(n)}_4(y,z) 
         + N^{(n)}_1(y,z)\,,\\
%%%%%%%%%%%%%%%%%%%%%%%%%%%%%%%%%
 {\partial  \over \partial y} M^{(n)}_2(y,z)\,&=\, \left[ {z \over 1-yz }\right]\,M^{(n)}_2(y,z) 
 + N^{(n)}_2(y,z)\,, \\
%%%%%%%%%%%%%%%%%%%%%%%%%%%%%%%%
 {\partial  \over \partial y} M^{(n)}_3(y,z)\,&=\, \left[ {1 \over 1-y  }\right]\,M^{(n)}_3(y,z) 
 + N^{(n)}_3(y,z)\,,\\
%%%%%%%%%%%%%%%%%%%%%%%%%%%%%%%%
 {\partial  \over \partial y} M^{(n)}_4(y,z)\,&=\, N^{(n)}_4(y,z)\,,
\end{align*}
where the $N^{(n)}_j(y,z)$ is the non-homogeneous term of the equation for $M_j$ at order $\epsilon^n$.
Note that the homogeneous part of the system is independent on the order of the expansion,
while the non-homogeneous terms do depend on $n$.

At order $n$ we start integrating the last three equations 
and fixing the boundary condition to determine the exact solution for 
$M_2^{(n)}$, $M_3^{(n)}$ and $M_4^{(n)}$.
The result of the integration must then be used
as input to derive a first order differential equation for $M_1^{(n)}$.
If we are able to integrate this equation and to fix the boundary condition, we can use the result
as input for the equations of $M_2^{(n+1)}$, $M_3^{(n+1)}$, $M_4^{(n+1)}$  proceeding bottom up
from $\epsilon^{-4}$ (the first relevant order for the expansion) to any desired order.

One complication arises. The homogeneous equation for $M_1^{(n)}(y,z)$ reads
\begin{align*}
{\partial  H^{(n)}_1(y,z) \over \partial y} \,&=\, 
{1 \over 2}\left[ {1 \over 2-y-z } - {1 \over 2+y+z} \right]\, H^{(n)}_1(y,z)\,,
\end{align*}
so that its solution contains a square root, which for $(y,z)>0$ and $y+z<1$ reads 
\begin{align*}
H_1^{(n)}(y,z) = {1 \over \sqrt{(2-y-z)(2+y+z)}}.
\end{align*}

Re-written in terms of $s$ and $Q^2$, this is exactly the square-root described in 
section~\ref{sec:TopoA}. The presence of this square-root tells us that
if we want to express the result in terms of $(y,z)$ we need to introduce a new set of 
generalized polylogarithms that may contain also non-rational integrating factors 
(see for example~\cite{Aglietti:2004tq}). 
The algebraic properties of these functions are much less well understood than those 
of GHPLs, and no numerical implementation exists for them. 
By studying $M_1(y,z)$ one easily realizes that the master has no divergences for 
$\epsilon \to 0$  so that its expansion starts only at order $\epsilon^0$
$$M_1(y,z,\epsilon) = M_1^{(0)}(y,z) + \mathcal{O}(\epsilon)\,.$$

From the $\epsilon$ dependence of the system in~\eqref{eq:M213d} we see that 
$M_1(y,z)$ can then enter  as non-homogeneous term in the equations for $M_2$, $M_3$ 
and $M_4$ only starting at order $\epsilon^2$.
Since we are interested in the expression for the masters up to transcendentality 
$4$, which corresponds in this case to computing them up to order $\epsilon^0$, 
the explicit value of $M_1^{(0)}(y,z)$ is never required.
Note that the same kind of cancellation takes place also for the integrals with larger number of denominators,
$\mathcal{I}^{(B)}_{215}$, $\mathcal{I}^{(B)}_{247,1}$ and $\mathcal{I}^{(B)}_{247,2}$. 
By direct inspection of their differential equations we see that both topologies contain 
only $M_2$, $M_3$ and $M_4$ as sub-topologies, while $M_1$  never contributes (at any order in $\epsilon\,$!).

In order to compute $M_1$ we proceed in the following way. 
We use the explicit results for $M_2$, $M_2$ and $M_3$ to obtain two 
linear first order differential equations for $M_1$ in $y$ and $z$.
These are then continued to the Minkowski region as outlined in section~\ref{sec:kinematicsB}.
We then re-interpret the equations given in $v,w$ as equation given in $v=\zeta$ and $w=w(\xi,\zeta)$ and obtain
\be
\frac{\partial M_1}{\partial \zeta} = \frac{\partial M_1 }{\partial v }\frac{\partial v}{\partial \zeta } + \frac{\partial M_1}{\partial \frac{1}{w} } \frac{\partial \frac{1}{w}}{\partial \zeta }\;.
\ee
Since we have the following relations in the physical region
\be
 v = \zeta \,, \qquad {1\over w} = {1 + \xi^2 \over \xi} - \zeta\,,
\ee
we find, using the analytically continued differential equations as input,
{\small
%\be\bsp
\begin{align}
\label{eq:2131DGL}
\frac{\partial M_1}{\partial \zeta}& =  \frac{1}{(\xi - \zeta)(\frac{1}{\xi} - \zeta)} 
 \left\{ 2G\left(-1, \frac{1}{\zeta}, 0; \frac{1+\xi^2}{\xi} - \zeta\right)
 - 2G\left(-1, -1, 0; \frac{1+\xi^2}{\xi} - \zeta\right)   \right. \nonumber \\
& + 2G\left(-1, -1, 0; \zeta\right) + 2G\left(0, -1, 0; \frac{1+\xi^2}{\xi} - \zeta\right) 
- 2G\left(0, -1, 0; \zeta\right) \nonumber \\ 
&- G\left(0, \frac{1}{\zeta}, 0; \frac{1+\xi^2}{\xi} - \zeta\right) 
 + G\left(0; \zeta\right)  \left[2G\left(-1, \frac{1}{\zeta}; \frac{1+\xi^2}{\xi} - \zeta\right) \right. \nonumber \\
 &\left. \qquad  \qquad \quad - 2G\left(-1; \frac{1+\xi^2}{\xi} - \zeta\right)G\left(-1, 0; \zeta\right) 
  - G\left(0, \frac{1}{\zeta}; \frac{1+\xi^2}{\xi} - \zeta\right)\right] \nonumber \\
  &+ \frac{\pi^2}{6}  \left[2G\left(-1; \frac{1+\xi^2}{\xi} - \zeta\right) - 2G\left(-1; \zeta\right) 
  - G\left(0; \frac{1+\xi^2}{\xi} - \zeta\right) + G\left(0; \zeta\right) \right] \nonumber \\
   &- 2\pi\,i\,\left[ G\left(-1; \frac{1+\xi^2}{\xi} - \zeta\right)G\left(-1; \zeta\right) 
   + G\left(-1, -1; \frac{1+\xi^2}{\xi} - \zeta\right) - G\left(-1, -1; \zeta\right) \right. \nonumber \\
 &  - 2G\left(-1, \frac{1}{\zeta}; \frac{1+\xi^2}{\xi} - \zeta\right) 
 - G\left(0, -1; \frac{1+\xi^2}{\xi} - \zeta\right) + G\left(0, -1; \zeta\right) \nonumber \\
&   + \left. \left.  G\left(0, \frac{1}{\zeta}; \frac{1+\xi^2}{\xi} - \zeta\right)\right]  
%- \epsilon\left(2\left(\frac{1+\xi^2}{\xi}-2 \zeta\right) M_1 + \ldots \right)
\right\}  +\mathcal{O}(\epsilon)  \;   .  
\end{align}
%\esp\ee
}

The $\mathcal{O}(\epsilon)$ part is not reproduced here but is nevertheless needed for fixing the boundary condition. 
The integration of the differential equation becomes trivial if all the $\zeta$ dependence of the GHPLs is 
in their argument.
With the procedure described in section~\ref{sec:algtools}, we can rewrite all the functions where this is not the case in terms of GHPLs of $\xi$ and $\zeta$.
For example, we obtain
\be\bsp
G\bigg(-1,\frac{1}{\zeta};  \frac{1+\xi^2}{\xi} - \zeta\bigg) =&
% \\ & 
G\left(0, \frac{1}{\xi}; \zeta\right) + G\left(0, \xi; \zeta\right) 
+ G\left(\frac{1}{I_\xi}, \frac{1}{\xi}; \zeta\right) + G\left(\frac{1}{I_\xi}, \xi; \zeta\right)\\
& - G\left(-1, \frac{1}{\xi}; \zeta\right) - G\left(-1, \xi; \zeta\right) + G\left(-1, \frac{1}{I_\xi}; \zeta\right)\\
& + G\left(-1; \zeta\right)\left(-G\left(0; \xi\right) + G\left(-c; \xi\right) 
+ G\left(-\bar{c}; \xi\right)\right) \;.
\esp \ee
In transforming the individual GHPLs independently we find that an apparently larger set of
functions is needed:
\be\bsp\label{eq:213indices}
G(a_1,\ldots,a_n;\zeta) \quad & \text{with} \;a_i 
\in \left\{-2, -1, 0, \frac{1}{\xi}, \xi, \frac{1}{I_\xi},\frac{1}{J_\xi}\right\}\;,\\
G(a_1,\ldots,a_n;\xi) \quad & \text{with} \;a_i \in \{-1, 0, \pm i, -c,-\bar{c} \}\;.
\esp\ee

The appearance of index $-2$ is surprising since, as explained in section~\ref{sec:TopoBGHPLs}, 
no denominator $1/(\zeta + 2)$ appears in the differential equations.
Nevertheless, putting everything together all GHPLs with index $-2$ cancel, as expected,
so that only the set described in section~\ref{sec:TopoBGHPLs} survives. 
This cancellation gives a confirmation of the consistency of our procedure.
Once equation~\eqref{eq:2131DGL} is put in this form, we can determine the primitive 
of $M_1$ by straightforward integration using the very definition of GHPLs.

In order to fix the boundary condition we require $M_1$ to be regular in $\zeta = \xi$. 
By multiplying equation (\ref{eq:2131DGL}) with $(\zeta - \xi)$ and taking the limit $\zeta \to \xi $ 
we obtain the value of $M_1(\xi,\zeta=\xi)$ from the $\mathcal{O}(\epsilon)$ part. 
The required limit identities can again be derived using the procedure from section~\ref{sec:algtools}.
As for the other master integrals, the explicit result is provided with the arXiv submission of this paper.

\section{Checks on the results}
\label{sec:checks}
Many non-trivial checks have been performed in order
to validate our result.
As already stated above, all triangle topologies had been already computed
in~\cite{Birthwright:2004kk,Chavez:2012kn}, in the more general case of 
three different external masses. We compared our results in the Euclidean region 
numerically to those in~\cite{Birthwright:2004kk,Chavez:2012kn}, finding perfect agreement.

We have used FIESTA~\cite{Smirnov:2008py} and SecDec2~\cite{Borowka:2012yc} 
in order to check numerically all the double-box topologies in 
\textbf{Topo A} and \textbf{Topo B}, except for $\I_{213,1}^{(B)}$, in the non-physical region, 
where all integrals are real, finding agreement in different phase space points.

% Since master $\I_{213,1}^{(B)}$ has been computed directly in the physical region,
% we could not use FIESTA, which cannot perform direct numerical evaluation where the 
% integrals are complex.
Recently a new version of SecDec has been released~\cite{Borowka:2013cma},
which has been successfully used to perform accurate numerical evaluation of
planar and non-planar double-boxes also in the physical region.
Using it we could then perform a full check in different phase space points of $\I_{213,1}^{(B)}$
finding again agreement with our result.

In the same way we could also check our analytic continuation procedure
evaluating with SecDec2 numerically most of the masters in \textbf{Topo A} and 
\textbf{Topo B} also in the physical region, even though the numerical evaluation in this
region is computationally much more demanding.

\section{Conclusions}
\label{sec:conc}
The precise interpretation of upcoming LHC results on vector boson pair production 
will require the computation of NNLO QCD corrections to this process, currently known 
only to NLO. Besides already known contributions with higher final state
multiplicity and a lower number of loops, this calculation requires the two-loop corrections 
to vector boson pair production matrix elements. Using the 
integration-by-parts (IBP) technique, the Feynman integrals 
appearing in these matrix elements can be expressed as a linear combination of 
a small set of so-called master integrals. 

In this paper, we considered the integrals relevant to 
the two-loop corrections for the production of massive equal-mass gauge bosons: $q\bar{q} \to VV$. 
For the application of the IBP technique, the integrals could be assigned to one of the 
three auxiliary topologies relevant to this process. Two of these auxiliary topologies contain only 
planar master integrals, which are the main focus of this paper. We derived differential equations 
in the external Mandelstam invariants for all integrals, starting at the integrand level. The integrals 
are then computed by solving these differential equations, matched to appropriate boundary conditions 
in special kinematical points. 

The master integrals are expressed in terms of generalized harmonic 
polylogarithms (GHPLs), which appear widely in analytical calculations of Feynman integrals. In the
course of deriving these integrals from the differential equations, three types of
manipulations have to be performed repeatedly: variable transformations, determination of 
limiting behavior in special points and analytical continuation. The 
computer algebra automation of these operations relies 
heavily on the algebraic properties of the GHPLs. The recently developed 
coproduct formalism~\cite{Duhr:2012fh,Anastasiou:2013srw} was used to perform 
most of the transformations on GHPLs.

We obtained analytical results for all massless planar two-loop four-point functions with 
two off-shell legs of equal invariant mass. The resulting expressions prior to analytical continuation 
to the physical region are fairly compact, and the pole parts of all integrals are 
documented in this paper. The finite pieces, as well as the analytical continuation of the integrals 
are more lengthy, and are enclosed with the arXiv submission of this paper. 
The newly derived master integrals will allow us to calculate the planar 
(for example leading-color) two-loop corrections to the amplitudes for $q\bar q \to VV$. This work, 
as well as the computation of the non-planar master integrals, is in progress.

\section*{Acknowledgements}
The authors would like to thank C.\ Duhr, R.\ Bonciani, A.\ von Manteuffel  and E.\ Remiddi
for numerous useful and enlightening 
discussions on many different issues throughout the project, and G.\ Heinrich for her 
assistance in the numerical validations of the results.
This research was supported in part by
the Swiss National Science Foundation (SNF) under contract
PDFMP2-135101 and  200020-138206, as well as  by the European Commission through the 
``LHCPhenoNet" Initial Training Network PITN-GA-2010-264564.

\appendix
\section{Master Integrals: Topo A}\label{A:MIsTopoA}
In this appendix we give the analytic expressions for the relevant genuine two-loop boxes
that appear in the reduction of \textbf{Topo A}, classified 
with respect to the number of denominators. 
We write explicitly the results up to weight three, while the full result up to weight four
can be found attached to the arXiv submission of this paper readable using Mathematica or FORM. 
All triangle topologies were derived previously in the 
literature~\cite{Gehrmann:2000zt,Gehrmann:2001ck,Birthwright:2004kk,Chavez:2012kn}; 
they are also included in the the arXiv submission, where they 
are expressed in the same functional basis as the box integrals derived here.

The common normalization factor of all master integrals is
\begin{equation}
 S_\epsilon = \left[ (4 \pi)^\epsilon \, {\Gamma(1+ \epsilon) \, \Gamma^2(1-\epsilon) \over \Gamma(1-2\epsilon)} \right]\,.
\end{equation}

\subsection*{5-denominator integrals}
\begin{equation}
\I_{174}^{(A)} = \left( {S_\epsilon} \over 16 \pi^2 \right)^2 \frac{( m^2 )^{-2\epsilon-1}}{(1-z)}\,
 \sum_{n=-2}^0\, \epsilon^n \, f_n^{(174)}(x,z) + \mathcal{O}(\epsilon)\,,
\end{equation}
with
{\small
\begin{align*}
 f_{-2}^{(174)}(x,z) &= G(0,z)\,,\\
 f_{-1}^{(174)}(x,z) &= - {\pi^2\over 6}
          - 2G(-1,x)G(0,z)
          + G(0,x)G(0,z)
          + 2G(0,z)
          - 2G(0,0,z)
          + G(1,0,z) \,, \\
 f_{0}^{(174)}(x,z) &=   \zeta_3
          - {\pi^2 \over 3 }
          - 1/6G( - 1/x,z)\pi^2
          - G( - 1/x,0,0,z)
          + 2G( - 1/x,1,0,z)
          - 1/2G( - x,z)\pi^2\\
          &- G( - x,0,0,z)
          + 2G( - x,1,0,z)
          + G(-1,x)\pi^2
          - 2G(-1,x)G( - 1/x,0,z)\\
          &- 2G(-1,x)G( - x,0,z)
          - 4G(-1,x)G(0,z)
          + 6G(-1,x)G(0,0,z)\\
          &- 2G(-1,x)G(1,0,z)
          + 8G(-1,-1,x)G(0,z)
          - 4G(-1,0,x)G(0,z)
          - 2/3G(0,x)\pi^2 \\
          &+ G(0,x)G( - 1/x,0,z)
          + G(0,x)G( - x,0,z)
         + 2G(0,x)G(0,z)
          + G(0,x)G(1,0,z)\\
         &+ 4G(0,z)
          + 2/3G(0,z)\pi^2
          - 2G(0,-1,x)G( - 1/x,z)
         + 2G(0,-1,x)G( - x,z)\\
         &- 4G(0,-1,x)G(0,z)
          + G(0,0,x)G( - 1/x,z)
          - G(0,0,x)G( - x,z)
         + 2G(0,0,x)G(0,z)\\
          &- 4G(0,0,z)
          - 3G(0,0,z)G(0,x)
          + 2G(0,0,-1,x)
          - G(0,0,0,x)
         + 5G(0,0,0,z)\\
         &- 4G(0,1,0,z)
          - 1/6G(1,z)\pi^2
          + 2G(1,0,z)
          - 2G(1,0,0,z)
          + G(1,1,0,z) \,.
\end{align*}
}

\begin{equation}
\I_{214}^{(A)} = \left( {S_\epsilon} \over 16 \pi^2 \right)^2 \frac{x\,( m^2 )^{-2\epsilon-1}}{(1+x)^2}\,
 \sum_{n=-3}^0\, \epsilon^n \, f_n^{(214)}(x,z) + \mathcal{O}(\epsilon)\,,
\end{equation}
with
{\small
\begin{align*}
 f_{-3}^{(214)}(x,z) &= 1\,,\\
 f_{-2}^{(214)}(x,z) &=  2
          - 2G(-1,x)
          + G(0,x)
          - G(0,z)\,,\\
 f_{-1}^{(214)}(x,z) &=  4 + {\pi^2 \over 3}
          - 4G(-1,x)
          + 2G(-1,x)G(0,z)
          + 4G(-1,-1,x)
          - 2G(-1,0,x)
          + 2G(0,x)\\
         & - G(0,x)G(0,z)
          - 2G(0,z)
          - 2G(0,-1,x)
          + G(0,0,x)
          + 3G(0,0,z)
          - 2G(1,0,z)\,, \\
 f_{0}^{(214)}(x,z) &=    8 - 6 \zeta_3 + {2 \over 3}\pi^2
          + 1/6G( - 1/x,z)\pi^2
          + G( - 1/x,0,0,z)
          - 2G( - 1/x,1,0,z)\\
          &+ 1/2G( - x,z)\pi^2
          + G( - x,0,0,z)
          - 2G( - x,1,0,z)
          - 8G(-1,x)
          - 2/3G(-1,x)\pi^2\\
         &+ 2G(-1,x)G( - 1/x,0,z)
          + 2G(-1,x)G( - x,0,z)
          + 4G(-1,x)G(0,z)\\
          &- 6G(-1,x)G(0,0,z)
          + 8G(-1,-1,x)
          - 4G(-1,-1,x)G(0,z)
          - 8G(-1,-1,-1,x)\\
          &+ 4G(-1,-1,0,x)
          - 4G(-1,0,x)
          + 2G(-1,0,x)G(0,z)
          + 4G(-1,0,-1,x)\\
          &- 2G(-1,0,0,x)
          + 4G(0,x)
          + 2/3G(0,x)\pi^2
          - G(0,x)G( - 1/x,0,z)
          - G(0,x)G( - x,0,z)\\
         &- 2G(0,x)G(0,z)
          - 4G(0,z)
          - G(0,z)\pi^2
          - 4G(0,-1,x)
          + 2G(0,-1,x)G( - 1/x,z)\\
         &- 2G(0,-1,x)G( - x,z)
          + 2G(0,-1,x)G(0,z)
          + 4G(0,-1,-1,x)
          - 2G(0,-1,0,x)\\
         &+ 2G(0,0,x)
          - G(0,0,x)G( - 1/x,z)
          + G(0,0,x)G( - x,z)
          - G(0,0,x)G(0,z)\\
          &+ 6G(0,0,z)
          + 3G(0,0,z)G(0,x)
          - 6G(0,0,-1,x)
          + 3G(0,0,0,x)
          - 9G(0,0,0,z)\\
          &+ 6G(0,1,0,z)
           - 4G(1,0,z)
          + 4G(1,0,0,z)\,.
\end{align*}
}

Sector 181 has 2 MIs, which read:
\begin{equation}
\I_{181,1}^{(A)} = \left( {S_\epsilon} \over 16 \pi^2 \right)^2 \frac{x\,( m^2 )^{-2\epsilon-1}}{(1-x^2)}\,
 \sum_{n=-1}^0\, \epsilon^n \, f_n^{(181,1)}(x,z) + \mathcal{O}(\epsilon)\,,
\end{equation}
with
{\small
\begin{align*}
 f_{-1}^{(181,1)}(x,z) &=   {\pi^2 \over 3}
          - 4G(0,-1,x)
          + 2G(0,0,x)\,, \\
 f_{0}^{(181,1)}(x,z) &=    4 \zeta_3
          + {2 \over 3} \pi^2
          + 1/6G( - 1/x,z)\pi^2
          + G( - 1/x,0,0,z)
          - 2G( - 1/x,1,0,z)\\
         & - 1/2G( - x,z)\pi^2
          - G( - x,0,0,z)
          + 2G( - x,1,0,z)
          + 2G(-1,x)G( - 1/x,0,z)\\
          &- 2G(-1,x)G( - x,0,z)
          - 2/3G(0,x)\pi^2
          - G(0,x)G( - 1/x,0,z)
          + G(0,x)G( - x,0,z)\\
         &- 8G(0,-1,x)
          + 2G(0,-1,x)G( - 1/x,z)
          + 2G(0,-1,x)G( - x,z)
          + 16G(0,-1,-1,x)\\
        & - 8G(0,-1,0,x)
          + 4G(0,0,x)
          - G(0,0,x)G( - 1/x,z)
          - G(0,0,x)G( - x,z)\\
         &+ 4/3G(1,x)\pi^2
          - 16G(1,0,-1,x)
          + 8G(1,0,0,x) \,.
\end{align*}
}
and
\begin{equation}
\I_{181,2}^{(A)} = \left( {S_\epsilon} \over 16 \pi^2 \right)^2 \frac{x\,( m^2 )^{-2\epsilon-2}}{z(1+x)^2}\,
 \sum_{n=-3}^0\, \epsilon^n \, f_n^{(181,2)}(x,z) + \mathcal{O}(\epsilon)\,,
\end{equation}
with
{\small
\begin{align*}
 f_{-3}^{(181,2)}(x,z) &=   {1 \over 4}\,, \\
 f_{-2}^{(181,2)}(x,z) &=   1/2G(0,x) - G(-1,x)
          - G(0,z)\,, \\
 f_{-1}^{(181,2)}(x,z) &=  {7 \over 12} \pi^2
          + 4G(-1,x)G(0,z)
          + 4G(-1,-1,x)
          - 2G(-1,0,x)
          - 2G(0,x)G(0,z)\\
          &- 2G(0,-1,x)
          + G(0,0,x)
          + 4G(0,0,z)
          - 3G(1,0,z)\,, \\
 f_{0}^{(181,2)}(x,z) &=    - 4 \zeta_3
          + 1/2G( - 1/x,z)\pi^2
          + 3G( - 1/x,0,0,z)
          - 6G( - 1/x,1,0,z)
          + 3/2G( - x,z)\pi^2\\
         &+ 3G( - x,0,0,z)
          - 6G( - x,1,0,z)
          - 7/3G(-1,x)\pi^2
          + 6G(-1,x)G( - 1/x,0,z)\\
         &+ 6G(-1,x)G( - x,0,z)
          - 16G(-1,x)G(0,0,z)
          - 16G(-1,-1,x)G(0,z)\\
        &- 16G(-1,-1,-1,x)
          + 8G(-1,-1,0,x)
          + 8G(-1,0,x)G(0,z)
          + 8G(-1,0,-1,x)\\
         &- 4G(-1,0,0,x)
          + 7/6G(0,x)\pi^2
          - 3G(0,x)G( - 1/x,0,z)
          - 3G(0,x)G( - x,0,z)\\
         &- 7/3G(0,z)\pi^2
          - 1/2G(1,z)\pi^2
          + 6G(0,-1,x)G( - 1/x,z)
          - 6G(0,-1,x)G( - x,z)\\
         &+ 8G(0,-1,x)G(0,z)
         + 8G(0,-1,-1,x)
          - 4G(0,-1,0,x)
          - 3G(0,0,x)G( - 1/x,z)\\
         &+ 3G(0,0,x)G( - x,z)
         - 4G(0,0,x)G(0,z)
          + 8G(0,0,z)G(0,x)
          - 4G(0,0,-1,x)\\
         &+ 2G(0,0,0,x)
         - 16G(0,0,0,z)
         + 12G(0,1,0,z)
          + 6G(1,0,0,z)
          + 3G(1,1,0,z)  \,.
\end{align*}
}

Sector 182 has 2 MIs, which read:
\begin{equation}
\I_{182,1}^{(A)} = \left( {S_\epsilon} \over 16 \pi^2 \right)^2 \frac{x\,( m^2 )^{-2\epsilon-1}}{(1+x+x^2+x\,z)}\,
 \sum_{n=-1}^0\, \epsilon^n \, f_n^{(181,1)}(x,z) + \mathcal{O}(\epsilon)\,,
\end{equation}
with
{\small
\begin{align*}
 f_{0}^{(182,1)}(x,z) &=  - 2 \zeta_3
          + 1/6G( - 1/x,z)\pi^2
          + G( - 1/x,0,0,z)
          - 2G( - 1/x,1,0,z)
          + 1/2G( - x,z)\pi^2\\
         & + G( - x,0,0,z)
          - 2G( - x,1,0,z)
          - 1/3G(-1,x)\pi^2
          + 2G(-1,x)G( - 1/x,0,z)\\
          &+ 2G(-1,x)G( - x,0,z)
          - 2G(-1,x)G(1,0,z)
          + 2/3G(0,x)\pi^2
          - G(0,x)G( - 1/x,0,z)\\
         & - G(0,x)G( - x,0,z)
          + G(0,x)G(1,0,z)
          + 2G(0,-1,x)G( - 1/x,z)\\
         &- 2G(0,-1,x)G( - x,z)
          - G(0,0,x)G( - 1/x,z)
          + G(0,0,x)G( - x,z)
          - 6G(0,0,-1,x)\\
         &+ 3G(0,0,0,x)
          - 1/6G(1,z)\pi^2
          - G(1,0,0,z)
          + G(1,1,0,z) \,.
\end{align*}
}
and
\begin{equation}
\I_{182,2}^{(A)} = \left( {S_\epsilon} \over 16 \pi^2 \right)^2 \frac{( m^2 )^{-2\epsilon-2}}{(1-z)}\,
 \sum_{n=-2}^0\, \epsilon^n \, f_n^{(181,2)}(x,z) + \mathcal{O}(\epsilon)\,,
\end{equation}
with
{\small
\begin{align*}
 f_{-2}^{(182,2)}(x,z) &=   G(0,z)\,, \\
 f_{-1}^{(182,2)}(x,z) &=   {\pi^2 \over 6}
          + 2G(-1,x)G(0,z)
          - G(0,x)G(0,z)
          - G(0,0,z)
          - G(1,0,z)\,, \\
 f_{0}^{(182,2)}(x,z) &=    - 2 \zeta_3
          + 1/2G( - 1/x,z)\pi^2
          + 3G( - 1/x,0,0,z)
          - 6G( - 1/x,1,0,z)
          + 3/2G( - x,z)\pi^2\\
         &+ 3G( - x,0,0,z)
          - 6G( - x,1,0,z)
          - 5/3G(-1,x)\pi^2
          + 6G(-1,x)G( - 1/x,0,z)\\
         &+ 6G(-1,x)G( - x,0,z)
          - 8G(-1,x)G(0,0,z)
          - 2G(-1,x)G(1,0,z)\\
         &- 8G(-1,-1,x)G(0,z)
          + 4G(-1,0,x)G(0,z)
          + 4/3G(0,x)\pi^2
          - 3G(0,x)G( - 1/x,0,z)\\
         &- 3G(0,x)G( - x,0,z)
          + G(0,x)G(1,0,z)
          - 7/6G(0,z)\pi^2
          + 6G(0,-1,x)G( - 1/x,z)\\
         &- 6G(0,-1,x)G( - x,z)
          + 4G(0,-1,x)G(0,z)
          - 3G(0,0,x)G( - 1/x,z)\\
         &+ 3G(0,0,x)G( - x,z)
          - 2G(0,0,x)G(0,z)
          + 4G(0,0,z)G(0,x)
          - 6G(0,0,-1,x)\\
         &+ 3G(0,0,0,x)
          - 2G(0,0,0,z)
          + 7G(0,1,0,z)
          - 1/6G(1,z)\pi^2
         + G(1,0,0,z)\\
         & + G(1,1,0,z)  \,.
\end{align*}
}

\subsection*{6-denominator integrals}
Two topologies with 6 denominators have one single MI:
\begin{equation}
\I_{215}^{(A)} = \left( {S_\epsilon} \over 16 \pi^2 \right)^2 \frac{x\,( m^2 )^{-2\epsilon-2}}{(1+x)^2(1-z)}\,
 \sum_{n=-3}^{-1}\, \epsilon^n \, f_n^{(215)}(x,z) + \mathcal{O}(\epsilon^0)\,,
\end{equation}
with
{\small
\begin{align*}
 f_{-3}^{(215)}(x,z) &=  - {1 \over 2} G(0,z) \,,\\
 f_{-2}^{(215)}(x,z) &=   - {\pi^2 \over 6}
          + G(0,0,z)
          + G(1,0,z) \,\\
 f_{-1}^{(215)}(x,z) &=    - 2 \zeta_3
          - 1/6G( - 1/x,z)\pi^2
          - G( - 1/x,0,0,z)
          + 2G( - 1/x,1,0,z)
          - 1/2G( - x,z)\pi^2\\
         &- G( - x,0,0,z)
          + 2G( - x,1,0,z)
          + 2/3G(-1,x)\pi^2
          - 2G(-1,x)G( - 1/x,0,z)\\
          &- 2G(-1,x)G( - x,0,z)
          + 2G(-1,x)G(0,0,z)
          + 4G(-1,-1,x)G(0,z)\\
          &- 2G(-1,0,x)G(0,z)
          - 1/2G(0,x)\pi^2
          + G(0,x)G( - 1/x,0,z)
          + G(0,x)G( - x,0,z)\\
         &- 2G(0,-1,x)G( - 1/x,z)
          + 2G(0,-1,x)G( - x,z)
          - 2G(0,-1,x)G(0,z)\\
         &+ G(0,0,x)G( - 1/x,z)
          - G(0,0,x)G( - x,z)
          + G(0,0,x)G(0,z)
          - G(0,0,z)G(0,x)\\
         &+ 2G(0,0,-1,x)
          - G(0,0,0,x)
          - G(0,0,0,z)
          - G(0,1,0,z)
          + 1/3G(1,z)\pi^2\\
          &- 2G(1,0,0,z)
          - 2G(1,1,0,z) + 1/6G(0,z)\pi^2\,.
\end{align*}

\begin{equation}
\I_{430}^{(A)} = \left( {S_\epsilon} \over 16 \pi^2 \right)^2 \frac{x\,( m^2 )^{-2\epsilon-2}}{(x + z(1+x+x^2))}\,
 \sum_{n=-1}^{-1}\, \epsilon^n \, f_n^{(430)}(x,z) + \mathcal{O}(\epsilon^0)\,,
\end{equation}
with
{\small
\begin{align*}
 f_{-1}^{(430)}(x,z) &=  \zeta_3
          + 1/6G( - 1/x,z)\pi^2
          + G( - 1/x,0,0,z)
          - 2G( - 1/x,1,0,z)
          + 1/2G( - x,z)\pi^2\\
         &+ G( - x,0,0,z)
          - 2G( - x,1,0,z)
          - 1/3G(-1,x)\pi^2
          + 2G(-1,x)G( - 1/x,0,z)\\
         &+ 2G(-1,x)G( - x,0,z)
          - 2G(-1,x)G(0,0,z)
          - 2G(-1,x)G(1,0,z)\\
         &- G(0,x)G( - 1/x,0,z)
          - G(0,x)G( - x,0,z)
          + G(0,x)G(1,0,z)
          - 1/2G(0,z)\pi^2\\
         &+ 2G(0,-1,x)G( - 1/x,z)
          - 2G(0,-1,x)G( - x,z)
          - G(0,0,x)G( - 1/x,z)\\
         &+ G(0,0,x)G( - x,z)
          + G(0,0,z)G(0,x)
          + 2G(0,0,-1,x)
          - G(0,0,0,x)\\
          &- 2G(0,0,0,z)
          + 3G(0,1,0,z)
          - 1/6G(1,z)\pi^2
          + G(1,1,0,z)\,.
\end{align*}

\subsection*{7-denominator integrals}
Finally, there are two independent topologies with 7 denominators. One has 3 MIs:
\begin{equation}
\I_{247,1}^{(A)} = \left( {S_\epsilon} \over 16 \pi^2 \right)^2 \frac{x^2\,( m^2 )^{-2\epsilon-3}}{z(1+x)^4}\,
 \sum_{n=-4}^{-1}\, \epsilon^n \, f_n^{(247,1)}(x,z) + \mathcal{O}(\epsilon^0)\,,
\end{equation}
with
{\small
\begin{align*}
 f_{-4}^{(247,1)}(x,z) &=  {1 \over 4} \,,\\
 f_{-3}^{(247,1)}(x,z) &=  1/2G(0,x) - G(-1,x)
          - G(0,z)\,, \\
 f_{-2}^{(247,1)}(x,z) &=   7 /12 \pi^2
          + 4G(-1,x)G(0,z)
          + 4G(-1,-1,x)
          - 2G(-1,0,x)
          - 2G(0,x)G(0,z)\\
         & - 2G(0,-1,x)
          + G(0,0,x)
          + 4G(0,0,z)
          - 2G(1,0,z)\,, \\
 f_{-1}^{(247,1)}(x,z) &=  - 2 \zeta_3
          + 1/3G( - 1/x,z)\pi^2
          + 2G( - 1/x,0,0,z)
          - 4G( - 1/x,1,0,z)
          + G( - x,z)\pi^2\\
         &+ 2G( - x,0,0,z)
          - 4G( - x,1,0,z)
          - 7/3G(-1,x)\pi^2
          + 4G(-1,x)G( - 1/x,0,z)\\
         &+ 4G(-1,x)G( - x,0,z)
          - 16G(-1,x)G(0,0,z)
          - 16G(-1,-1,x)G(0,z)\\
         &- 16G(-1,-1,-1,x)
          + 8G(-1,-1,0,x)
          + 8G(-1,0,x)G(0,z)
          + 8G(-1,0,-1,x)\\
         &- 4G(-1,0,0,x)
          + 7/6G(0,x)\pi^2
          - 2G(0,x)G( - 1/x,0,z)
          - 2G(0,x)G( - x,0,z)\\
         &- 7/3G(0,z)\pi^2
          + 4G(0,-1,x)G( - 1/x,z)
          - 4G(0,-1,x)G( - x,z)\\
         &+ 8G(0,-1,x)G(0,z)
          + 8G(0,-1,-1,x)
          - 4G(0,-1,0,x)
          - 2G(0,0,x)G( - 1/x,z)\\
         &+ 2G(0,0,x)G( - x,z)
          - 4G(0,0,x)G(0,z)
          + 8G(0,0,z)G(0,x)
          - 4G(0,0,-1,x)\\
         &+ 2G(0,0,0,x)
          - 16G(0,0,0,z)
          + 10G(0,1,0,z)
          - 2/3G(1,z)\pi^2
          + 4G(1,0,0,z)\\
         &+ 4G(1,1,0,z) \,. 
\end{align*}
}

\begin{equation}
\I_{247,2}^{(A)} = \left( {S_\epsilon} \over 16 \pi^2 \right)^2 \frac{x^2\,( m^2 )^{-2\epsilon-2}}{(1-x)(1+x)^3}\,
 \sum_{n=-2}^{-1}\, \epsilon^n \, f_n^{(247,2)}(x,z) + \mathcal{O}(\epsilon^0)\,,
\end{equation}
with
{\small
\begin{align*}
 f_{-2}^{(247,2)}(x,z) &=   - {\pi^2 \over 3 }
          + 4G(0,-1,x)
          - 2G(0,0,x) \,, \\
 f_{-1}^{(247,2)}(x,z) &=  - 6 \zeta_3
          - 1/3G( - 1/x,z)\pi^2
          - 2G( - 1/x,0,0,z)
          + 4G( - 1/x,1,0,z)
          + G( - x,z)\pi^2\\
         &+ 2G( - x,0,0,z)
          - 4G( - x,1,0,z)
          - 4G(-1,x)G( - 1/x,0,z)
          + 4G(-1,x)G( - x,0,z)\\
         &+ G(0,x)\pi^2
          + 2G(0,x)G( - 1/x,0,z)
          - 2G(0,x)G( - x,0,z)
          - 4G(0,-1,x)G( - 1/x,z)\\
         &- 4G(0,-1,x)G( - x,z)
          - 24G(0,-1,-1,x)
          + 12G(0,-1,0,x)
          - 2G(1,x)\pi^2\\
          &+ 2G(0,0,x)G( - 1/x,z)
          + 2G(0,0,x)G( - x,z)
           + 24G(1,0,-1,x)
          - 12G(1,0,0,x)\,.
\end{align*}
}

\begin{align*}
\I_{247,3}^{(A)} = \left( {S_\epsilon} \over 16 \pi^2 \right)^2 \frac{x^2\,( m^2 )^{-2\epsilon-2}}{(1+x)^4}\, &\left[ 
 \sum_{n=-4}^{-1}\, \epsilon^n \, f_n^{(247,3)}(x,z) \right. 
+ \left. \frac{1}{z}\, \sum_{n=-2}^{-1}\, \epsilon^n \, g_n^{(247,3)}(x,z)\,\right]
+ \mathcal{O}(\epsilon^0)\,, 
\end{align*}

with
{\small
\begin{align*}
 f_{-4}^{(247,3)}(x,z) &=   -{1 \over 4 }\,, \\
f_{-3}^{(247,3)}(x,z) &=   - 1/2G(0,x) + G(-1,x)
          + G(0,z)\,, \\
 f_{-2}^{(247,3)}(x,z) &=   - 7/12\pi^2
          - 4G(-1,x)G(0,z)
          - 4G(-1,-1,x)
          + 2G(-1,0,x)
          + 2G(0,x)G(0,z)\\
         &+ 2G(0,-1,x)
          - G(0,0,x)
          - 4G(0,0,z)
          + 3G(1,0,z)\,, \\
 f_{-1}^{(247,3)}(x,z) &=  2 \zeta_3
          - 1/3G( - 1/x,z)\pi^2
          - 2G( - 1/x,0,0,z)
          + 4G( - 1/x,1,0,z)
          - G( - x,z)\pi^2\\
         &- 2G( - x,0,0,z)
          + 4G( - x,1,0,z)
          + 7/3G(-1,x)\pi^2
          - 4G(-1,x)G( - 1/x,0,z)\\
         &- 4G(-1,x)G( - x,0,z)
          + 16G(-1,x)G(0,0,z)
          - 4G(-1,x)G(1,0,z)\\
         &+ 16G(-1,-1,x)G(0,z)
          + 16G(-1,-1,-1,x)
          - 8G(-1,-1,0,x)\\
         &- 8G(-1,0,x)G(0,z)
          - 8G(-1,0,-1,x)
          + 4G(-1,0,0,x)
          - 7/6G(0,x)\pi^2\\
         &+ 2G(0,x)G( - 1/x,0,z)
          + 2G(0,x)G( - x,0,z)
          + 2G(0,x)G(1,0,z)
          + 7/3G(0,z)\pi^2\\
         &- 4G(0,-1,x)G( - 1/x,z)
          + 4G(0,-1,x)G( - x,z)
          - 8G(0,-1,x)G(0,z)\\
         &- 8G(0,-1,-1,x)
          + 4G(0,-1,0,x)
          + 2G(0,0,x)G( - 1/x,z)
          - 2G(0,0,x)G( - x,z)\\
         &+ 4G(0,0,x)G(0,z)
          - 8G(0,0,z)G(0,x)
          + 4G(0,0,-1,x)
          - 2G(0,0,0,x)\\
         &+ 16G(0,0,0,z)
          - 14G(0,1,0,z)
          - 1/6G(1,z)\pi^2
          - 8G(1,0,0,z)
          + G(1,1,0,z) \,.
\end{align*}
}
and 
{\small
\begin{align*}
 g_{-2}^{(247,3)}(x,z) &=  - G(1,0,z) \,, \\
 g_{-1}^{(247,3)}(x,z) &=  4G(-1,x)G(1,0,z)
          - 2G(0,x)G(1,0,z)
          + 4G(0,1,0,z)
          + 5/6G(1,z)\pi^2\\
         &+ 4G(1,0,0,z)
          - 5G(1,1,0,z) \,.
\end{align*}
}

The other has 2 MIs:
\begin{equation}
\I_{446,1}^{(A)} = \left( {S_\epsilon} \over 16 \pi^2 \right)^2 \frac{x\,( m^2 )^{-2\epsilon-3}}{z^2(1+x)^2}\,
 \sum_{n=-4}^{-1}\, \epsilon^n \, f_n^{(446,1)}(x,z) + \mathcal{O}(\epsilon^0)\,,
\end{equation}
with
{\small
\begin{align*}
 f_{-4}^{(446,1)}(x,z) &=   {1 \over 4} \,,\\
 f_{-3}^{(446,1)}(x,z) &=    1/2G(0,x) - G(-1,x)
          - G(0,z)\,, \\
 f_{-2}^{(446,1)}(x,z) &= {\pi^2 \over 3}
          + 4G(-1,x)G(0,z)
          + 4G(-1,-1,x)
          - 2G(-1,0,x)
          - 2G(0,x)G(0,z)\\
          &- 2G(0,-1,x)
          + G(0,0,x)
          + 4G(0,0,z)
          - 2G(1,0,z) \,,\\
 f_{-1}^{(446,1)}(x,z) &=  {\zeta_3 \over 2} 
          - 4/3G(-1,x)\pi^2
          - 16G(-1,x)G(0,0,z)
          + 8G(-1,x)G(1,0,z)\\
          &- 16G(-1,-1,x)G(0,z)
          - 16G(-1,-1,-1,x)
          + 8G(-1,-1,0,x)\\
          &+ 8G(-1,0,x)G(0,z)
          + 8G(-1,0,-1,x)
          - 4G(-1,0,0,x)
          + 2/3G(0,x)\pi^2\\
          &- 4G(0,x)G(1,0,z)
          - 4/3G(0,z)\pi^2
          + 8G(0,-1,x)G(0,z)
          + 8G(0,-1,-1,x)\\
         &- 4G(0,-1,0,x)
          - 4G(0,0,x)G(0,z)
          + 8G(0,0,z)G(0,x)
          - 4G(0,0,-1,x)\\
         &+ 2G(0,0,0,x)
          - 16G(0,0,0,z)
          + 8G(0,1,0,z)
          + 2/3G(1,z)\pi^2
          + 14G(1,0,0,z)\\
         &- 4G(1,1,0,z) \,.
\end{align*}
}
\begin{align}
\I_{446,2}^{(A)} = \left( {S_\epsilon} \over 16 \pi^2 \right)^2 \frac{( m^2 )^{-2\epsilon-2} }{ (1-z)}\,
&\left[\, \frac{1}{z} \sum_{n=-3}^0\, \epsilon^n \, f_n^{(446,2)}(x,z) \right.\\
& \left.- \frac{(1+x)^2}{(x+z(1+x+x^2))}\, 
 \sum_{n=-1}^{-1}\, \epsilon^n \, f_n^{(430)}(x,z) \right] + \mathcal{O}(\epsilon^0)\,,
\end{align}
with
{\small
\begin{align*}
 f_{-3}^{(446,2)}(x,z) &=   -1/2 G(0,z)\,,\\
 f_{-2}^{(446,2)}(x,z) &=   {\pi^2 \over 4}
          + 2G(-1,x)G(0,z)
          - G(0,x)G(0,z)
          + 7/2G(0,0,z)
          - 3/2G(1,0,z)\,,\\
 f_{-1}^{(446,2)}(x,z) &=   - \zeta_3
          + 1/3G( - 1/x,z)\pi^2
          + 2G( - 1/x,0,0,z)
          - 4G( - 1/x,1,0,z)
          + G( - x,z)\pi^2\\
          &+ 2G( - x,0,0,z)
          - 4G( - x,1,0,z)
          - 4/3G(-1,x)\pi^2
          + 4G(-1,x)G( - 1/x,0,z)\\
          &+ 4G(-1,x)G( - x,0,z)
          - 14G(-1,x)G(0,0,z)
          - 8G(-1,-1,x)G(0,z)\\
          &+ 4G(-1,0,x)G(0,z)
          + 2/3G(0,x)\pi^2
          - 2G(0,x)G( - 1/x,0,z)
          - 2G(0,x)G( - x,0,z)\\
          &- 23/12G(0,z)\pi^2
          + 4G(0,-1,x)G( - 1/x,z)
          - 4G(0,-1,x)G( - x,z)\\
          &+ 4G(0,-1,x)G(0,z)
          - 2G(0,0,x)G( - 1/x,z)
          + 2G(0,0,x)G( - x,z)\\
          &- 2G(0,0,x)G(0,z)
          + 7G(0,0,z)G(0,x)
          - 37/2G(0,0,0,z)
          + 23/2G(0,1,0,z)\\
          &- 1/12G(1,z)\pi^2
          + 21/2G(1,0,0,z)
          + 1/2G(1,1,0,z)\,.
\end{align*}
}

\section{Master Integrals: Topo B}\label{A:MIsTopoB}
In this appendix we provide the analytic expressions for the relevant two-loop boxes
that appear in the reduction of \textbf{Topo B}, classified 
with respect to the number of denominators. 
We give the explicit results up to weight three for all MIs except $\I_{213,1}^{(B)}$, for which the
expansion starts only at weight four. As already extensively discussed throughout the paper,
the computation of this integral required a special treatment and we include here the full
expression in order to show what the result looks like. 
The full results for all MIs up to weight four, including the already known triangle topologies,
can be found attached to the arXiv submission of this paper in Mathematica and FORM format.
As for the first topology the common normalization factor of all master integrals is
\begin{equation}
 S_\epsilon = \left[ (4 \pi)^\epsilon \, {\Gamma(1+ \epsilon) \, \Gamma^2(1-\epsilon) \over \Gamma(1-2\epsilon)} \right]\,.
\end{equation}

\subsection*{5-denominator integrals}
We start listing the two topologies with one single master integral:
\begin{equation}
\I_{174}^{(B)} = \left( {S_\epsilon} \over 16 \pi^2 \right)^2 \frac{( m^2 )^{-2\epsilon-1}}{(1-z)}\,
 \sum_{n=-2}^0\, \epsilon^n \, f_n^{(174)}(y,z) + \mathcal{O}(\epsilon)\,,
\end{equation}
with
{\small
\begin{align*}
 f_{-2}^{(174)}(y,z) &= G(0,z)\,,\\
 f_{-1}^{(174)}(y,z) &= - {\pi^2 \over 6}
          - G(1/z,0,y)
          + 2 G(0,z)
          - G(0,z) G(1/z,y)
          - G(0,0,z) \\
          &+ G(1,0,z)
          + G(1,0,y)\,, \\
 f_{0}^{(174)}(y,z) &= - {11 \over 4} \zeta_3
          - { \pi^2 \over 3} 
          - { \pi^2 \over 2}  \ln{2}
          - G(2 - z,1/z,0,y)
          - { 1/ 2} G(2 - z,y)  \pi^2
          + 2 G(2 - z,1,0,y)\\
          &+ 2 G(1/z,1/z,0,y)
          + {2 / 3} G(1/z,y)\pi^2
          - 2 G(1/z,0,y)
          + 2 G(1/z,0,0,y)\\
          &- 3 G(1/z,1,0,y)
          + 4 G(0,z)
          - 1/6 G(0,z)\pi^2 
          - G(0,z) G(2 - z,1/z,y)\\
          &+ 2 G(0,z) G(1/z,1/z,y)
          - 2 G(0,z) G(1/z,y)
          - 2 G(0,0,z)
          + G(0,0,z) G(1/z,y)\\
          &+ G(0,0,0,z)
          + G(0,1,0,z)
          - 1/6 G(1,z) \pi^2
          + 2 G(1,0,z)
          + 2 G(1,0,z) G(2 - z,y)\\
          &- 3 G(1,0,z) G(1/z,y)
          + 2 G(1,0,y)
          - G(1,0,0,z)
          - 2 G(1,0,0,y)\\
          &+ G(1,1,0,z)
          - 1/2 G(2,z) \pi^2
          + 2 G(2,1,0,z)\,.
\end{align*}
}

\begin{equation}
\I_{182}^{(B)} = \left( {S_\epsilon} \over 16 \pi^2 \right)^2 \frac{( m^2 )^{-2\epsilon-1}}{(y+z-2)}\,
 \sum_{n=-2}^{-1}\, \epsilon^n \, f_n^{(182)}(y,z) + \mathcal{O}(\epsilon^0)\,,
\end{equation}
with
{\small
\begin{align*}
 f_{-2}^{(182)}(y,z) &= - 1/2 \pi^2
          - G(1/z,0,y)
          - G(0,z) G(1/z,y)
          + 2 G(1,0,z)
          + 2 G(1,0,y)\,,\\
 f_{-1}^{(182)}(y,z) &= - {7 \over 2} \zeta_3
          - \pi^2 \ln{2}
          - 2 G(2 - z,1/z,0,y)
          - G(2 - z,y) \pi^2
          + 4 G(2 - z,1,0,y)\\
          &+ 2 G(1/z,1/z,0,y)
          + G(1/z,y) \pi^2
          + 2 G(1/z,0,0,y)
          - 4 G(1/z,1,0,y)\\
          &- 2 G(0,z) G(2 - z,1/z,y)
          + 2 G(0,z) G(1/z,1/z,y)
          + 2 G(0,0,z) G(1/z,y)\\
          &+ 4 G(1,0,z) G(2 - z,y)
          - 4 G(1,0,z) G(1/z,y)
          - 4 G(1,0,0,z)
          - 4 G(1,0,0,y)\\
          &- G(2,z) \pi^2
          + 4 G(2,1,0,z)\,.
\end{align*}
}

Sector $213$ contains 4 MIs which read:
\begin{equation}
\I_{213,1}^{(B)} = \left( {S_\epsilon} \over 16 \pi^2 \right)^2 ( m^2 )^{-2\epsilon-1}\,\frac{\xi}{(1-\xi^2)}\,
 \left[ f_0^{(213,1)}(\xi,\zeta) + i \,\pi g_0^{(213,1)}(\xi,\zeta) \,\right] + \mathcal{O}(\epsilon^1)\,,
\end{equation}
with
{\small
\begin{align*}
f_{0}^{(213,1)}(\xi,\zeta)  &= 
%%%%%%%%%%%%%%%
{1 \over 6} \Big\{ G(1/\xi,\zeta) - G(\xi,\zeta) \Big\}
 \Big[ - 12 \zeta_3 + 12G( - i, - c, -i,\xi) + 12G( -i, - c,i,\xi)\\ 
     &- 12G( -i, - c,0,\xi) + 12G( -i, - \bar{c}, -i,\xi) + 12G( -i, - \bar{c},i,\xi) - 12
      G( -i, - \bar{c},0,\xi)\\ 
      &+ 21G( -i,\xi)\pi^2 - 12G( -i,0, -i,\xi) - 12G( - I
      ,0,i,\xi) + 12G( -i,0,0,\xi) \\
      &- 12G( - c, - c, -i,\xi) - 12G( - c, - c
      ,i,\xi) + 12G( - c, - c,0,\xi) - 12G( - c, - \bar{c}, -i,\xi) \\
      &- 12G( - c,- \bar{c},i,\xi) + 12G( - c, - \bar{c},0,\xi) - 44G( - c,\xi)\pi^2 + 12G( - c,0,
       -i,\xi) \\
       &+ 12G( - c,0,i,\xi) - 12G( - c,0,0,\xi) - 12G( - \bar{c}, - c, -i,
      \xi) - 12G( - \bar{c}, - c,i,\xi) \\
       &+ 12G( - \bar{c}, - c,0,\xi) - 12G( - \bar{c}, - \bar{c},
       -i,\xi) - 12G( - \bar{c}, - \bar{c},i,\xi) + 12G( - \bar{c}, - \bar{c},0,\xi) \\
       &- 44G( - \bar{c},\xi)
      \pi^2 + 12G( - \bar{c},0, -i,\xi) + 12G( - \bar{c},0,i,\xi) - 12G( - \bar{c},0,0,\xi) \\
       &+ 12G(i, - c, -i,\xi) + 12G(i, - c,i,\xi) - 12G(i, - c,0,\xi) + 12G(i, - \bar{c}, -i,\xi) \\
       &+ 12G(i, - \bar{c},i,\xi) - 12G(i, - \bar{c},0,\xi) + 21G(i,\xi)\pi^2 - 
       12G(i,0, -i,\xi) - 12G(i,0,i,\xi) \\
       &+ 12G(i,0,0,\xi) + 23G(0,\xi)\pi^2 \Big]
      %%%%%%%%%%%%%%% 
       + {1 \over 2} \Big\{ G(1/\xi,1/J_\xi,\zeta) - G(\xi,1/J_\xi,\zeta) \Big\}
       \Big[ 7\pi^2 \\
        &+ 4G( - c, -i,\xi) + 4G( - c,i,\xi) - 4G( - c,0,\xi) + 4
      G( - \bar{c}, -i,\xi) + 4G( - \bar{c},i,\xi)\\
      & - 4G( - \bar{c},0,\xi) - 4G(0, -i,\xi) - 4G(
      0,i,\xi) + 4G(0,0,\xi) \Big]
      %%%%%%%%%%%%%%%%
       +  {2\over 3} \Big\{  -G(1/\xi,1/I_\xi,\zeta) \\
       &+ G(1/\xi,-1,\zeta) + G(\xi,1/I_\xi,\zeta) - G(\xi,-1,\zeta) \Big\}
       \Big[ 11\pi^2 + 3G( - c, -i,\xi) + 3G( - c,i,\xi)\\ 
       &- 3G( - c,0,\xi) + 3 G( - \bar{c}, -i,\xi) + 3G( - \bar{c},i,\xi) - 3G( - \bar{c},0,\xi) - 3G(0, -i,\xi) \\
      &- 3G(0,i,\xi) + 3G(0,0,\xi) \Big] \\
       %%%%%%%%%%%%%%%%%%%%
      &-{1 \over 72} \Big[ 65\pi^4 + 576G(0, -i, - c, -i,\xi) + 576G(0, -i, - c,i,\xi) \\&- 
      576G(0, -i, - c,0,\xi) + 576G(0, -i, - \bar{c}, -i,\xi) + 576G(0, -i, - \bar{c},i,\xi) \\ 
      &- 576G(0, -i, - \bar{c},0,\xi) + 1008G(0, -i,\xi)\pi^2 - 576G(0, - i,0, -i,\xi) \\
      &- 576G(0, -i,0,i,\xi) + 576G(0, -i,0,0,\xi) - 432G(0, - c,- c, -i,\xi) \\
       &- 432G(0, - c, - c,i,\xi) + 432G(0, - c, - c,0,\xi) - 
      432G(0, - c, - \bar{c}, -i,\xi) \\
      &- 432G(0, - c, - \bar{c},i,\xi) + 432G(0, - c,- \bar{c},0,\xi) - 1584G(0, - c,\xi)\pi^2 \\
       &+ 432G(0, - c,0, -i,\xi) + 432G(0,- c,0,i,\xi) - 432G(0, - c,0,0,\xi) \\
       &- 432G(0, - \bar{c}, - c, -i,\xi) - 432
      G(0, - \bar{c}, - c,i,\xi) + 432G(0, - \bar{c}, - c,0,\xi) \\
      &- 432G(0, - \bar{c}, - \bar{c},-i,\xi) - 432G(0, - \bar{c}, - \bar{c},i,\xi) 
        + 432G(0, - \bar{c}, - \bar{c},0,\xi) \\
       &- 1584G(0, - \bar{c},\xi)\pi^2 + 432G(0, - \bar{c},0, -i,\xi) + 432G(0, - \bar{c},0,i,\xi) \\
      &- 432G(0, - \bar{c},0,0,\xi)+ 576G(0,i, - c, -i,\xi) + 576G(0,i, - c,i,\xi) \\
      &- 576G(0,i, - c,0,\xi) + 576G(0,i, - \bar{c}, -i,\xi) + 576G(0,i, - \bar{c},i,\xi) \\
      &- 576G(0,i, - \bar{c},0,\xi) + 1008G(0,i,\xi)\pi^2 - 576G(0,i,0, -i,\xi) \\
      &- 576G(0,i,0,i,\xi) + 576G(0,i,0,0,\xi) + 576G(0,-1, - c, -i,\xi) \\
      &+ 576G(0,-1, - c,i,\xi) - 576G(0,-1, - c,0,\xi) + 576G(0,-1, - \bar{c}, -i,\xi) \\
      &+ 576G(0,-1, - \bar{c},i,\xi) - 576G(0,-1, - \bar{c},0,\xi) + 912G(0,-1,\xi)\pi^2\\ 
      &- 576G(0,-1,0, -i,x) - 576G(0,-1,0,i,\xi) + 576G(0,-1,0,0,\xi) \\
      &- 432G(0,0, - c, -i,\xi) - 432G(0,0, - c,i,\xi) + 432G(0,0, - c,0,\xi) \\
      &- 432G(0,0, - \bar{c}, -i,\xi) - 432G(0,0, - \bar{c},i,\xi) + 432G(0,0, - \bar{c},0,\xi) \\
      &+ 120G(0,0,\xi)\pi^2 + 432G(0,0,0, -i,\xi) + 432G(0,0,0,i,\xi) - 432G(0,0,0,0,\xi) \Big]\\
      %%%%%%%%%%%%%%%%%%%%%%
       &+ \Big[  G(0,\xi) - G( -i,\xi) - G(i,\xi) \Big]  
          \Big\{  2G(1/\xi,1/I_\xi,1/I_\xi,\zeta) - 2G(1/\xi,1/I_\xi,1/\xi,\zeta)  \\
          &- 2G(1/\xi,1/I_\xi,\xi,\zeta) + G(1/\xi,1/J_\xi,1/\xi,\zeta)
          - 2G(1/\xi,1/J_\xi,1/I_\xi,\zeta) \\
          &+ G(1/\xi,1/J_\xi,\xi,\zeta) + 2G(1/\xi,-1,1/\xi,\zeta)
          - 2G(1/\xi,-1,1/I_\xi,\zeta) + 2G(1/\xi,-1,\xi,\zeta) \\
          &- G(1/\xi,0,1/\xi,\zeta) - G(1/\xi,0,\xi,\zeta) + 2G(\xi,1/I_\xi,1/\xi,\zeta) - 2G(\xi,1/I_\xi,1/I_\xi,\zeta) \\
          &+ 2G(\xi,1/I_\xi,\xi,\zeta) - G(\xi,1/J_\xi,1/\xi,\zeta) + 2G(\xi,1/J_\xi,1/I_\xi,\zeta) - G(\xi,1/J_\xi,\xi,\zeta) \\
          &- 2G(\xi,-1,1/\xi,\zeta) + 2G(\xi,-1,1/I_\xi,\zeta) - 2G(\xi,-1,\xi,\zeta) + G(\xi,0,1/\xi,\zeta) \\
          &+ G(\xi,0,\xi,\zeta) \Big\}
         %%%%%%%%%%%%%%%%%%%%%%%
       + 2G(1/\xi,1/I_\xi,1/\xi,1/J_\xi,\zeta) + 2G(1/\xi,1/I_\xi,1/\xi,0,\zeta) \\
       &- 2G(1/\xi,1/I_\xi,1/I_\xi,1/J_\xi,\zeta) + 2G(1/\xi,1/I_\xi,\xi,1/J_\xi,\zeta) + 2G(1/\xi,1/I_\xi,\xi,0,\zeta) \\
       &- 2G(1/\xi,1/I_\xi,-1,0,\zeta) - G(1/\xi,1/J_\xi,1/\xi,1/J_\xi,\zeta) - G(1/\xi,1/J_\xi,1/\xi,0,\zeta) \\
       &+ 2G(1/\xi,1/J_\xi,1/I_\xi,1/J_\xi,\zeta) - G(1/\xi,1/J_\xi,\xi,1/J_\xi,\zeta) - G(1/\xi,1/J_\xi,\xi,0,\zeta) \\
       &- 2G(1/\xi,-1,1/\xi,1/J_\xi,\zeta) - 2G(1/\xi,-1,1/\xi,0,\zeta) + 2G(1/\xi,-1,1/I_\xi,1/J_\xi,\zeta) \\
       &- 2G(1/\xi,-1,\xi,1/J_\xi,\zeta)- 2G(1/\xi,-1,\xi,0,\zeta) + 2G(1/\xi,-1,-1,0,\zeta) \\
       &+ G(1/\xi,0,1/\xi,1/J_\xi,\zeta)+ G(1/\xi,0,1/\xi,0,\zeta) + G(1/\xi,0,\xi,1/J_\xi,\zeta) \\
       &+ G(1/\xi,0,\xi,0,\zeta) - 23/6G(1/\xi,0,\zeta)\pi^2 - 2G(1/\xi,0,-1,0,\zeta) - 2G(\xi,1/I_\xi,1/\xi,1/J_\xi,\zeta) \\
       &- 2G(\xi,1/I_\xi,1/\xi,0,\zeta) + 2G(\xi,1/I_\xi,1/I_\xi,1/J_\xi,\zeta) - 2G(\xi,1/I_\xi,\xi,1/J_\xi,\zeta) \\
       &- 2G(\xi,1/I_\xi,\xi,0,\zeta) + 2G(\xi,1/I_\xi,-1,0,\zeta) + G(\xi,1/J_\xi,1/\xi,1/J_\xi,\zeta) \\
       &+ G(\xi,1/J_\xi,1/\xi,0,\zeta) - 2G(\xi,1/J_\xi,1/I_\xi,1/J_\xi,\zeta) + G(\xi,1/J_\xi,\xi,1/J_\xi,\zeta) \\
       &+ G(\xi,1/J_\xi,\xi,0,\zeta) + 2G(\xi,-1,1/\xi,1/J_\xi,\zeta) + 2G(\xi,-1,1/\xi,0,\zeta) \\
       &- 2G(\xi,-1,1/I_\xi,1/J_\xi,\zeta) + 2G(\xi,-1,\xi,1/J_\xi,\zeta) + 2G(\xi,-1,\xi,0,\zeta) \\
       &- 2G(\xi,-1,-1,0,\zeta) - G(\xi,0,1/\xi,1/J_\xi,\zeta) - G(\xi,0,1/\xi,0,\zeta)
          - G(\xi,0,\xi,1/J_\xi,\zeta)\\& - G(\xi,0,\xi,0,\zeta) + 23/6G(\xi,0,\zeta)\pi^2 + 2G(\xi,0,-1,0,\zeta)\,,
\end{align*}
}
and
{\small
\begin{align*}
g_{0}^{(213,1)}(\xi,\zeta) &=
       {1 \over 3} \Big\{ G(1/\xi,\zeta) - G(\xi,\zeta) \Big\} \Big[ \pi^2 + 6G( -i, - c,\xi) + 6G( -i, - \bar{c},\xi) 
       - 6G( -i,0,\xi) \\&- 6G( - c, - c,\xi) 
      - 6G( - c, - \bar{c},\xi) + 6G( - c,0,\xi) - 6G( - \bar{c}, - c,\xi) - 6G( - \bar{c}, - \bar{c},\xi) \\
      &+ 6G( - \bar{c},0,\xi) + 6G(i, - c,\xi) + 6G(i, - \bar{c},\xi) - 6G(i,0,\xi) \Big]  \\
      %%%%%%%%%%%%%%%%%
      &+ \Big[ G(0,\xi)  - G( - c,\xi) - G( - \bar{c},\xi)\Big]  \Big\{ 2G(1/\xi,1/I_\xi,\zeta) - 2G(
         1/\xi,1/J_\xi,\zeta) \\&- 2G(1/\xi,-1,\zeta) - 2G(\xi,1/I_\xi,\zeta) + 2G(\xi,1/J_\xi,\zeta) + 2
         G(\xi,-1,\zeta) \Big\}\\
       &-{1 \over 3} \Big[ 24G(0, -i, - c,\xi) + 24G(0, -i, - \bar{c},\xi) - 24G(0, -i,0,\xi) - 
      18G(0, - c, - c,\xi) \\
      &- 18G(0, - c, - \bar{c},\xi) + 18G(0, - c,0,\xi) - 18
      G(0, - \bar{c}, - c,\xi) - 18G(0, - \bar{c}, - \bar{c},\xi) \\
      &+ 18G(0, - \bar{c},0,\xi) + G(0,\xi)
      \pi^2 + 24G(0,i, - c,\xi) + 24G(0,i, - \bar{c},\xi) - 24G(0,i,0,\xi) \\
      &+ 24G(0,-1, - c,\xi) + 24G(0,-1, - \bar{c},\xi) - 24G(0,-1,0,\xi) - 18G(0,0, - c,\xi) \\
      &- 18G(0,0, - \bar{c},\xi) + 18G(0,0,0,\xi) \Big] 
      + 4G(1/\xi,1/I_\xi,1/\xi,\zeta) - 2G(1/\xi,1/I_\xi,1/I_\xi,\zeta) \\
      &+ 4G(1/\xi,1/I_\xi,\xi,\zeta) 
      - 2G(1/\xi,1/I_\xi,-1,\zeta) - 2G(1/\xi,1/J_\xi,1/\xi,\zeta) \\
      &+ 2G(1/\xi,1/J_\xi,1/I_\xi,\zeta) 
         - 2G(1/\xi,1/J_\xi,\xi,\zeta) - 4G(1/\xi,-1,1/\xi,\zeta) \\
         &+ 2G(1/\xi,-1,1/I_\xi,\zeta) - 4G(1/\xi,-1,\xi,\zeta) + 2G(1/\xi,-1,-1,\zeta) + 2G(1/\xi,0,1/\xi,\zeta)\\
          &+ 2G(1/\xi,0,\xi,\zeta) - 2G(1/\xi,0,-1,\zeta) - 4G(\xi,1/I_\xi,1/\xi,\zeta) + 2G(\xi,
         1/I_\xi,1/I_\xi,\zeta) \\
         &- 4G(\xi,1/I_\xi,\xi,\zeta) + 2G(\xi,1/I_\xi,-1,\zeta) + 2G(\xi,1/J_\xi,
         1/\xi,\zeta) - 2G(\xi,1/J_\xi,1/I_\xi,\zeta) \\
         &+ 2G(\xi,1/J_\xi,\xi,\zeta) + 4G(\xi,-1,1/\xi,\zeta)
          - 2G(\xi,-1,1/I_\xi,\zeta) + 4G(\xi,-1,\xi,\zeta) \\&- 2G(\xi,-1,-1,\zeta) - 2G(\xi,0,1/\xi,\zeta) 
          - 2G(\xi,0,\xi,\zeta) + 2G(\xi,0,-1,\zeta) \,.
\end{align*}
}

\begin{equation}
\I_{213,2}^{(B)} = \left( {S_\epsilon} \over 16 \pi^2 \right)^2 \frac{( m^2 )^{-2\epsilon-2}}{(1-yz)}\,
 \sum_{n=-2}^0\, \epsilon^n \, f_n^{(213,2)}(y,z) + \mathcal{O}(\epsilon)\,,
\end{equation}
with
{\small
\begin{align*}
 f_{-2}^{(213,2)}(y,z) &= G(0,z) + G(0,y)\,, \\
 f_{-1}^{(213,2)}(y,z) &=  - {5 \over 6} \pi^2
          - 3 G(1/z,0,y)
          - 3 G(0,z) G(1/z,y)
          - 2 G(0,0,z)
          - 2 G(0,0,y)\\
          &+ 4 G(1,0,z)
          + 4 G(1,0,y)\,,\\ 
 f_{0}^{(213,2)}(y,z) &=  - 6\zeta_3
          - 2\pi^2 \ln{2}
          - 4G(2 - z,1/z,0,y)
          - 2G(2 - z,y)\pi^2
          + 8G(2 - z,1,0,y)\\
          &+ 9G(1/z,1/z,0,y)
          + 5/2G(1/z,y)\pi^2
          + 6G(1/z,0,0,y)
          - 12G(1/z,1,0,y)\\
          &- G(0,1/z,0,y)
          - 1/6G(0,z)\pi^2
          - 4G(0,z)G(2 - z,1/z,y)
          + 9G(0,z)G(1/z,1/z,y)\\
          &- G(0,z)G(0,1/z,y)
          - 2G(0,z)G(1,1/z,y)
          - 1/6G(0,y)\pi^2
          + 6G(0,0,z)G(1/z,y)\\
          &+ 4G(0,0,0,z)
          + 4G(0,0,0,y)
          + 2G(0,1,0,z)
          + 2G(0,1,0,y)
          - 2G(1,1/z,0,y)\\
          &+ 1/3G(1,z)\pi^2
          + 1/3G(1,y)\pi^2
          + 8G(1,0,z)G(2 - z,y)
          - 12G(1,0,z)G(1/z,y)\\
          &+ 2G(1,0,z)G(1,y)
          - 8G(1,0,0,z)
          - 8G(1,0,0,y)
          - 2G(1,1,0,z)
          - 2G(1,1,0,y)\\
          &- 2G(2,z)\pi^2
          + 8G(2,1,0,z)\,.
\end{align*}
}
\begin{equation}
\I_{213,3}^{(B)} = \left( {S_\epsilon} \over 16 \pi^2 \right)^2 \frac{( m^2 )^{-2\epsilon-2}}{(1-z)}\,
 \sum_{n=-2}^0\, \epsilon^n \, f_n^{(213,3)}(y,z) + \mathcal{O}(\epsilon)\,,
\end{equation}
with
{\small
\begin{align*}
 f_{-2}^{(213,3)}(y,z) &= 1/2 G(0,z)\,,\\
 f_{-1}^{(213,3)}(y,z) &=  {\pi^2 \over 6}
          + 1/2 G(1/z,0,y)
          + 1/2 G(0,z) G(1/z,y)
          - G(0,0,z)
          - G(1,0,z)
          - 1/2 G(1,0,y) \\ 
 f_{0}^{(213,3)}(y,z) &=  {11\over4} \zeta_3
          + { \pi^2 \over 2} \ln{2}
          + G(2 - z,1/z,0,y)
          + 1/2G(2 - z,y)\pi^2
          - 2G(2 - z,1,0,y)\\
          &- 3/2G(1/z,1/z,0,y)
          - 5/12G(1/z,y)\pi^2
          - G(1/z,0,0,y)
          + 2G(1/z,1,0,y)\\
          &+ 1/6G(0,z)\pi^2
          + G(0,z)G(2 - z,1/z,y)
          - 3/2G(0,z)G(1/z,1/z,y)\\
          &+ G(0,z)G(1,1/z,y)
          - G(0,0,z)G(1/z,y)
          + 2G(0,0,0,z)
          - 2G(0,1,0,z)\\
          &+ G(1,1/z,0,y)
          - 1/3G(1,z)\pi^2
          + 1/12G(1,y)\pi^2
          - 2G(1,0,z)G(2 - z,y)\\
          &+ 2G(1,0,z)G(1/z,y)
          - G(1,0,z)G(1,y)
          + 2G(1,0,0,z)
          + G(1,0,0,y)\\
          &+ 2G(1,1,0,z)
          - 1/2G(1,1,0,y)
          + 1/2G(2,z)\pi^2
          - 2G(2,1,0,z)\,. 
\end{align*}
}

Finally the last master is equal to the previous one under the exchange $y \leftrightarrow z$:

\begin{equation}
\I_{213,4}^{(B)} = \left( {S_\epsilon} \over 16 \pi^2 \right)^2 \frac{( m^2 )^{-2\epsilon-2}}{(1-y)}\,
 \sum_{n=-2}^0\, \epsilon^n \, f_n^{(213,3)}(z,y) + \mathcal{O}(\epsilon)\,.
\end{equation}

\subsection*{6-denominator Integrals}
The only irreducible topology with 6 denominators has one MI which reads:
\begin{equation}
\I_{215}^{(B)} = \left( {S_\epsilon} \over 16 \pi^2 \right)^2 \frac{( m^2 )^{-2\epsilon-2}}{y(1-z)}\,
 \sum_{n=-2}^{-1}\, \epsilon^n \, f_n^{(215)}(y,z) + \mathcal{O}(\epsilon^0)\,,
\end{equation}
with
{\small
\begin{align*}
 f_{-2}^{(215)}(y,z) &=   1/2 G(1/z,0,y)
          + 1/2 G(0,z)G(1/z,y)
          - 1/2 G(1,0,y)\,,\\
 f_{-1}^{(215)}(y,z) &= - 1/2 G(1/z,1/z,0,y)
          - 1/4 G(1/z,y)\pi^2
          + G(1/z,1,0,y)
          - G(0,1/z,0,y)\\
          &- 1/2 G(0,z)G(1/z,1/z,y)
          - G(0,z)G(0,1/z,y)
          + 2G(0,z)G(1,1/z,y)\\
          &- G(0,0,z)G(1/z,y)
          + G(0,1,0,y)
          + 2G(1,1/z,0,y)
          + 5/12G(1,y)\pi^2\\
          &+ G(1,0,z)G(1/z,y)
          - 2G(1,0,z)G(1,y)
          - 5/2G(1,1,0,y)\,.
\end{align*}
}

\subsection*{7-denominator Integrals}
There is one irreducible topology with 7 denominators with two MIs which read:
\begin{equation}
\I_{247,1}^{(B)} = \left( {S_\epsilon} \over 16 \pi^2 \right)^2 \frac{( m^2 )^{-2\epsilon-3}}{y(1-yz)}\,
 \sum_{n=-2}^{-1}\, \epsilon^n \, f_n^{(247,1)}(y,z) + \mathcal{O}(\epsilon^0)\,,
\end{equation}
with
{\small
\begin{align*}
 f_{-2}^{(247,1)}(y,z) &=   {\pi^2 \over 3}
          - 2G(1/z,0,y)
          - 2G(0,z)G(1/z,y)\,,\\
 f_{-1}^{(247,1)}(y,z) &=   - 7 \zeta_3
          + 7G(1/z,1/z,0,y)
          + 3/2G(1/z,y)\pi^2
          + 4G(1/z,0,0,y)
          - 8G(1/z,1,0,y)\\
          &+ 2G(0,1/z,0,y)
          + 7G(0,z)G(1/z,1/z,y)
          + 2G(0,z)G(0,1/z,y)\\
          &- 6G(0,z)G(1,1/z,y)
          - 2/3G(0,y)\pi^2
          + 4G(0,0,z)G(1/z,y)
          + 2G(0,1,0,y)\\
          &- 6G(1,1/z,0,y)
          - 2/3G(1,y)\pi^2
          - 8G(1,0,z)G(1/z,y)
          + 6G(1,0,z)G(1,y)\\
          &- 6G(1,0,0,y)
          + 4G(1,1,0,y)\,.
\end{align*}
}

\begin{align*}
\I_{247,2}^{(B)} = \left( {S_\epsilon} \over 16 \pi^2 \right)^2 ( m^2 )^{-2\epsilon-2} &\Bigg[ 
\frac{(1-z)}{(1-y)(1-yz)}\,
 \sum_{n=-2}^{-1}\, \epsilon^n \, f_n^{(247,2)}(y,z)  \\
&+
\frac{1}{y(1-y)}\,
 \sum_{n=-2}^{-1}\, \epsilon^n \, g_n^{(247,2)}(y,z)  
\Bigg] + \mathcal{O}(\epsilon^0)\,,
\end{align*}
with
{\small
\begin{align*}
 f_{-2}^{(247,2)}(y,z) &=  {\pi^2 \over 2}
          + 3/2G(0,0,y)
          - 3/2G(1,0,z)
          - 3/2G(1,0,y)\,, \\
 f_{-1}^{(247,2)}(y,z) &=   {15 \over  4}\zeta_3
          + {3 \over 2} \pi^2 \ln{2}
          + 3G(2 - z,1/z,0,y)
          + 3/2G(2 - z,y)\pi^2
          - 6G(2 - z,1,0,y)\\
         &- G(1/z,y)\pi^2
          - 3G(1/z,0,0,y)
          + 3G(1/z,1,0,y)
          + 3G(0,z)G(2 - z,1/z,y)\\
         &- 3G(0,z)G(1,1/z,y)
          - 1/4G(0,y)\pi^2
          - 9/2G(0,0,0,y)
          + 3/2G(0,1,0,y)\\
          &- 3G(1,1/z,0,y)
          + 1/4G(1,z)\pi^2
          - 1/4G(1,y)\pi^2
          - 6G(1,0,z)G(2 - z,y)\\
          &+ 3G(1,0,z)G(1/z,y)
          + 3G(1,0,z)G(1,y)
          + 3G(1,0,0,z)
          + 15/2G(1,0,0,y)\\
          &- 3/2G(1,1,0,z)
          + 3/2G(1,1,0,y)
          + 3/2G(2,z)\pi^2
          - 6G(2,1,0,z)\,,
\end{align*} }
and
{\small
\begin{align*}
 g_{-2}^{(247,2)}(y,z) &=  - {\pi^2 \over 4 }
          + 3/2G(1/z,0,y)
          + 3/2G(0,z)G(1/z,y)\,,\\
 g_{-1}^{(247,2)}(y,z) &=           { 9 \over 2 }\zeta_3
          - 9/2G(1/z,1/z,0,y)
          - 11/12G(1/z,y)\pi^2
          - 2G(1/z,0,0,y)
          + 5G(1/z,1,0,y)\\
          &- G(0,1/z,0,y)
          - 9/2G(0,z)G(1/z,1/z,y)
          - G(0,z)G(0,1/z,y)
          + 4G(0,z)G(1,1/z,y)\\
          &+ 1/2G(0,y)\pi^2
          - 3G(0,0,z)G(1/z,y)
          - 2G(0,1,0,y)
          + 4G(1,1/z,0,y)
          + 1/2G(1,y)\pi^2\\
          &+ 5G(1,0,z)G(1/z,y)
          - 4G(1,0,z)G(1,y)
          + 2G(1,0,0,y)
          - 3G(1,1,0,y)\,.
 \end{align*}
}

\bibliographystyle{JHEP}   
\bibliography{Biblio}     

\providecommand{\href}[2]{#2}\begingroup\raggedright\begin{thebibliography}{10}

\bibitem{Accomando:2005xp}
E.~Accomando and A.~Kaiser, {\it {Electroweak corrections and anomalous triple
  gauge-boson couplings in $W^{+} W^{-}$ and $W^\pm Z$ production at the LHC}},
   {\em Phys.Rev.} {\bf D73} (2006) 093006,
  [\href{http://xxx.lanl.gov/abs/hep-ph/0511088}{{\tt hep-ph/0511088}}].

\bibitem{Accomando:2004de}
E.~Accomando, A.~Denner, and A.~Kaiser, {\it {Logarithmic electroweak
  corrections to gauge-boson pair production at the LHC}},  {\em Nucl.Phys.}
  {\bf B706} (2005) 325--371,
  [\href{http://xxx.lanl.gov/abs/hep-ph/0409247}{{\tt hep-ph/0409247}}].

\bibitem{Accomando:2005ra}
E.~Accomando, A.~Denner, and C.~Meier, {\it {Electroweak corrections to $W
  \gamma$ and $Z \gamma$ production at the LHC}},  {\em Eur.Phys.J.} {\bf C47}
  (2006) 125--146, [\href{http://xxx.lanl.gov/abs/hep-ph/0509234}{{\tt
  hep-ph/0509234}}].

\bibitem{Bierweiler:2013dja}
A.~Bierweiler, T.~Kasprzik, and J.~H. {K\"uhn}, {\it {Vector-boson pair
  production at the LHC to $\mathcal{O}(\alpha^3)$ accuracy}},
  \href{http://xxx.lanl.gov/abs/1305.5402}{{\tt arXiv:1305.5402}}.

\bibitem{Ohnemus:1992jn}
J.~Ohnemus, {\it {Order $\alpha_s$ calculations of hadronic $W^\pm \gamma$ and
  $Z \gamma$ production}},  {\em Phys.Rev.} {\bf D47} (1993) 940--955.

\bibitem{Baur:1993ir}
U.~Baur, T.~Han, and J.~Ohnemus, {\it {QCD corrections to hadronic $W \gamma$
  production with nonstandard $W W \gamma$ couplings}},  {\em Phys.Rev.} {\bf
  D48} (1993) 5140--5161, [\href{http://xxx.lanl.gov/abs/hep-ph/9305314}{{\tt
  hep-ph/9305314}}].

\bibitem{Baur:1997kz}
U.~Baur, T.~Han, and J.~Ohnemus, {\it {QCD corrections and anomalous couplings
  in $Z \gamma$ production at hadron colliders}},  {\em Phys.Rev.} {\bf D57}
  (1998) 2823--2836, [\href{http://xxx.lanl.gov/abs/hep-ph/9710416}{{\tt
  hep-ph/9710416}}].

\bibitem{Dixon:1998py}
L.~J. Dixon, Z.~Kunszt, and A.~Signer, {\it {Helicity amplitudes for
  O($alpha_s$) production of $W^{+} W^{-}$, $W^\pm Z$, $Z Z$, $W^\pm \gamma$,
  or $Z \gamma$ pairs at hadron colliders}},  {\em Nucl.Phys.} {\bf B531}
  (1998) 3--23, [\href{http://xxx.lanl.gov/abs/hep-ph/9803250}{{\tt
  hep-ph/9803250}}].

\bibitem{Dixon:1999di}
L.~J. Dixon, Z.~Kunszt, and A.~Signer, {\it {Vector boson pair production in
  hadronic collisions at order $\alpha_s$ : Lepton correlations and anomalous
  couplings}},  {\em Phys.Rev.} {\bf D60} (1999) 114037,
  [\href{http://xxx.lanl.gov/abs/hep-ph/9907305}{{\tt hep-ph/9907305}}].

\bibitem{Catani:2011qz}
S.~Catani, L.~Cieri, D.~de~Florian, G.~Ferrera, and M.~Grazzini, {\it {Diphoton
  production at hadron colliders: a fully-differential QCD calculation at
  NNLO}},  {\em Phys.Rev.Lett.} {\bf 108} (2012) 072001,
  [\href{http://xxx.lanl.gov/abs/1110.2375}{{\tt arXiv:1110.2375}}].

\bibitem{DelDuca:2003uz}
V.~Del~Duca, F.~Maltoni, Z.~Nagy, and Z.~Trocsanyi, {\it {QCD radiative
  corrections to prompt diphoton production in association with a jet at hadron
  colliders}},  {\em JHEP} {\bf 0304} (2003) 059,
  [\href{http://xxx.lanl.gov/abs/hep-ph/0303012}{{\tt hep-ph/0303012}}].

\bibitem{Gehrmann:2013aga}
T.~Gehrmann, N.~Greiner, and G.~Heinrich, {\it {Photon isolation effects at NLO
  in gamma gamma + jet final states in hadronic collisions}},  {\em JHEP} {\bf
  1306} (2013) 058, [\href{http://xxx.lanl.gov/abs/1303.0824}{{\tt
  arXiv:1303.0824}}].

\bibitem{Dittmaier:2007th}
S.~Dittmaier, S.~Kallweit, and P.~Uwer, {\it {NLO QCD corrections to WW+jet
  production at hadron colliders}},  {\em Phys.Rev.Lett.} {\bf 100} (2008)
  062003, [\href{http://xxx.lanl.gov/abs/0710.1577}{{\tt arXiv:0710.1577}}].

\bibitem{Dittmaier:2009un}
S.~Dittmaier, S.~Kallweit, and P.~Uwer, {\it {NLO QCD corrections to
  $pp/p\bar{p} \to WW+jet+X$ including leptonic W-boson decays}},  {\em
  Nucl.Phys.} {\bf B826} (2010) 18--70,
  [\href{http://xxx.lanl.gov/abs/0908.4124}{{\tt arXiv:0908.4124}}].

\bibitem{Binoth:2009wk}
T.~Binoth, T.~Gleisberg, S.~Karg, N.~Kauer, and G.~Sanguinetti, {\it {NLO QCD
  corrections to ZZ+ jet production at hadron colliders}},  {\em Phys.Lett.}
  {\bf B683} (2010) 154--159, [\href{http://xxx.lanl.gov/abs/0911.3181}{{\tt
  arXiv:0911.3181}}].

\bibitem{Campanario:2009um}
F.~Campanario, C.~Englert, M.~Spannowsky, and D.~Zeppenfeld, {\it {NLO-QCD
  corrections to W $\gamma$ j production}},  {\em Europhys.Lett.} {\bf 88}
  (2009) 11001, [\href{http://xxx.lanl.gov/abs/0908.1638}{{\tt
  arXiv:0908.1638}}].

\bibitem{Campanario:2010hv}
F.~Campanario, C.~Englert, and M.~Spannowsky, {\it {Precise predictions for
  (non-standard) $W \gamma$ + jet production}},  {\em Phys.Rev.} {\bf D83}
  (2011) 074009, [\href{http://xxx.lanl.gov/abs/1010.1291}{{\tt
  arXiv:1010.1291}}].

\bibitem{Campanario:2010hp}
F.~Campanario, C.~Englert, S.~Kallweit, M.~Spannowsky, and D.~Zeppenfeld, {\it
  {NLO QCD corrections to WZ+jet production with leptonic decays}},  {\em JHEP}
  {\bf 1007} (2010) 076, [\href{http://xxx.lanl.gov/abs/1006.0390}{{\tt
  arXiv:1006.0390}}].

\bibitem{Bern:2001df}
Z.~Bern, A.~De~Freitas, and L.~J. Dixon, {\it {Two loop amplitudes for gluon
  fusion into two photons}},  {\em JHEP} {\bf 0109} (2001) 037,
  [\href{http://xxx.lanl.gov/abs/hep-ph/0109078}{{\tt hep-ph/0109078}}].

\bibitem{Gehrmann:2011ab}
T.~Gehrmann and L.~Tancredi, {\it {Two-loop QCD helicity amplitudes for $q\bar
  q \to W^\pm \gamma$ and $q\bar q \to Z^0 \gamma$}},  {\em JHEP} {\bf 1202}
  (2012) 004, [\href{http://xxx.lanl.gov/abs/1112.1531}{{\tt
  arXiv:1112.1531}}].

\bibitem{Gehrmann:2013vga}
T.~Gehrmann, L.~Tancredi, and E.~Weihs, {\it {Two-loop QCD helicity amplitudes
  for $g\,g \to Z\,g$ and $g\,g \to Z\,\gamma $}},  {\em JHEP} {\bf 1304}
  (2013) 101, [\href{http://xxx.lanl.gov/abs/1302.2630}{{\tt
  arXiv:1302.2630}}].

\bibitem{Chachamis:2008yb}
G.~Chachamis, M.~Czakon, and D.~Eiras, {\it {W Pair Production at the LHC. I.
  Two-loop Corrections in the High Energy Limit}},  {\em JHEP} {\bf 0812}
  (2008) 003, [\href{http://xxx.lanl.gov/abs/0802.4028}{{\tt
  arXiv:0802.4028}}].

\bibitem{Tkachov:1981wb}
F.~Tkachov, {\it {A Theorem on Analytical Calculability of Four Loop
  Renormalization Group Functions}},  {\em Phys.Lett.} {\bf B100} (1981)
  65--68.

\bibitem{Chetyrkin:1981qh}
K.~Chetyrkin and F.~Tkachov, {\it {Integration by Parts: The Algorithm to
  Calculate beta Functions in 4 Loops}},  {\em Nucl.Phys.} {\bf B192} (1981)
  159--204.

\bibitem{Laporta:2001dd}
S.~Laporta, {\it {High precision calculation of multiloop Feynman integrals by
  difference equations}},  {\em Int.J.Mod.Phys.} {\bf A15} (2000) 5087--5159,
  [\href{http://xxx.lanl.gov/abs/hep-ph/0102033}{{\tt hep-ph/0102033}}].

\bibitem{Anastasiou:2004vj}
C.~Anastasiou and A.~Lazopoulos, {\it {Automatic integral reduction for higher
  order perturbative calculations}},  {\em JHEP} {\bf 0407} (2004) 046,
  [\href{http://xxx.lanl.gov/abs/hep-ph/0404258}{{\tt hep-ph/0404258}}].

\bibitem{Smirnov:2008iw}
A.~Smirnov, {\it {Algorithm FIRE -- Feynman Integral REduction}},  {\em JHEP}
  {\bf 0810} (2008) 107, [\href{http://xxx.lanl.gov/abs/0807.3243}{{\tt
  arXiv:0807.3243}}].

\bibitem{Studerus:2009ye}
C.~Studerus, {\it {Reduze-Feynman Integral Reduction in C++}},  {\em
  Comput.Phys.Commun.} {\bf 181} (2010) 1293--1300,
  [\href{http://xxx.lanl.gov/abs/0912.2546}{{\tt arXiv:0912.2546}}].

\bibitem{vonManteuffel:2012np}
A.~von Manteuffel and C.~Studerus, {\it {Reduze 2 - Distributed Feynman
  Integral Reduction}},  \href{http://xxx.lanl.gov/abs/1201.4330}{{\tt
  arXiv:1201.4330}}.

\bibitem{Kotikov:1990kg}
A.~Kotikov, {\it {Differential equations method: New technique for massive
  Feynman diagrams calculation}},  {\em Phys.Lett.} {\bf B254} (1991) 158--164.

\bibitem{Remiddi:1997ny}
E.~Remiddi, {\it {Differential equations for Feynman graph amplitudes}},  {\em
  Nuovo Cim.} {\bf A110} (1997) 1435--1452,
  [\href{http://xxx.lanl.gov/abs/hep-th/9711188}{{\tt hep-th/9711188}}].

\bibitem{Caffo:1998du}
M.~Caffo, H.~Czyz, S.~Laporta, and E.~Remiddi, {\it {The Master differential
  equations for the two loop sunrise selfmass amplitudes}},  {\em Nuovo Cim.}
  {\bf A111} (1998) 365--389,
  [\href{http://xxx.lanl.gov/abs/hep-th/9805118}{{\tt hep-th/9805118}}].

\bibitem{Gehrmann:1999as}
T.~Gehrmann and E.~Remiddi, {\it {Differential equations for two loop four
  point functions}},  {\em Nucl.Phys.} {\bf B580} (2000) 485--518,
  [\href{http://xxx.lanl.gov/abs/hep-ph/9912329}{{\tt hep-ph/9912329}}].

\bibitem{Gehrmann:2000zt}
T.~Gehrmann and E.~Remiddi, {\it {Two loop master integrals for $\gamma^*
  \rightarrow$ 3 jets: The Planar topologies}},  {\em Nucl.Phys.} {\bf B601}
  (2001) 248--286, [\href{http://xxx.lanl.gov/abs/hep-ph/0008287}{{\tt
  hep-ph/0008287}}].

\bibitem{Gehrmann:2001ck}
T.~Gehrmann and E.~Remiddi, {\it {Two loop master integrals for $\gamma^*
  \rightarrow$ 3 jets: The Nonplanar topologies}},  {\em Nucl.Phys.} {\bf B601}
  (2001) 287--317, [\href{http://xxx.lanl.gov/abs/hep-ph/0101124}{{\tt
  hep-ph/0101124}}].

\bibitem{Bonciani:2008az}
R.~Bonciani, A.~Ferroglia, T.~Gehrmann, D.~Maitre, and C.~Studerus, {\it
  {Two-Loop Fermionic Corrections to Heavy-Quark Pair Production: The
  Quark-Antiquark Channel}},  {\em JHEP} {\bf 0807} (2008) 129,
  [\href{http://xxx.lanl.gov/abs/0806.2301}{{\tt arXiv:0806.2301}}].

\bibitem{Bonciani:2009nb}
R.~Bonciani, A.~Ferroglia, T.~Gehrmann, and C.~Studerus, {\it {Two-Loop Planar
  Corrections to Heavy-Quark Pair Production in the Quark-Antiquark Channel}},
  {\em JHEP} {\bf 0908} (2009) 067,
  [\href{http://xxx.lanl.gov/abs/0906.3671}{{\tt arXiv:0906.3671}}].

\bibitem{Bonciani:2010mn}
R.~Bonciani, A.~Ferroglia, T.~Gehrmann, A.~Manteuffel, and C.~Studerus, {\it
  {Two-Loop Leading Color Corrections to Heavy-Quark Pair Production in the
  Gluon Fusion Channel}},  {\em JHEP} {\bf 1101} (2011) 102,
  [\href{http://xxx.lanl.gov/abs/1011.6661}{{\tt arXiv:1011.6661}}].

\bibitem{vonManteuffel:2013uoa}
A.~von Manteuffel and C.~Studerus, {\it {Massive planar and non-planar double
  box integrals for light $N_f$ contributions to $gg \to tt$}},
  \href{http://xxx.lanl.gov/abs/1306.3504}{{\tt arXiv:1306.3504}}.

\bibitem{Brown:2008um}
F.~Brown, {\it {The Massless higher-loop two-point function}},  {\em
  Commun.Math.Phys.} {\bf 287} (2009) 925--958,
  [\href{http://xxx.lanl.gov/abs/0804.1660}{{\tt arXiv:0804.1660}}].

\bibitem{Duhr:2012fh}
C.~Duhr, {\it {Hopf algebras, coproducts and symbols: an application to Higgs
  boson amplitudes}},  {\em JHEP} {\bf 1208} (2012) 043,
  [\href{http://xxx.lanl.gov/abs/1203.0454}{{\tt arXiv:1203.0454}}].

\bibitem{Anastasiou:2013srw}
C.~Anastasiou, C.~Duhr, F.~Dulat, and B.~Mistlberger, {\it {Soft triple-real
  radiation for Higgs production at N$^3$LO}},
  \href{http://xxx.lanl.gov/abs/1302.4379}{{\tt arXiv:1302.4379}}.

\bibitem{Vermaseren:2000nd}
J.~Vermaseren, {\it {New features of FORM}},
  \href{http://xxx.lanl.gov/abs/math-ph/0010025}{{\tt math-ph/0010025}}.

\bibitem{mathematica}
W.~Research, {\em {Mathematica}}.
\newblock Wolfram Reserach, Champaign, Illinois, USA, 9.0~ed., 2013.

\bibitem{Henn:2013pwa}
J.~M. Henn, {\it {Multiloop integrals in dimensional regularization made
  simple}},  {\em Phys.Rev.Lett.} {\bf 110} (2013) 251601, [\href{http://xxx.lanl.gov/abs/1304.1806}{{\tt arXiv:1304.1806}}].

\bibitem{Tausk:1999vh}
J.~Tausk, {\it {Nonplanar massless two loop Feynman diagrams with four on-shell
  legs}},  {\em Phys.Lett.} {\bf B469} (1999) 225--234,
  [\href{http://xxx.lanl.gov/abs/hep-ph/9909506}{{\tt hep-ph/9909506}}].

\bibitem{Gehrmann:2002zr}
T.~Gehrmann and E.~Remiddi, {\it {Analytic continuation of massless two loop
  four point functions}},  {\em Nucl.Phys.} {\bf B640} (2002) 379--411,
  [\href{http://xxx.lanl.gov/abs/hep-ph/0207020}{{\tt hep-ph/0207020}}].

\bibitem{Zagier}
D.~Zagier, {\it {Polylogarithms, Dedekind zeta functions and the algebraic
  $K$-theory of fields}},  in {\em Arithmetic Algebraic Geometry}
  (J.~G.v.d.Geer, F.Oort, ed.), vol.~Prog. Math. 89, pp.~391--430.,
  {Birkh\"auser}, 1991.

\bibitem{Goncharov}
A.~B. Goncharov, {\it {Geometry of configurations, polylogarithms, and motivic
  cohomology}},  {\em Adv. Math.} {\bf 114} (1995), no.~2 197--318.

\bibitem{Goncharov:2010jf}
A.~B. Goncharov, M.~Spradlin, C.~Vergu, and A.~Volovich, {\it {Classical
  Polylogarithms for Amplitudes and Wilson Loops}},  {\em Phys.Rev.Lett.} {\bf
  105} (2010) 151605, [\href{http://xxx.lanl.gov/abs/1006.5703}{{\tt
  arXiv:1006.5703}}].

\bibitem{Goncharov:2005sla}
A.~Goncharov, {\it {Galois symmetries of fundamental groupoids and
  noncommutative geometry}},  {\em Duke Math.J.} {\bf 128} (2005) 209,
  [\href{http://xxx.lanl.gov/abs/math/0208144}{{\tt math/0208144}}].

\bibitem{Duhr:2011zq}
C.~Duhr, H.~Gangl, and J.~R. Rhodes, {\it {From polygons and symbols to
  polylogarithmic functions}},  {\em JHEP} {\bf 1210} (2012) 075,
  [\href{http://xxx.lanl.gov/abs/1110.0458}{{\tt arXiv:1110.0458}}].

\bibitem{Remiddi:1999ew}
E.~Remiddi and J.~Vermaseren, {\it {Harmonic polylogarithms}},  {\em
  Int.J.Mod.Phys.} {\bf A15} (2000) 725--754,
  [\href{http://xxx.lanl.gov/abs/hep-ph/9905237}{{\tt hep-ph/9905237}}].

\bibitem{Vollinga:2004sn}
J.~Vollinga and S.~Weinzierl, {\it {Numerical evaluation of multiple
  polylogarithms}},  {\em Comput.Phys.Commun.} {\bf 167} (2005) 177,
  [\href{http://xxx.lanl.gov/abs/hep-ph/0410259}{{\tt hep-ph/0410259}}].

\bibitem{2001math.3059G}
A.~Goncharov, {\it {Multiple polylogarithms and mixed Tate motives}},
  \href{http://xxx.lanl.gov/abs/math/0103059}{{\tt math/0103059}}.

\bibitem{Brown:2011ik}
F.~Brown, {\it {On the decomposition of motivic multiple zeta values}},
  \href{http://xxx.lanl.gov/abs/1102.1310}{{\tt arXiv:1102.1310}}.

\bibitem{Aglietti:2004tq}
U.~Aglietti and R.~Bonciani, {\it {Master integrals with 2 and 3 massive
  propagators for the 2 loop electroweak form-factor - planar case}},  {\em
  Nucl.Phys.} {\bf B698} (2004) 277--318,
  [\href{http://xxx.lanl.gov/abs/hep-ph/0401193}{{\tt hep-ph/0401193}}].

\bibitem{Bonciani:2003te}
R.~Bonciani, P.~Mastrolia, and E.~Remiddi, {\it {Vertex diagrams for the QED
  form-factors at the two loop level}},  {\em Nucl.Phys.} {\bf B661} (2003)
  289--343, [\href{http://xxx.lanl.gov/abs/hep-ph/0301170}{{\tt
  hep-ph/0301170}}].

\bibitem{Bonciani:2003cj}
R.~Bonciani, A.~Ferroglia, P.~Mastrolia, E.~Remiddi, and J.~van~der Bij, {\it
  {Planar box diagram for the N$_F$ = 1 two loop QED virtual corrections to
  Bhabha scattering}},  {\em Nucl.Phys.} {\bf B681} (2004) 261--291,
  [\href{http://xxx.lanl.gov/abs/hep-ph/0310333}{{\tt hep-ph/0310333}}].

\bibitem{Birthwright:2004kk}
T.~Birthwright, E.W.N.~Glover, and P.~Marquard, {\it {Master integrals for massless
  two-loop vertex diagrams with three offshell legs}},  {\em JHEP} {\bf 0409}
  (2004) 042, [\href{http://xxx.lanl.gov/abs/hep-ph/0407343}{{\tt
  hep-ph/0407343}}].

\bibitem{Chavez:2012kn}
F.~Chavez and C.~Duhr, {\it {Three-mass triangle integrals and single-valued
  polylogarithms}},  {\em JHEP} {\bf 1211} (2012) 114,
  [\href{http://xxx.lanl.gov/abs/1209.2722}{{\tt arXiv:1209.2722}}].

\bibitem{Smirnov:2008py}
A.~Smirnov and M.~Tentyukov, {\it {Feynman Integral Evaluation by a Sector
  decomposiTion Approach (FIESTA)}},  {\em Comput.Phys.Commun.} {\bf 180}
  (2009) 735--746, [\href{http://xxx.lanl.gov/abs/0807.4129}{{\tt
  arXiv:0807.4129}}].

\bibitem{Borowka:2012yc}
S.~Borowka, J.~Carter, and G.~Heinrich, {\it {Numerical Evaluation of
  Multi-Loop Integrals for Arbitrary Kinematics with SecDec 2.0}},  {\em
  Comput.Phys.Commun.} {\bf 184} (2013) 396--408,
  [\href{http://xxx.lanl.gov/abs/1204.4152}{{\tt arXiv:1204.4152}}].

\bibitem{Borowka:2013cma}
S.~Borowka and G.~Heinrich, {\it {Massive non-planar two-loop four-point
  integrals with SecDec 2.1}},  \href{http://xxx.lanl.gov/abs/1303.1157}{{\tt
  arXiv:1303.1157}}.

\end{thebibliography}\endgroup

\end{document}